\documentclass[12pt,epsf]{article}
\def\hybrid{\topmargin -20pt    \oddsidemargin 0pt
        \headheight 0pt \headsep 0pt
        \textwidth 6.5in        
        \textheight 9in         
        \marginparwidth .875in
        \parskip 5pt plus 1pt   \jot = 1.5ex}

\hybrid
\makeatletter
\@addtoreset{equation}{section}
\makeatother

\newcommand{\cC}{{\cal C}}

\newcommand{\cK}{{\cal K}}

\newcommand{\cM}{{\cal M}}
\newcommand{\cN}{{\cal N}}

\newcommand{\hf}{\frac12}

\newcommand{\bea}{\begin{eqnarray}}
\newcommand{\eea}{\end{eqnarray}}
\newcommand{\be}{\begin{equation}}
\newcommand{\ee}{\end{equation}}
\newcommand{\bt}{\begin{tabular}}
\newcommand{\et}{\end{tabular}}
\newcommand{\ba}{\begin{array}}
\newcommand{\ea}{\end{array}}

\newcommand{\bmat}{\left(\begin{array}}
\newcommand{\emat}{\end{array}\right)}

\newcommand{\intmod}[1]{\left\lceil{#1}\right\rceil}
\newcommand{\Tr}{\mathop{\rm Tr}}

\def\beq{\begin{equation}}
\def\eeq{\end{equation}}
\def\beqa{\begin{eqnarray}}
\def\eeqa{\end{eqnarray}}

\def\NPB#1#2#3{Nucl. Phys. B{#1} (19#2) #3}
\def\PLB#1#2#3{Phys. Lett. B{#1} (19#2) #3}

\def\PRD#1#2#3{Phys. Rev. D{#1} (19#2) #3}

\def\JGP#1#2#3{J. Geom. Phys. {#1} (19#2) #3}

\def\bfm#1{\mbox{\boldmath$#1$}}

\def\N#1{$\cN=#1$}
\def\D#1{$D=#1$}

\def\yzero{\smash{\hbox{$y\kern-4pt\raise1pt\hbox{${}^\circ$}$}}}
\def\eps{\epsilon}

\def\s2{\frac{1}{\sqrt2}}

\def\bZ{{\mathbf Z}}
\def\ZNM{$\IZ_N\times\IZ_M$\ }
\def\half{\frac12}
\def\th{^{\rm th}}
\def\ssum#1#2{{\textstyle\sum\limits_{#1}^{#2}}}
\def\sprod#1#2{{\textstyle\prod\limits_{#1}^{#2}}}

\def\aR{\alpha^{\rm R}}
\def\aNS{\alpha^{\rm NS}}
\def\sus{\hbox{(susy)}}
\def\nsus{\hbox{(non-susy)}}
\def\spin{\hbox{spinors: }}
\def\scal{\hbox{scalars: }}
\def\smult{\cN=1\hbox{ mult.: }}

\def\ie{i.e.\ }
\def\eg{e.g.\ }

\def\IB{\relax{\rm I\kern-.18em B}}
\def\ID{\relax{\rm I\kern-.18em D}}
\def\IE{\relax{\rm I\kern-.18em E}}
\def\IF{\relax{\rm I\kern-.18em F}}
\def\IH{\relax{\rm I\kern-.18em H}}
\def\II{\relax{\rm I\kern-.18em I}}
\def\IK{\relax{\rm I\kern-.18em K}}
\def\IL{\relax{\rm I\kern-.18em L}}
\def\IM{\relax{\rm I\kern-.18em M}}
\def\IN{\relax{\rm I\kern-.18em N}}
\def\IP{\relax{\rm I\kern-.18em P}}
\def\IR{\relax{\rm I\kern-.18em R}}
\def\IT{\relax{\rm I\kern-.42em T}}
\def\IZ{\relax{\hbox{\raisebox{.38ex}
    {\scriptsize\bfseries\slshape /}\kern-.40em\_\kern-.28em\rm Z}}}
\def\Iz{\relax{\hbox{\raisebox{.38ex}
    {\tiny\bfseries\slshape /}\kern-.25em\raisebox{.65ex}
    {\tiny\bfseries\slshape /}\kern-.43em\_\kern-.26em\rm Z}}}
\def\inbar{\vrule height1.5ex width.8pt depth-0.2pt}
\def\inbarhi{\vrule height1.55ex width.5pt depth-.85ex}
\def\inbarlo{\vrule height.8ex width.5pt depth0ex}
\def\IC{\relax{\rm C\kern-.48em \inbar\kern.48em}}
\def\IO{\relax{\rm O\kern-.56em \inbar\kern.56em}}
\def\IQ{\relax{\rm Q\kern-.56em \inbar\kern.56em}}
\def\IS{\relax{\rm S\kern-.37em \inbarhi\kern.08em\inbarlo\kern.29em}}

\def \one{\relax{\rm 1\kern-.26em I}}

\def\Dsl{\,\raise.15ex\hbox{/}\mkern-13.5mu D} 

\def\cp#1{\relax\ifmmode {\IP\kern-2pt{}_{#1}}\else $\IP\kern-2pt{}_{#1}$\fi}

\newcommand{\drawsquare}[2]{\hbox{%
\rule{#2pt}{#1pt}\hskip-#2pt
\rule{#1pt}{#2pt}\hskip-#1pt
\rule[#1pt]{#1pt}{#2pt}}\rule[#1pt]{#2pt}{#2pt}\hskip-#2pt
\rule{#2pt}{#1pt}}

\newcommand{\Yfun}{{\raisebox{-.5pt}{\drawsquare{6.5}{0.4}}}}
\newcommand{\Ysym}{{\raisebox{-.5pt}{\drawsquare{6.5}{0.4}}\hskip-0.4pt%
     \raisebox{-.5pt}{\drawsquare{6.5}{0.4}}}}
\newcommand{\Yasym}{{\raisebox{-3.5pt}{\drawsquare{6.5}{0.4}}\hskip-6.9pt%
     \raisebox{3pt}{\drawsquare{6.5}{0.4}}}}
\def \Yfunb{\overline\Yfun}
\def \Ysymb{\overline\Ysym}
\def \Yasymb{\overline\Yasym}

\def \nonsusy{non\discretionary{-}{}{-}super\-sym\-met\-ric\ }

\def \nonab{non\discretionary{-}{}{-}Abel\-ian\ }
\def \nonvan{non\discretionary{-}{}{-}van\-ish\-ing\ }
\def \noneq{non\discretionary{-}{}{-}equi\-va\-lent\ }
\def \nonze{non\discretionary{-}{}{-}ze\-ro\ }
\def \nontriv{non\discretionary{-}{}{-}triv\-i\-al\ }

\begin{document}

\pagestyle{empty}
\renewcommand{\thefootnote}{\fnsymbol{footnote}}
\rightline{FTUAM-00/19, IFT-UAM/CSIC-00-24}
\rightline{\tt hep-th/0008173}
\vspace{0.5cm}
\begin{center}
\LARGE{\bf $\bZ_N\times\bZ_M$ orientifolds with 
        and without discrete torsion\\[20mm]}

\large{M.~Klein, R.~Rabad\'an\\[5mm]}

\small{Departamento de F\'{\i}sica Te\'orica C-XI
       and Instituto de F\'{\i}sica Te\'orica  C-XVI,\\[-0.3em]
       Universidad Aut\'onoma de Madrid,
       Cantoblanco, 28049 Madrid, Spain.\footnote{E-mail: 
       m.klein@damtp.cam.ac.uk, rabadan@delta.ft.uam.es}\\[20mm]}

\small{\bf Abstract} \\[7mm]
\end{center}

\begin{center}
\begin{minipage}[h]{14.0cm}

{\small
We discuss compact four-dimensional $\bZ_N\times\bZ_M$ type IIB orientifolds. 
We take a systematic approach to classify the possible models and construct 
them explicitely. The supersymmetric orientifolds of this type have
already been constructed some time ago. We find that there exist several
consistent orientifolds for each of the discrete groups $\bZ_2\times\bZ_2$, 
$\bZ_2\times\bZ_4$, $\bZ_4\times\bZ_4$, $\bZ_2\times\bZ_6$, 
$\bZ_2\times\bZ_6'$ and $\bZ_6\times\bZ_6$ if anti-$D5$-branes are
introduced. Supersymmetry is broken by the open strings ending on
antibranes. 
The rank of the gauge group is reduced by a factor two if the
underlying orbifold space has discrete torsion. 
}

\end{minipage}
\end{center}
\newpage

\setcounter{page}{1}
\pagestyle{plain}
\renewcommand{\thefootnote}{\arabic{footnote}}
\setcounter{footnote}{0}


\section{Introduction}

Type IIB orientifolds \cite{sagnotti,gp} provide a promising framework
for semi-realistic model building. \D4, \N1 type IIB orientifolds are
well studied by now \cite{bl,abpss,ks,z,afiv,k,cpw,bgk,kr2}. In
contrast to heterotic orbifolds, the action of the orbifold group on the
gauge degrees of freedom of orientifolds is rather constrained and in
many cases even uniquely determined due to the RR tadpole cancellation
conditions. For some discrete groups that lead to \D4, \N1 heterotic
orbifolds there is no consistent supersymmetric orientifold at all.
Some of the tadpole conditions vanish
in the limit where the internal orbifold space gets very large. This
allows for many additional consistent orientifolds. An important subset
of non-compact orientifolds were classified in \cite{kr1}. 
The action of the orientifold group $\Gamma\times\{\one,\Omega\}$ on 
the Chan-Paton matrices is specified by choosing a projective real 
representation $\gamma$ of the orbifold group $\Gamma$ \cite{bi,kr1}. 
If $\Gamma$ contains elements $g$ of even order, \ie the smallest 
positive integer $N$, such that $g^N=e$, is even, where $e$ is the 
neutral element of $\Gamma$, then there are two inequivalent choices for the 
representation matrix: $\gamma_g^N = \mu\,\one$, with $\mu=\pm1$. 
Orientifold models with $\mu=+1$ ($\mu=-1$) have been called to have (no) 
vector structure in \cite{blpssw}. If $\Gamma$ contains two generators 
$g$, $h$ of even even order, then it is possible to have non-commuting
$\gamma$ matrices: $\gamma_g\gamma_h=\eps\,\gamma_h\gamma_g$. Orientifolds
with $\eps=-1$ have discrete torsion in the sense of \cite{v,df}.
Thus, to each orbifold group $\Gamma$ with two generators of even order, 
there correspond eight non-compact orientifold models, characterised by
the three signs $\mu_g$, $\mu_h$ and $\eps$. If the internal space is 
compact and only $D$-branes with positive RR charge are present, then
only the choice $\mu_g=\mu_h=\eps=-1$ is consistent \cite{bl,z}. It is 
not possible to cancel the tadpoles for the remaining seven models.

A striking feature of orientifolds is the existence of a natural mechanism
to break supersymmetry by introducing antibranes \cite{ads,au}. The negative
RR charge changes the GSO projection for open strings stretched between
branes and antibranes, and in the antibrane-antibrane R sector it changes
the $\Omega$-projection. As a consequence, the fermionic spectrum generically 
differs from the bosonic spectrum in the antibrane sectors of the orientifold.
This has led to orientifold models that come very close to the \nonsusy 
Standard Model \cite{aiq}. The presence of two types of $D$-branes, with 
positive and negative RR charge, gives rise to many new consistent 
orientifolds. Indeed, the tadpoles of most of the non-compact models 
constructed in \cite{kr1} can also be cancelled in the compact case if 
antibranes are introduced. The \nonsusy $\IZ_2\times\IZ_2$ orientifold 
models have been analysed in \cite{aaads}. A supersymmetric version of
these orientifolds is possible if $D5$-branes with negative NSNS and RR 
charge inside $D9$-branes with positive charges are introduced \cite{kr2}.

The aim of this paper is to classify the possible \ZNM orientifolds 
with \N1 supersymmetry in the closed string sector and
to determine their spectra explicitely. This can be viewed as a 
generalisation of the work of \cite{fiq} to type IIB orientifolds.
For discrete groups $\IZ_N\times\IZ_M$, with $N$ and $M$ odd, there is
only one compact orientifold: the supersymmetric $\IZ_3\times\IZ_3$ 
constructed in the second ref.\ of \cite{ks}. For each of the two 
discrete groups $\IZ_2\times\IZ_3$ and $\IZ_3\times\IZ_6$, there are 
two orientifold models, with and without vector structure. 
The cases without vector structure contain only positively
charged $D$-branes. They have been constructed in \cite{z}.
The $\IZ_2\times\IZ_3\cong\IZ_6'$ orientifold with vector structure
has been analysed in \cite{kr2}. 
Therefore, we restrict ourselves to even $N$ and $M$ in this article.

As stated above, there are eight orientifold models for each discrete group.
However only six of them are inequivalent for $\IZ_2\times\IZ_4$ and
$\IZ_2\times\IZ_6$, and only four of them are inequivalent for 
$\IZ_2\times\IZ_2$, $\IZ_4\times\IZ_4$, $\IZ_2\times\IZ_6'$ and
$\IZ_6\times\IZ_6$. The orbifold groups containing a $\IZ_4$ factor are
special in some respects. It is not possible to build the corresponding
orientifolds if only positively charged $D$-branes are present \cite{z}.
The reason is that there exists no projective representation of 
$\IZ_N\times\IZ_4$ on the $D9$-branes and $D5_i$-branes, $i=1,2,3$, that
respects all the consistency constraints. We will see below that the
the tadpole amplitudes only factorise if the charges $\alpha_i$ of the
three sets of $D5_i$-branes satisfy $\alpha_1\alpha_2\alpha_3=-1$. 
Thus, all consistent $\IZ_N\times\IZ_4$ orientifolds necessarily involve
$D5_i$-branes with negative RR charge. Surprisingly, we find that even
if antibranes are introduced, only the models with vector structure in
the $\IZ_4$ factor have a solution to the tadpole equations.\footnote{This
statement is valid if no Wilson lines are present. There might well be
consistent $\bZ_N\times\bZ_4$ orientifolds without vector structure in
the $\bZ_4$ factor when appropriate Wilson lines are added. In this
article we only consider models without Wilson lines.}

There are many phase factors (more precisely, signs) that appear
in type IIB orientifolds. In section \ref{phasefac}, we explain their
significance and classify the inequivalent models. In particular, we
discuss the relations between vector structure, discrete torsion,
$D$-brane charges and $O$-plane charges.
In section \ref{construct}, we review and generalise a straightforward
method to construct supersymmetric and \nonsusy type IIB orientifolds. 
The closed string spectrum is determined using the orbifold cohomology 
\cite{vw,klein,kr1}. To obtain the open string spectrum, we generalise 
the algorithm developed in \cite{kr2} based on quiver theory \cite{dm} 
and on the formalism of \cite{bi}. We also give the general formulae for
the tadpole cancellation conditions including all the possible sign
factors. The algorithm described provides a useful tool for orientifold
model building, especially if implemented in a computer algebra program.
The complete spectra of the models discussed are presented in section 
\ref{descript}.
In two appendices, we explain how to obtain the Hodge numbers of orbifolds
with and without discrete torsion using the cohomology and the closed string
spectrum using the shift formalism.

\section{Phase factors in orientifolds}
\label{phasefac}

Due to the fact that the $\gamma$ matrices form a {\it projective}
representation of the orbifold group $\Gamma$ \cite{df}, there appear 
many phase factors in the expressions relevant to type IIB orientifolds. 
These phases are normally chosen to take some convenient values.

In this section, we want to answer the question how many inequivalent
choices there are. To this end, we first include all possible phase factors
and then identify the factor systems that lead to equivalent models.

\subsection{Discrete torsion on orbifolds and orientifolds}
\label{disctor}

We briefly review some basic results about discrete torsion.\footnote{For
a recent geometric treatment of discrete torsion, see \cite{sharpe}.}
In the closed
string sector of orbifold theories, discrete torsion appears as phases 
$\beta_{g,h}$ in the one-loop partition function \cite{v}:
\be \label{partfunc}
Z={1\over|\Gamma|}\sum_{g,h\in\Gamma}\beta_{g,h}\,Z(g,h),
\ee
where $Z(g,h)$ is the contribution of the $(g,h)$-twisted sector 
and $\Gamma$ is the orbifold group. The discrete torsion phases must
satisfy
\be \label{betacon}
\beta_{g,g}=1,\qquad \beta_{g,h}=\beta_{h,g}^{-1},\qquad
\beta_{g,hk}=\beta_{g,h}\beta_{g,k}\qquad \forall g,h,k\in\Gamma,
\ee
to be consistent with modular invariance \cite{v}. If we restrict ourselves 
to Abelian orbifold groups and to complex three-dimensional internal spaces, 
then discrete torsion is only possible for $\Gamma=\IZ_N\times\IZ_M$. 
Let $g_1,g_2$ be the generators of $\IZ_N,\IZ_M$ respectively, $p=\gcd(N,M)$ 
and $\eps=e^{2\pi im/p}$, with $m=1,\ldots,p$. If we choose 
$\beta_{g_1,g_2}=\eps$, then all the other phases $\beta_{g,h}$ are 
fixed by (\ref{betacon}). Introducing the notation $(a,b)=g_1^ag_2^b$ 
for the elements of $\Gamma$, they read
\be \label{beta_gh}
\beta_{(a,b),(c,d)}=\eps^{ad-bc}.
\ee
In the following, we will characterise the discrete torsion either by the
parameter $\eps$ of (\ref{beta_gh}) or by the number $s$, defined
as the smallest positive integer such that $\eps^s=1$.

The matrices $\gamma_g$ that represent the action of the elements $g$ of 
the orbifold group $\Gamma$ on the Chan-Paton indices of the open strings 
form a projective representation: 
\be  \label{gamma_gh}
\gamma_g\gamma_h=\alpha_{g,h}\gamma_{gh},
\ee
where $\alpha_{g,h}$ are arbitrary \nonze complex numbers. They are called 
the factor system of the projective representation of $\Gamma$. Discrete
torsion in the open string sector means that the matrices $\gamma_g$ do not
commute. More precisely \cite{df,g}, this non-commutativity is controlled by 
the phases $\beta_{g,h}$ defined above
\be  \label{gamg_gamh}
\gamma_g\gamma_h=\beta_{g,h}\gamma_{h}\gamma_g.
\ee
Two matrices $\gamma_g$ and $\hat\gamma_g$ are considered projectively 
equivalent if there exists a \nonze complex number $\rho_g$ such that 
$\hat\gamma_g=\rho_g\gamma_g$. In general, the projective representations
$\gamma$ and $\hat\gamma$ have
different factor systems, but they have the same discrete torsion phases
$\beta_{g,h}$. From (\ref{gamma_gh}) and (\ref{gamg_gamh}), we find
\be  \label{beta_alpha}
\beta_{g,h}=\alpha_{g,h}\alpha_{h,g}^{-1}.
\ee
To perform explicit calculations, it is convenient to choose a specific
factor system. We choose the $\alpha_{g,h}$ in (\ref{gamma_gh}) such that
\be  \label{fs_choice}
(\gamma_{g_1})^a(\gamma_{g_2})^b=\gamma_{g_1^ag_2^b}, \qquad
a=1,\ldots,N,\quad b=1,\ldots,M,
\ee
where $g_1$, $g_2$ are the generators of $\IZ_N\times\IZ_M$. This choice
is possible for each equivalence class of factor systems. Moreover, all
the remaining phases $\alpha_{g,h}$ are then determined by the parameter
$\eps$. We find
\be  \label{gam_relation}
\gamma_{g_1^ag_2^b}\gamma_{g_1^cg_2^d}
  =\eps^{-bc}\,\gamma_{g_1^{a+c}g_2^{b+d}}.
\ee

We conclude that for a $\IZ_N$ orbifold all the possible choices of factor
systems $\alpha_{g,h}$ in (\ref{gamg_gamh}) are equivalent. For 
$\IZ_N\times\IZ_M$ orbifolds, there are $p/2+1$ (if $p$ is even) or
$(p+1)/2$ (if $p$ is odd) \noneq choices, parametrised by 
$\eps=e^{2\pi im/p}$, $m=0,\ldots,\intmod{p/2}$, where $p=\gcd(N,M)$.
(Note that $\eps$ and $\eps^{-1}$ lead to equivalent orbifolds.)

In orientifold models, the matrices $\gamma_g$ have to satisfy the
additional constraint \cite{afiv,kr1}
\be \label{gamma_real}
\gamma_\Omega\gamma_g^*\gamma_\Omega^{-1} \ =\  c\,\delta_g\,\gamma_g, 
\ee
where $\gamma_\Omega=c\,\gamma_\Omega^\top$ is the matrix that
represents the action of the world-sheet parity $\Omega$ on the Chan-Paton
indices and $\delta_g$ is a new phase, defined by
\be  \label{def_delta}
\gamma_{\Omega g}^{-1}\gamma_{\Omega g}^\top = \delta_g(\gamma_g^\top)^2.
\ee
It is easy to see that one can always choose $\delta_g=c\ \forall\,g$.
Consider the redefinition $\gamma_g\:\to\:\hat\gamma_g=\rho_g\gamma_g$,
where the phases $\rho_g$ satisfy $\rho_g\rho_h=\rho_{gh}$. This
redefinition leaves (\ref{gamma_gh}) and (\ref{fs_choice}) invariant and 
therefore does not modify the factor system $\alpha_{g,h}$. Now, choosing 
$\rho_g=\sqrt{c\,\delta_g}$ and using $c=\pm1$, we find from 
(\ref{gamma_real}) that $\hat\delta_g=c$.

The condition (\ref{gamma_real}) together with $c\,\delta_g=1$ tells us
that the matrices $\gamma_g$ form a {\it real} or {\it pseudo-real} projective
representation of $\Gamma$ \cite{bi,kr1}.

In general, there are several sets of $Dp$-branes, $p=9,5_1,5_2,5_3$,
and correspondingly several sets of $\gamma$ matrices, $\gamma_{g,p}$,
$\gamma_{\Omega,p}$. It turns out (\eg when considering tadpole cancellation)
that the discrete torsion parameter $\eps$ cannot be chosen independently
for the different sets of $\gamma$ matrices. There is only one $\eps$ which
is the same for all $p$. Furthermore, the coefficients $c_p$, defined by
$\gamma_{\Omega,p}^\top=c_p\,\gamma_{\Omega,p}$, cannot be chosen at will.
Tadpole cancellation requires\footnote{if no anti-$D9$-branes are present} 
$c_9=1$ and a consistent coupling of the strings stretching between 9- and
$5_i$-branes implies \cite{bl}
\be  \label{cp_constr}
c_{5_1}=c_{5_2}=c_{5_3}=-c_9.
\ee
These constraints can be circumvented in models similar to the one proposed
by the authors of \cite{dp,bz}. However, such models are only consistent if 
all matter at integer mass levels in the $95_i$ sectors is projected out. 
We will therefore discard this possibility in the following.

As a consequence of the reality condition (\ref{gamma_real}), the discrete 
torsion parameter $\eps$ can only take the real values $\pm 1$ in the 
orientifold case \cite{kr1}. At this point, we might conclude that there
are only two \noneq orientifold models corresponding to each discrete
group $\IZ_N\times\IZ_M$. But we will see in the next subsection that there
are more possibilities.

\subsection{Vector structure}
\label{vecstruc}

For $\Gamma=\IZ_N$, the identity $g_1^N=e$, where $g_1$ is the generator and
$e$ the neutral element of $\IZ_N$, translates to
\be  \label{def_mu}
(\gamma_{g_1})^N=\mu\,\one,\qquad {\rm with}\ \mu\in\IC^*\ {\rm (orbifold)}
\quad {\rm or}\quad \mu\in\IR^*\ {\rm (orientifold)}.
\ee
In the orbifold case, this can be brought to the form 
$(\hat\gamma_{g_1})^N=\one$ by redefining the $\gamma$ matrices.
In the orientifold case, such a redefinition is only possible for
odd $N$ ($\gamma_{g_1}\,\to\,\hat\gamma_{g_1}={\rm sgn}(\mu)|\mu|^{-1/N}
\gamma_{g_1}$). If $N$ is even, the two cases $\mu=+1$ and $\mu=-1$ 
are not equivalent because $(\gamma_{g_1})^N=-\one$ cannot be brought
to the form $(\hat\gamma_{g_1})^N=\one$ by multiplying $\gamma_{g_1}$ 
with a real number. For the factor system of (\ref{fs_choice}), this
means that $\gamma_{g_1^N}=\mu\,\gamma_e$. If $\mu=-1$, then the matrices
$\gamma_g$ represent the elements of $\Gamma$ only 2:1. This class of
orientifolds has been called to have no vector structure in \cite{blpssw}.

For $\IZ_N\times\IZ_M$ orientifolds, the notion of vector structure can
be defined for each of the two factors of the discrete group separately.%
\footnote{Strictly speaking, a $\bZ_N\times\bZ_M$ orientifold only has
vector structure in the sense of \cite{blpssw,w} if
$(\gamma_{g_1})^N=(\gamma_{g_2})^M=\one$ and $\eps=1$. We use the word
`vector structure' in a generalised sense.}
There are four inequivalent models, characterised by $\mu_{3,p}=\pm1$,
$\mu_{1,p}\pm1$, where
\be  \label{def_mu_i}
(\gamma_{g_1,p})^N=\mu_{3,p}\,\one,\qquad 
(\gamma_{g_2,p})^M=\mu_{1,p}\,\one.
\ee
The indices $i=1,3$ on $\mu_{i,p}$ refer to the fact that, in our conventions,
$g_1$ fixes the third complex plane and $g_2$ fixes the first complex
plane. The index $p=9,5_1,5_2,5_3$ indicates which set of $Dp$-branes
is considered. We will see below that the signs $\mu_{i,p}$ cannot be
chosen independently for each set of $Dp$-branes. Indeed, the choice of
$\mu_{1,9}$ and $\mu_{3,9}$ fixes all other $\mu_{i,p}$.

We conclude that to each $\IZ_N\times\IZ_M$ orientifold there correspond 
eight distinct models characterised by the three signs $\mu_{1,9}$, 
$\mu_{3,9}$ and $\eps$.

\subsection{Phases in the open string one-loop amplitudes}
The three open string one-loop amplitudes that contribute to the tadpoles ---
the cylinder, the Klein bottle and the M\"obius strip ---
contain additional signs. However, the requirement of tadpole cancellation
relates these signs to $\mu_{1,9}$, $\mu_{3,9}$ and $\eps$.

In the following, we label the elements of $\IZ_N\times\IZ_M$ by 
the two-vector $\bar k=(a,b)$, $a=0,\ldots,N-1$, $b=0,\ldots,M-1$.
We define $s_i=\sin(\pi\bar k\cdot\bar v_i)$, 
$c_i=\cos(\pi\bar k\cdot\bar v_i)$ 
and $\tilde s_i=\sin(2\pi\bar k\cdot\bar v_i)$, as in section \ref{tadpoles}
below.

The RR part of the cylinder amplitude can be written in the form
\cite{kr1,kr2}:
\be  \label{cylinder}
\cC\ =\ \sum_{\bar k=(0,0)}^{(N-1,M-1)}\cC_{(\bar k)}\ =\ 
        \sum_{\bar k=(0,0)}^{(N-1,M-1)} \frac{1}{8s_1s_2s_3} 
          \left[\Tr\gamma_{\bar k,9} 
          + 4\ssum{i=1}{3}\alpha_i\,s_j s_k\Tr\gamma_{\bar k,5_i}
         \right]^2,
\ee
where we suppressed the volume dependence.\footnote{In sectors with fixed
planes, this formula has to be modified as explained in appendix B of 
\cite{kr2}.} The indices $j$, $k$ take values such that $(ijk)$ is a 
permutation of $(123)$. The signs $\alpha_i$ weight the $95_i$ sectors
relative to the 99 and $5_i5_i$ sectors. They are related to the $D$-brane
charges. If the $D9$-branes have positive RR charge, then $\alpha_i$ is the
RR charge of the $D5_i$-branes.

The RR part of the Klein bottle amplitude can be written in the form:
\be  \label{Klein_bottle}
\cK = \sum_{\bar k=(0,0)}^{(N-1,M-1)}\cK_{(\bar k)}
    = \sum_{\bar k=(0,0)}^{(N-1,M-1)}\left[ 
      16\,\sprod{i=1}{3} \frac{2 c_i^2}{\tilde{s_i}}
      -16\,\ssum{i=1}{3} \eps_i\,\frac{2 c_i^2}{\tilde s_i} \right].
\ee
The sign $\epsilon_i$ weights the $i\th$ order-two twisted sector relative
to the untwisted sector.\footnote{These signs were introduced in the 
twisted Klein bottle amplitude by the authors of \cite{aaads}. Note that the 
$\eps_i$ of \cite{aaads} are $-\eps_i$ in our notation.} It is related to the 
$O$-plane charges. If the $O9$-planes have negative RR charge, then $\eps_i$ 
is (half) the RR charge of the $O5_i$-planes.

The RR part of the M\"obius strip amplitude can be written in the form:
\bea  \label{Moebius_strip}
\cM &= &\sum_{\bar k=(0,0)}^{(N-1,M-1)} \cM_{(\bar k)} \\
    &= &\sum_{\bar k=(0,0)}^{(N-1,M-1)} \left[
     - 8\,\sprod{i=1}{3}\frac{1}{2s_i}  
        \Tr\left(\gamma^{-1}_{\Omega{\bar k},9}
                 \gamma^\top_{\Omega{\bar k},9}\right)
       -8\,\ssum{i=1}{3}\frac{2 c_j c_k}{s_i}\,(-\alpha_i) 
        \Tr\left(\gamma^{-1}_{\Omega{\bar k},5_i}
                 \gamma^\top_{\Omega{\bar k},5_i}\right)  \right]. \nonumber
\eea

From the interpretation of these open string one-loop amplitudes as
closed string tree-level exchange between $D$-branes and $O$-planes,
it is clear that the tadpoles must factorise, \ie $\cC+\cK+\cM$ can 
be written as a sum of squares. More precisely, we need
\be  \label{factorise}
\cC_{(2\bar k)}+\cK_{(\bar k)}+\sum_{i=1}^3\cK_{(\bar k+\bar k_i)}
+\cM_{(\bar k)}+\sum_{i=1}^3\cM_{(\bar k+\bar k_i)}\ =\ [\ldots]^2,
\ee
where $\bar k_i$ denotes the order-two twist that fixes the $i\th$
complex plane.
This implies a relation between the $\eps_i$ and the vector structures
and discrete torsion, defined above. We find \cite{kr1}
\be  \label{fac_cond}
\eps_i=\mu_{i,9}=\mu_{i,5_i}=-\mu_{i,5_j},\quad i=1,3,\ j\neq i,\qquad
\eps_2=\eps_1\eps_3\,\eps^{-MN/4}.
\ee
A further constraint on the signs arises from the requirement of
tadpole cancellation in the untwisted sector:
\be  \label{eps_alpha}
\eps_i=-\alpha_i,\quad i=1,2,3.
\ee
This means that the tadpoles can only be cancelled if the RR charge of
the $O5_i$-planes ($\eps_i$) is opposite to the RR charge of the 
$D5_i$-branes ($\alpha_i$).

We conclude that the signs $\alpha_i$, $\eps_i$ are not independent
parameters. As stated above, there are eight distinct $\IZ_N\times\IZ_M$
orientifolds. They can be characterised by $\mu_{1,9}$, $\mu_{3,9}$ and 
$\eps$. Note that the three signs $\eps_i$ are only independent if $N$ 
and $M$ are not multiples of four.

\section{Construction of the models}
\label{construct}
We consider compact orientifolds of the form 
$T^6/(\Gamma\times\{\one,\Omega\})$, with $\Gamma=\IZ_N\times\IZ_M$, 
$N,M$ even. The six-torus is defined as $T^6=\IC^3/\Lambda$, with 
$\Lambda$ a factorisable lattice, \ie it is the direct sum of three 
two-dimensional lattices. The world-sheet symmetry $\Omega$ is of the form
\be \label{OmJT}
\Omega\ =\ \Omega_0\, J\, T,
\ee
where $\Omega_0$ is the world-sheet parity and the operator $J$ exchanges the
$k\th$ and the $(N-k)\th$ twisted sector \cite{p}. The operator $T$ acts as 
$-\eps_i\bfm1$ on the $g$-twisted states if $g$ is an order-two twist that 
fixes the $i\th$ complex plane and as the identity on the remaining states.%
\footnote{The analogous operator in \D6 was discussed by the authors of 
\cite{dp,p} when analysing a new $\bZ_2$ orientifold.} The signs $\eps_i$ are
defined in (\ref{Klein_bottle}).

The action of $\Gamma$ on the coordinates $(z_1,z_2,z_3)$ of $\IC^3$ can be 
characterised by the twist vectors $v=(v_1,v_2,v_3)$, $w=(w_1,w_2,w_3)$:
\be  \label{twistvec}
g_1:\ z_i \longrightarrow e^{2\pi iv_i} z_i,\qquad
g_2:\ z_i \longrightarrow e^{2\pi iw_i} z_i,
\ee
where $g_1, g_2$ are the generators of $\Gamma$ and 
$\sum_{i=1}^3v_i=\sum_{i=1}^3w_i=0$ to ensure \N1 supersymmetry of the
closed string sector in \D4. 
Not all possible shifts correspond to a symmetry of some lattice. Indeed, 
there is a finite number of $\IZ_N\times\IZ_M$ orbifolds \cite{fiq}. For even 
$N$ and $M$, there are only 6 models: $\IZ_2\times\IZ_2$, $\IZ_2\times\IZ_4$, 
$\IZ_4\times\IZ_4$, $\IZ_2\times\IZ_6$, $\IZ_6\times\IZ_6$, with twist vectors
$v=1/N(1,-1,0)$, $w=1/M(0,1,-1)$, and $\IZ_2\times\IZ_6'$ with twist 
vectors $v=1/N (1,-1,0)$, $w=1/M (-2,1,1)$. We chose the twist vectors such
that $g_1^{N/2}$ fixes the third complex plane and $g_2^{M/2}$ fixes the
first complex plane.

Supersymmetric compact $\IZ_N\times\IZ_M$ orientifolds have been discussed 
in \cite{bl,ks,z,k}. The author of \cite{z} found that the models with 
discrete groups $\IZ_2\times\IZ_4$ and $\IZ_4\times\IZ_4$ are not consistent 
because the algebra of $\gamma_{\Omega,p}$ and $\gamma_{g,p}$, where $p=9,5_i$,
seems to imply that $\gamma_{R,9}$ is antisymmetric, where $R$ is an 
order-two twist. It turns out that there is no projective representation
of $\IZ_N\times\IZ_M$, with $\gamma_{R,9}$ antisymmetric, if $N$ and/or 
$M$ is a multiple of four \cite{z}. However, the condition that $\gamma_{R,9}$ 
be antisymmetric can be escaped if the action of the operator $T$ (\ie the
signs $\eps_i$) in (\ref{OmJT}) is modified. We will see that there are 
solutions to the tadpole equations for all orbifold groups \ZNM if one chooses 
the appropriate signs $\eps_i$.

In this section, we sketch the basic steps to construct $\IZ_N\times\IZ_M$ 
orientifolds. We explain how to obtain the closed string spectrum, 
the open string spectrum and the tadpole cancellation conditions. This is
very similar to, and in fact a straightforward generalisation of, the method
presented in \cite{kr2}.

\subsection{Closed string spectrum}
\label{closed}

The closed string spectrum can be obtained from the cohomology of the internal
orbifold space \cite{vw,klein,kr1,kr2}.
In appendix \ref{app_hodge}, we explain in detail how to obtain these 
numbers and the explicit contribution from each twisted sector. 
The results are shown in tables \ref{ZNM_Hodge} and \ref{ZNM_Hodge_B}.
For completeness, we also give the Hodge numbers
of $\IZ_3\times\IZ_3$ and $\IZ_3\times\IZ_6$. An alternative method to
obtain the closed string spectrum is presented in appendix B.

\begin{table}[pht!]
\renewcommand{\arraystretch}{1.25}
$$\begin{array}{|c|c|c||c|c|c|c||c|}
\hline
\Gamma &\eps & &\rm untw. &\hbox{order-two} 
       &\rule[-8mm]{0mm}{18mm}
        \parbox{18mm}{fixed pl.\\[-0.5ex] but not\\[-0.5ex] order-two} 
       &\parbox{15mm}{no\\[-0.5ex] fixed pl.} 
       &\rm total\\
\hline\hline
\IZ_2 \times \IZ_2 &1 &h^{1,1}  &3  &48 &-  &-  &51\\
&&h^{2,1}                       &3  &0  &-  &-  &3\\ \cline{2-8}
&-1 &h^{1,1}                    &3  &0  &-  &-  &3\\
&&h^{2,1}                       &3  &48 &-  &-  &51\\ 
\hline
\IZ_2 \times \IZ_4 &1 &h^{1,1}  &3 &34  &8  &16 &61\\
&&h^{2,1}                       &1  &0  &0  &0  &1\\ \cline{2-8}
&-1 &h^{1,1}                    &3  &18 &0  &0  &21\\
&&h^{2,1}                       &1  &0  &8  &0  &9\\ 
\hline
\IZ_2 \times \IZ_6 &1 &h^{1,1}  &3 &22  &10 &16 &51\\
&&h^{2,1}                       &1  &0  &2  &0  &3\\ \cline{2-8}
&-1 &h^{1,1}                    &3  &0  &8  &8  &19\\
&&h^{2,1}                       &1  &14 &4  &0  &19\\ 
\hline
\IZ_2 \times \IZ_6' &1 &h^{1,1} &3 &18  &0  &15 &36\\
&&h^{2,1}                       &0  &0  &0  &0  &0\\ \cline{2-8}
&-1 &h^{1,1}                    &3  &0  &0  &12 &15\\
&&h^{2,1}                       &0  &15 &0  &0  &15\\ 
\hline
\IZ_3 \times \IZ_3 &1 &h^{1,1}  &3  &-  &54 &27 &84\\
&&h^{2,1}                       &0  &-  &0  &0  &0\\ \cline{2-8}
&e^{\pm2\pi i/3} &h^{1,1}       &3  &-  &0  &0  &3\\
&&h^{2,1}                       &0  &-  &27 &0  &27\\ 
\hline
\IZ_3 \times \IZ_6 &1 &h^{1,1}  &3  &4  &36 &30 &73\\
&&h^{2,1}                       &0  &1  &0  &0  &1\\ \cline{2-8}
&e^{\pm2\pi i/3} &h^{1,1}       &3  &4  &0  &6  &13\\
&&h^{2,1}                       &0  &1  &12 &0  &13\\ 
\hline
\end{array}$$
\caption{\label{ZNM_Hodge} Contribution from the different sectors to
   the Hodge numbers of $\bZ_N\times\bZ_M$ orbifolds with and without 
   discrete torsion.}
\end{table}

\begin{table}[ht]
\renewcommand{\arraystretch}{1.25}
$$\begin{array}{|c|c|c||c|c|c|c||c|}
\hline
\Gamma &\eps & &\rm untw. &\hbox{order-two} 
       &\rule[-8mm]{0mm}{18mm}
        \parbox{18mm}{fixed pl.\\[-0.5ex] but not\\[-0.5ex] order-two} 
       &\parbox{15mm}{no\\[-0.5ex] fixed pl.} 
       &\rm total\\
\hline\hline
\IZ_4 \times \IZ_4 &1 &h^{1,1}  &3  &27 &24 &36 &90\\
&&h^{2,1}                       &0  &0  &0  &0  &0\\ \cline{2-8}
&-1 &h^{1,1}                    &3  &27 &0  &12 &42\\
&&h^{2,1}                       &0  &0  &0  &0  &0\\ \cline{2-8} 
&\pm i &h^{1,1}                 &3  &3  &0  &0  &6\\
&&h^{2,1}                       &0  &0  &12 &0  &12\\ 
\hline
\IZ_6 \times \IZ_6 &1 &h^{1,1}  &3  &12 &30 &39 &84\\
&&h^{2,1}                       &0  &0  &0  &0  &0\\ \cline{2-8}
&-1 &h^{1,1}                    &3  &0  &24 &24 &51\\
&&h^{2,1}                       &0  &3  &0  &0  &3\\ \cline{2-8} 
&e^{\pm2\pi i/3} &h^{1,1}       &3  &12 &0  &12 &27\\
&&h^{2,1}                       &0  &0  &3  &0  &3\\ \cline{2-8} 
&e^{\pm2\pi i/6} &h^{1,1}       &3  &0  &0  &6  &9\\
&&h^{2,1}                       &0  &3  &6  &0  &9\\ 
\hline
\end{array}$$
\caption{\label{ZNM_Hodge_B} Contribution from the different sectors to
   the Hodge numbers of $\bZ_N\times\bZ_M$ orbifolds with and without 
   discrete torsion. (continued)}
\end{table}

Let us remark that the Hodge numbers of tables \ref{ZNM_Hodge} and 
\ref{ZNM_Hodge_B} directly give the closed string
spectrum of the orbifold. It is obtained by dimensionally reducing to \D4 
the massless spectrum of type IIB supergravity in \D{10}: the metric
$g$, the NSNS 2-form $B$, the dilaton $\phi$, the
RR-forms $C^{(0)}$, $C^{(2)}$, $C^{(4)}$. (We only give the bosons,
the fermions are related to them by supersymmetry.)
The Lorentz indices of the 10D fields are contracted with the differential 
forms of the internal space.
The resulting spectrum has \N2 supersymmetry in \D4. 
For a general configuration one finds (see e.g.\ \cite{kr1,lf}): 
\begin{itemize}
\item a gravity multiplet (consisting of the graviton, a vector and fermions)
\item a double tensor multiplet (consisting of two 2-forms, two scalars and
      fermions) 
\item $h^{1,1}$ tensor multiplets (consisting of a 2-form, three scalars and
      fermions)
\item $h^{2,1}$ vector multiplets (consisting of a vector, two scalars and
      fermions)
\end{itemize}

To obtain the orientifold spectrum, one has to perform the $\Omega$-projection.
In the following, we will denote an element $g=g_1^ag_2^b$ of 
$\IZ_N\times\IZ_M$ by the two-vector $\bar k=(a,b)$, \ie $(0,0)$ is the neutral
element, $(1,0)$ is the generator of $\IZ_N$, $(0,1)$ the generator of $\IZ_M$,
etc. As explained in \cite{kr1,kr2}, it is convenient to split the sectors 
into three different types:

(i) The untwisted sector ($\bar k=(0,0)$), it is invariant under $J$ and $T$.
The bosonic fields in $D=4$ are found contracting the Lorentz indices
of the $\Omega$-even 10D fields $g_{\mu\nu}$, $\phi$, $C^{(2)}_{\mu\nu}$ with
the harmonic forms corresponding to $h^{0,0}$, $h^{3,0}$, $h^{1,1}_{\rm untw}$,
$h^{2,1}_{\rm untw}$.

(ii) The order-two sectors ($\bar k_1=(0,M/2)$, $\bar k_2=(N/2,M/2)$, 
$\bar k_3=(N/2,0)$), they are invariant under $J$ but acquire
an extra sign $-\eps_i$ under the action of $T$. In the sector $\bar k_i$, 
one has to distinguish the two cases $\eps_i=\pm1$. If $\eps_i=-1$, then
the $\Omega$-even 10D fields $g_{\mu\nu}$, $\phi$, $C^{(2)}_{\mu\nu}$ are
contracted with $h^{1,1}_{\bar k_i}$, $h^{2,1}_{\bar k_i}$. If $\eps_i=+1$,
then the $\Omega$-odd 10D fields $B_{\mu\nu}$, $C^{(4)}_{\mu\nu\rho\sigma}$ are
contracted with $h^{1,1}_{\bar k_i}$, $h^{2,1}_{\bar k_i}$. 

(iii) The remaining sectors. To get the fields in $D=4$, one forms linear
combinations of the harmonic forms that belong to the $k\th$ and $(N-k)\th$
twisted sector. The $J$-even forms are contracted with the $\Omega$-even 10D 
fields and the $J$-odd forms are contracted with the $\Omega$-odd 10D fields.

The spectrum fits into $\cN=1$ supermultiplets. In total, one finds:

(i) the gravity multiplet, a linear multiplet, $(h^{1,1}+h^{2,1})_{\rm untw}$ 
chiral multiplets.

(ii) for each $i=1,2,3$: $h^{1,1}_{\bar k_i}$+$h^{2,1}_{\bar k_i}$ 
     chiral multiplets if $\eps_i=-1$ and 
     $h^{1,1}_{\bar k_i}$ linear multiplets and 
     $h^{2,1}_{\bar k_i}$ vector multiplets if $\eps_i=+1$.

(iii) if the $\bar k$-twisted sector has fixed planes, then there are 
$\hf h^{1,1}_{\bar k}$ linear multiplets, $\hf h^{2,1}_{\bar k}$ vector 
multiplets and $\hf (h^{1,1}_{\bar k}+h^{2,1}_{\bar k})$ chiral
multiplets, else one has $h^{1,1}_{\bar k}$ linear multiplets.

As explained in \cite{kr1}, orbifolds with non-real discrete torsion have
twisted sectors $\bar k$, where $h^{2,1}_{\bar k}\neq h^{1,2}_{\bar k}$.
The world-sheet symmetry $\Omega$, eq.\ (\ref{OmJT}), is not a symmetry 
of such orbifolds. Therefore, consistent orientifolds can only be constructed
from orbifolds with discrete torsion $\eps=\pm1$.

\subsection{Open string spectrum}
\label{open}

There are 32 $D9$-branes and 32 $D5_i$-branes for each $i=1,2,3$. The index
$i$ indicates that the $5_i$-branes fill the four non-compact directions and 
the $i\th$ complex plane. 

The action of $\Gamma$ on the Chan-Paton indices of the open strings is
described by a projective representation $\gamma^{(p)}$ that associates a 
$(32\times32)$-matrix $\gamma_{g,p}$ to each element $g$ of $\Gamma$, where
$p=9,5_i$ denotes the type of $D$-brane the open strings end on:
\bea  \label{def_gamma}
\gamma^{(p)}:\ \Gamma &\longrightarrow &GL(32,\IC) \\
                    g &\longmapsto     &\gamma_{g,p} \nonumber
\eea
Because of the orientifold projection, this representation must be real or
pseudo-real. In general, $\gamma^{(p)}$ can be decomposed in irreducible 
blocks of real ($R^r$), pseudo-real ($R^p$) and complex ($R^c$) representations
\cite{bi}:
\be  \label{gamma_rep}
\gamma^{(p)}=\left(\bigoplus_{l_1}n^r_{l_1}R^r_{l_1}\right)\oplus
             \left(\bigoplus_{l_2}n^p_{l_2}R^p_{l_2}\right)\oplus 
         \left(\bigoplus_{l_3}n^c_{l_3}(R^c_{l_3}\oplus\bar R^c_{l_3})\right).
\ee
In this expression, the notation $n_lR_l$ is short for $R_l\otimes\one_{n_l}$,
\ie $n_l$ is the number of copies of the irreducible representation (irrep) 
$R_l$ in $\gamma^{(p)}$ \cite{bi}. 

Let us first consider the 99 and $5_i5_i$ sectors. The projection on invariant
states of the Chan-Paton matrices $\lambda^{(0)}$ that correspond to gauge
bosons (NS sector) and their fermionic partners (R sector) in $D=4$ imposes 
the constraints (see \eg \cite{afiv,au})
\be  \label{gauge_proj}
\lambda^{(0)}=\gamma_{g,p}\lambda^{(0)}\gamma_{g,p}^{-1},\quad
\forall\,g\in\Gamma,\qquad
\lambda^{(0)}=-\alpha_p\,\gamma_{\Omega,p}{\lambda^{(0)}}^\top
               \gamma_{\Omega,p}^{-1},
\ee
where $\alpha_p$ is the charge of the $Dp$-branes. More precisely,
if we consider the fermionic (bosonic) components of $\lambda^{(0)}$
in (\ref{gauge_proj}), then $\alpha_p$ is the RR (NSNS) charge. 
The sign $\alpha_p$ in (\ref{gauge_proj}) can be explained as follows.
In \cite{s}, it was found that the action of $\Omega$ acquires an additional
minus sign in the R sector if the RR charge of the $Dp$-branes is reversed.
This change in $\Omega$ manifests itself in the M\"obius strip amplitude
which is proportional to the RR charge of the $Dp$-branes. It is easy to
see that a sign flip in the NSNS charge leads to an additional minus sign
for the action of $\Omega$ in the NS sector. 

Let us make some further comments on $\aNS_p$.
Supersymmetry breaking is a consequence of the fact that the RR charge
of the antibranes differs from their NSNS charge. In \cite{kr2}, it was
shown that a supersymmetric version of the orientifolds discussed in
\cite{aaads} is possible if instead of antibranes one introduces a new
type of $D$-branes with negative NSNS and RR charge. These objects ---
called $D5^-$-branes in \cite{kr2} --- live inside $D9$-branes with positive
NSNS and RR charge. It is not yet clear, whether $D5^-$-branes exist in
string theory. However, we would like to stress that once the possibility
of a sign flip in the NSNS charge is accepted everything fits nicely 
together. Orientifolds involving $D5^-$-branes have the same fermionic
spectrum as their \nonsusy analogues based on antibranes. The sign flip
in the NSNS charge leads to modified projection conditions for the bosons
which results in a supersymmetric spectrum.

The constraints (\ref{gauge_proj}) are easily solved. 
One finds that the gauge group on the $Dp$-branes is \cite{bi}
\be  \label{Gp_sym}
G_{(p)}\ =\ \prod_{l_1}SO(n^r_{l_1})\times\prod_{l_2}USp(n^p_{l_2})\times
            \prod_{l_3}U(n^c_{l_3})
\ee
if $\gamma_{\Omega,p}=\aNS_p\,\gamma_{\Omega,p}^\top$ and
\be  \label{Gp_asym}
G_{(p)}\ =\ \prod_{l_1}USp(n^r_{l_1})\times\prod_{l_2}SO(n^p_{l_2})\times
            \prod_{l_3}U(n^c_{l_3})
\ee
if $\gamma_{\Omega,p}=-\aNS_p\,\gamma_{\Omega,p}^\top$. The fermions 
transform in the same representation as the gauge bosons if the gauge group 
is unitary or $\aR_p=\aNS_p$, else they transform in the symmetric
(antisymmetric) representation if the the gauge bosons transform in the 
antisymmetric (symmetric) representation.

In our case, $\Gamma=\IZ_N\times\IZ_M$. There are $NM/s^2$ irreps, all of
them $s$-dimensional, where $s$ is the smallest positive integer such that 
$\eps^s=1$ \cite{beren}. In the case without discrete torsion, $\eps=s=1$,
they are of the form
\bea  \label{irrep_nodt}
&&R_{k,l}(g_1)=e^{\pi i (2k+\eta_1)/N},\qquad 
  R_{k,l}(g_2)=e^{\pi i (2l+\eta_2)/M},\\
&&{\rm with}\quad k=0,\ldots,N-1,\ l=0,\ldots,M-1, \nonumber
\eea
where $\eta_{1/2}=0$ or 1, depending on whether the first/second factor of the
discrete group has vector structure or not. One has $\eta_1=(1-\mu_{3,p})/2$,
$\eta_2=(1-\mu_{1,p})/2$, where the $\mu_{i,p}$ are defined in 
(\ref{def_mu_i}).
In the case with discrete torsion, $\eps=-1$, $s=2$, the irreps are of the form
(see \eg \cite{beren})
\bea  \label{irrep_dt}
&&R_{k,l}(g_1)=e^{\pi i (2k+\eta_1)/N}\,\left(\ba{cc}1&0\\0&-1\ea\right),
 \qquad 
 R_{k,l}(g_2)=e^{\pi i (2l+\eta_2)/M}\,\left(\ba{cc}0&1\\1&0\ea\right),\\ 
&&{\rm with}\quad k=0,\ldots,N/2-1,\ l=0,\ldots,M/2-1. \nonumber
\eea 

The action of $\Gamma$ on the internal $\IC^3$ is described by a
representation $R_{C^3}$:
\bea  \label{def_RC3}
\gamma:\ R_{C^3} &\longrightarrow &SU(3) \\
               g &\longmapsto     &R_{C^3}(g) \nonumber
\eea
We write $R_{C^3}=R^{(1)}_{C^3}\oplus R^{(2)}_{C^3}\oplus R^{(3)}_{C^3}$
(this decomposition is possible whenever $\Gamma$ is Abelian),
where $R^{(i)}_{C^3}$ corresponds to the action of $R_{C^3}$ on the $i\th$
coordinate of $\IC^3$. In the notation of eq.\ (\ref{twistvec}), we have
\be  \label{R_C3}
R_{C^3}^{(i)}(g_1)=e^{2\pi iv_i},\qquad
R_{C^3}^{(i)}(g_2)=e^{2\pi iw_i}.
\ee

The matter fields corresponding to the $i\th$ complex plane are obtained from
the projections (see \eg \cite{afiv,au})
\bea  \label{matter_proj}
\lambda^{(i)}=R_{C^3}^{(i)}(g)\gamma_{g,p}\lambda^{(i)}\gamma_{g,p}^{-1},
\quad\forall\,g\in\Gamma, &\quad
&\lambda^{(i)}=\alpha_p\,R_{\Omega}^{(i)}\gamma_{\Omega,p}{\lambda^{(i)}}^\top
              \gamma_{\Omega,p}^{-1}, \\
{\rm with}\ R_{\Omega}^{(i)}=\left\{\begin{array}{ll}
                                -1\ &{\rm if\ }p=9\ {\rm or\ }p=5_i\\
                                +1\ &{\rm if\ }p=5_j\ {\rm and\ }j\neq i
                                    \end{array}\right.
&&{\rm and}\quad \alpha_p=\hbox{charge of $Dp$-brane}. \nonumber
\eea
In general, the fermionic spectrum (R sector) may differ from the bosonic 
spectrum (NS sector). In the R sector (NS sector) $\alpha_p$ is the RR (NSNS)
charge of the $Dp$-branes. 
These equations can be solved using quiver theory \cite{dm}, as explained
in appendix A of \cite{kr2}. 

Let us now consider the $95_i$ and $5_i5_j$ sectors. The projection of the 
open strings $\lambda^{(95_i)}$, stretching from 9-branes to $5_i$-branes, on 
$\Gamma$-invariant states reads (see \eg \cite{afiv,au}):
\bea  \label{mixed_proj}
&& \hbox{R sector:}\quad
     \lambda^{(95_i)}=(R_{C^3}^{(i)}(g))^{-\aR_i/2}\,
     \gamma_{g,9}\lambda^{(95_i)}\gamma_{g,5_i}^{-1}, \nonumber\\
&& \hbox{NS sector:}\quad
     \lambda^{(95_i)}=(R_{C^3}^{(j)}(g))^{\aR_i/2}
                      (R_{C^3}^{(k)}(g))^{\aNS_i/2}\,
     \gamma_{g,9}\lambda^{(95_i)}\gamma_{g,5_i}^{-1},
\eea
where $\aR_i$ ($\aNS_i$) is the RR (NSNS) charge of 
the $D5_i$-branes and $(ijk)$ is a cyclic permutation of $(123)$.\footnote{In
contrast to \cite{au}, we choose all fermions to have positive chiralities,
the fermions with negative chiralities being their antiparticles.}
In the supersymmetric case $\aR_i=\aNS_i$, the 
conditions on the fermionic and bosonic spectrum coincide as they should.%
\footnote{Note that $R_{C^3}^{(i)}(g)R_{C^3}^{(j)}(g)R_{C^3}^{(k)}(g)=1$
is a consequence of $R_{C^3}^{(1)}(g)\oplus R_{C^3}^{(2)}(g)\oplus 
R_{C^3}^{(3)}(g)\in SU(3)$ in (\ref{def_RC3}).}
$\Omega$ relates the $95_i$ sector with the $5_i9$ sector and does not 
impose extra conditions on $\lambda^{(95_i)}$. Similarly, the projection
on the open strings $\lambda^{(5_i5_j)}$, stretching from $5_i$-branes to 
$5_j$-branes, on $\Gamma$-invariant states reads 
\bea  \label{mixed_pro55}
&& \hbox{R sector:}\quad
     \lambda^{(5_i5_j)}=(R_{C^3}^{(k)}(g))^{-\aR_i\aR_j/2}\,
     \gamma_{g,5_i}\lambda^{(5_i5_j)}\gamma_{g,5_j}^{-1}, \nonumber\\
&& \hbox{NS sector:}\quad
     \lambda^{(5_i5_j)}=(R_{C^3}^{(i)}(g))^{\aR_i\aR_j/2}
                      (R_{C^3}^{(j)}(g))^{\aNS_i\aNS_j/2}\,
     \gamma_{g,5_i}\lambda^{(5_i5_j)}\gamma_{g,5_j}^{-1},
\eea
The spectrum is easiest obtained using quiver theory, as explained
in appendix A of \cite{kr2}.

\subsection{Tadpoles}
\label{tadpoles}

In this subsection we give the tadpole cancellation conditions including
all the possible signs that may appear in the different contributions.%
\footnote{We restrict ourselves to the RR tadpoles. The formulae are valid 
if $D9$-branes and $D5_i$-branes or anti-$D5_i$-branes are present. The
NSNS tadpoles have exactly the same form if the signs $\alpha_i$, 
$\eps_i$ are now interpreted as the NSNS charges of the $D5_i$-branes, 
$O5_i$-planes respectively.} 
It is straightforward to obtain these conditions from the eqs.\ 
(\ref{cylinder}) -- (\ref{Moebius_strip}) using the methods described 
in \cite{kr1,kr2,pru}.

The elements of $\IZ_N\times\IZ_M$ are labelled by the two-vector 
$\bar k=(a,b)$, $a=0,\ldots,N-1$, $b=0,\ldots,M-1$.
We define $s_i=\sin(\pi\bar k\cdot\bar v_i)$, 
$c_i=\cos(\pi\bar k\cdot\bar v_i)$ 
and $\tilde s_i=\sin(2\pi\bar k\cdot\bar v_i)$,
where $\bar v=(v,w)$ combines the two twist vectors of (\ref{twistvec}).
Let us comment on the various signs appearing in the tadpole 
contributions (see also section \ref{phasefac}).
\begin{itemize}
\item In the cylinder amplitude, $\alpha_i$ weights the $95_i$ sector relative
      to the 99 and $5_i5_i$ sectors. It gives the RR charge of the the
      $D5_i$-branes.

\item In the Klein bottle amplitude, $\epsilon_i$ is related to the choice
      between the standard and alternative $\Omega$-projection, eq.\ 
      (\ref{OmJT}). It gives (half) the RR charge of the $O5_i$-planes.

\item The signs $\mu_{i,p}$ and $c_p$ are related to the symmetry properties
      of the $\gamma$ matrices. They are defined by 
      $(\gamma_{(1,0),p})^N=\mu_{3,p}\,\one$,
      $(\gamma_{(0,1),p})^M=\mu_{1,p}\,\one$ and 
      $\gamma_{\Omega,p}^\top=c_p\,\gamma_{\Omega,p}$.

\item The discrete torsion parameter $\eps$ can only take the values $\pm1$.
      It is defined in section \ref{disctor}.

\item The sign $\beta_i$ depends on the discrete torsion (\ref{beta_gh}) via
      $\beta_i=\beta_{\bar k_i,\bar k}$, where $\bar k_i$ is the order-two
      twist that fixes the $i\th$ complex plane.
\end{itemize}

Factorisation of the twisted tadpoles requires \cite{kr1}
\be  \label{constr_fac}
\eps_i=\mu_{i,9}=\mu_{i,5_i}=-\mu_{i,5_j},\quad i=1,3,\ j\neq i,\qquad
\eps_1\eps_2\eps_3=\eps^{-MN/4}.
\ee
The untwisted tadpoles can only be cancelled if
\be  \label{alpha_epsilon}
\alpha_i=-\epsilon_i,\qquad i=1,2,3.
\ee
This is the statement that the $D$-brane charges must be opposite to the 
$O$-plane charges in order to cancel the RR tadpoles. If the RR flux can
escape to infinity (non-compact cases), this condition can be absent.

Furthermore, it is easy to see \cite{gp} that the action of $\Omega^2$ on
the oscillator part of strings stretching from 9-branes to $5_i$-branes is
related to $c_9$ and $c_{5_i}$ by 
\be  \label{Om95}
\Omega^2|_{95_i}\ =\ c_9c_{5_i}\ =-1
\ee 
The last identity follows from an argument given in \cite{gp}. In special 
orientifold models with no massless matter in the $95_i$ sectors, it is 
possible to have $\Omega^2|_{95_i}=+1$, as shown in \cite{dp,bz}. 
But in this article, we will stick to the standard $\Omega$-action. 
Moreover, in order to avoid the introduction of anti-$D9$-branes, 
we choose $c_9=1$.

We are still free to choose the signs $\mu_{1,9}$, $\mu_{3,9}$ and $\eps$. 
All the other signs are then fixed by the above conditions:
\bea  \label{all_signs} 
&&\Omega^2|_{95_j}=-c_9=c_{5_j}=-1,\qquad j=1,2,3 \nonumber\\
&&\mu_{i,9}=\mu_{i,5_i}=-\mu_{i,5_j}=-\alpha_i=\eps_i,\qquad i=1,3,\ j\neq i 
   \nonumber\\
&&-\alpha_2=\eps_2=\eps_1\eps_3\,\eps^{-MN/4}.
\eea

The tadpole cancellation conditions are:

a) untwisted sector
\be  \label{tad_untw}
  \Tr\gamma_{(0,0),9} = 32, \qquad
  \alpha_i\,\Tr\gamma_{(0,0),5_i} = 32\,c_{5_i}\,\eps_i,\qquad i=1,2,3.
\ee

b) twisted sectors without fixed tori, 
   \ie $\bar k\cdot\bar v_i\ne0\ {\rm mod\ }\IZ$:
\begin{itemize}
\item odd $\bar k$:
\be  \label{tad_oddk}  
   \Tr\gamma_{\bar k,9} + \sum_{i=1}^3 4\,\alpha_i\,s_j s_k
                                          \Tr\gamma_{\bar k,5_i}=0,
\ee
where $(ijk)$ is a permutation of $(123)$.

\item even $\bar k=2\bar k'$:
\be  \label{tad_evenk}  
  \Tr\gamma_{2\bar k',9} + \sum_{i=1}^3 4\,\alpha_i\tilde s_j\tilde s_k 
                                           \Tr\gamma_{2\bar k',5_i}
   = 32\,\eps^{-k_1'k_2'}\,(c_1c_2c_3 - \sum_{i=1}^3\eps_i\beta_i\,c_is_js_k),
\ee
where $s_i$, $c_i$, $\tilde s_i$ and $\beta_i$ are evaluated with the argument 
$\bar k'=(k_1',k_2')$.
\end{itemize}

c) twisted sectors with fixed tori, 
   \ie $\bar k\cdot\bar v_i=0\ {\rm mod\ }\IZ$:
\begin{itemize}
\item odd $\bar k$:
  \bea  \label{tad_fixodd}
     \Tr\gamma_{\bar k,9} + 4\,\alpha_i\,s_j s_k \Tr\gamma_{\bar k,5_i} &= &0
     \nonumber\\
     \alpha_j \Tr\gamma_{\bar k,5_j} &= &\pm\alpha_k \Tr\gamma_{\bar k,5_k}
  \eea
The sign in the second line depends on whether $\bar k\cdot\bar v_i$ is even
(upper sign) or odd (lower sign).

\item even $\bar k=2\bar k'$, with $\bar k'\cdot\bar v_i=0$:
  \bea  \label{tad_fixeven1}  
     \Tr\gamma_{2\bar k',9} + 4\,\alpha_i\,\tilde s_j\tilde s_k
                                            \Tr\gamma_{2\bar k',5_i}
         &= &32\,\eps^{-k_1'k_2'}\,(c_jc_k - \eps_i\beta_i\,s_js_k) \\
     \alpha_j \Tr\gamma_{2\bar k',5_j} - \alpha_k \Tr\gamma_{2\bar k',5_k}
         &= &-8\,\eps^{-k_1'k_2'}\,(\eps_j\beta_j-\eps_k\beta_k) \nonumber
  \eea

\item even $\bar k=2\bar k'$, with $\bar k'\cdot\bar v_i=\pm\half$:
  \bea  \label{tad_fixeven2}  
     \Tr\gamma_{2\bar k',9} + 4\,\alpha_i\,\tilde s_j\tilde s_k
                                            \Tr\gamma_{2\bar k',5_i} 
           &= &\mp32\,\eps^{-k_1'k_2'}\,(\eps_j\beta_j\,c_j^2
                                        +\eps_k\beta_k\,c_k^2)\\
     \alpha_j \Tr\gamma_{2\bar k',5_j} + \alpha_k \Tr\gamma_{2\bar k',5_k}
         &= &\pm8\,\eps^{-k_1'k_2'}\,(1-\eps_i\beta_i) \nonumber
  \eea
\end{itemize}

From (\ref{all_signs}), we see that the orientifolds corresponding to the
discrete groups \ZNM fall into two classes: (i) neither $N$ nor $M$ is a
multiple of four, (ii) $N$ and/or $M$ is a multiple of four. If we want to
introduce only standard $D5_i$-branes with positive RR charge 
(\ie $\alpha_i=1$, $i=1,2,3$), then the unique solution to the 
type (i) orientifolds is characterised by $\mu_{1,9}=\mu_{3,9}=-1$ 
and $\eps=-1$. These are the \N1 supersymmetric models discussed in 
\cite{bl,z,k}. They have discrete torsion in the sense of \cite{df,kr1} 
which is evident from the fact that the rank of their gauge groups is 
reduced by a factor two. 
There are no models of type (ii) with only positively charged
$D5_i$-branes \cite{z}. However, many more consistent \ZNM orientifolds
are possible if one allows for antibranes. In these orientifolds, 
supersymmetry is broken by the open strings ending on antibranes.

The $D5_i$-branes may be distributed over different points in the $j\th$
and $k\th$ internal torus. Of course the tadpole equations depend on the 
location of the 5-branes and not all configurations are consistent. 
The Klein bottle contribution to the tadpoles of the $\bar k$-twisted sector
consists of an untwisted part $\cK_0(\bar k)$ and three twisted parts 
$\cK_i(\bar k)$ corresponding to the three order-two twists 
$\bar k_1=(0,M/2)$, $\bar k_2=(N/2,M/2)$, $\bar k_3=(N/2,0)$.
The $\cK_0$ contribution gives the term proportional to $c_1c_2c_3$ in 
(\ref{tad_evenk}), the $\cK_i$ contributions give the terms proportional to 
$\eps_i\beta_i\,c_is_js_k$. (Note that these terms are combined with the 
cylinder contribution to the $2\bar k$-twisted sector.)
At each point of the internal space, the contribution $\cK_0(\bar k)$ is only 
present if this point is fixed under $\bar k$ and the contribution 
$\cK_i(\bar k)$ is only present if this point is fixed under $\bar k+\bar k_i$.
Thus, the above tadpole cancellation conditions are strictly valid only at 
the points that are fixed under the whole $\IZ_N\times\IZ_M$. At points
which are only fixed under some subgroup of $\IZ_N\times\IZ_M$, these 
conditions have to be modified accordingly.

One has to analyse the tadpoles at all the fixed points in each twisted sector.
If the twisted sector under consideration has fixed tori extended in the 
$i\th$ direction, one has to analyse the tadpoles at each fixed point in the 
remaining two directions (\ie the points where the fixed tori are located).
At each fixed point all the $D9$-branes contribute but only those $D5_i$-branes
that have this point inside their world-volume. The Klein bottle contribution 
to this fixed point is determined as explained in the preceding paragraph.
We will see how this works in the examples below. In the $\IZ_2\times\IZ_4$
and $\IZ_4\times\IZ_4$ models it is not possible to put all the $D5_i$-branes 
at the origin.

\section{Description of the models}
\label{descript}

All the models contain 32 $D9$-branes and three sets of 32 $D5_i$-branes
wrapping the $i\th$ internal torus, $i=1,2,3$. In general there are eight
different models for each orbifold group $\IZ_N\times\IZ_M$, with $N$ and
$M$ even. They can be
characterised by the three signs $\mu_{1,9}$, $\mu_{3,9}$ (vector structures)
and $\eps$ (discrete torsion). If neither $N$ nor $M$ is a multiple of four
(type (i) models), one can alternatively choose the signs $\alpha_1$, 
$\alpha_2$, $\alpha_3$ (RR charges of $D5_i$-branes), whereas only two 
of the three $\alpha_i$'s are independent if $N$ and/or $M$ is a multiple 
of four (type (ii) models), eq.\ (\ref{all_signs}). If $N=M$, only four of 
the eight possible models are inequivalent, corresponding to the number of
negative $\alpha_i$'s. As explained in the paragraph below eq.\ 
(\ref{tad_fixeven2}), 
the orientifolds involving only standard $D9$-branes and $D5_i$-branes, 
\ie $(\alpha_1,\alpha_2,\alpha_3)=(+1,+1,+1)$, are quite restricted. Only 
the type (i) models are possible. All of them are supersymmetric and have 
discrete torsion. They have been constructed by the authors of \cite{bl,z,k}. 
More possibilities arise if one introduces antibranes \cite{ads,au}.
The $\IZ_2\times\IZ_2$ orientifolds with supersymmetry broken on 
antibranes have been discussed in \cite{aaads}. In this section, we 
construct the remaining $\IZ_N\times\IZ_M$ orientifolds for all possible
values of $(\alpha_1,\alpha_2,\alpha_3)$.
Interestingly enough, some of the models are inconsistent (in the
absence of Wilson lines) because it is impossible to cancel the
twisted tadpoles at all fixed points. The inconsistent models are 
$\IZ_2\times\IZ_4$ with $(\alpha_1,\alpha_2,\alpha_3)=(+1,-1,+1)$ 
and $(+1,+1,-1)$ and $\IZ_4\times\IZ_4$ for all 
$(\alpha_1,\alpha_2,\alpha_3)\neq(-1,-1,-1)$.
This inconsistency does not depend on the discrete torsion. As we will see
it is related to the vector structure in each $\IZ_4$ generator. 
Most probably, there exists a solution to the tadpole conditions for
the above models if appropriate Wilson lines are added. But we did not
consider models with Wilson lines in this article.

If $D5^-$-branes with negative NSNS and RR charge exist, then replacing
the anti-$D5$-branes by $D5^-$-branes in the \nonsusy models discussed below
leads to a supersymmetric version for each of these models.
The fermionic spectrum of the supersymmetric orientifolds coincides
with the fermionic spectrum of the orientifolds containing antibranes.
The complete spectrum is obtained by replacing in the antibrane sectors 
$USp$ gauge group factors by $SO$ and changing the bosons such that they 
form \N1 multiplets with the corresponding fermions.

\subsection{$\bZ_2\times\bZ_2$, $v=\frac12(1,-1,0)$, $w=\frac12(0,1,-1)$}
\label{Z2Z2}

These orientifolds have been constructed in \cite{bl,aaads}. The eight
possible models can be characterised by the RR charges $(\alpha_1,\alpha_2,
\alpha_3)$ of the $D5_i$-branes. Of course, two models $(\alpha_i,\alpha_j,
\alpha_k)$ and $(\alpha_{i'},\alpha_{j'},\alpha_{k'})$ are equivalent
if $(i'j'k')$ is a permutation $(ijk)$. There are only four inequivalent 
$\IZ_2\times\IZ_2$ orientifolds \cite{aaads}, corresponding to $n=0,1,2,3$
negative $\alpha_i$'s. From (\ref{all_signs}), we find that
$\alpha_1\alpha_2\alpha_3=-\eps$. Thus, the models with even (odd) $n$
have (no) discrete torsion. 

Using the method described in the previous section, it is very easy to
construct the two $\IZ_2\times\IZ_2$ orientifolds with discrete torsion.
From (\ref{irrep_dt}), we find that there is a unique projective irreducible
representation, which is pseudo-real if $(\mu_3,\mu_1)=(-1,-1)$ and real
else. From (\ref{all_signs}) and (\ref{Gp_sym}), (\ref{Gp_asym}), one then
finds that the $(+1,+1,+1)$ model has gauge group $USp(n_0)$ on the 
$D9$-branes and gauge group $USp(n_i)$ on the $D5_i$-branes, whereas
the $(+1,-1,-1)$ model has gauge group $SO(n_0)$ on the $D9$-branes,
gauge group $SO(n_1)$ on the $D5_1$-branes and gauge group $USp(n_j)$
on the $D5_j$-branes, $j=2,3$.
The tadpole conditions (\ref{tad_fixodd}) for the three twisted sectors
$(0,1)$, $(1,1)$, $(1,0)$ are trivially satisfied because the $\gamma$
matrices are of the form (\ref{irrep_dt}), \ie they are traceless.
The untwisted tadpole conditions (\ref{tad_untw}) imply that the rank
of all gauge groups is 8, \ie $n_0=n_1=n_2=n_3=16$. We consider the
situation where all the $D5_i$-branes are located at the origin in 
the internal space.

\begin{table}[t]
\renewcommand{\arraystretch}{1.25}
$$\begin{array}{|c|c|}
\hline
\multicolumn{2}{|c|}{\IZ_2\times\IZ_2,\ 
      (\alpha_1,\alpha_2,\alpha_3)=(+1,+1,+1),\ \eps=-1}
   \\ \hline\hline
\multicolumn{2}{|c|}{\hbox{open string spectrum}}\\ \hline
\rm sector &\hbox{gauge group / matter fields}\\ \hline
99,\ 5_i5_i  &USp(16)\\
             &3\,\Yasym\\
\hline
95_i,\ 5_i5_j    & (\Yfun,\Yfun)\\ 
\hline\hline
\multicolumn{2}{|c|}{\hbox{closed string spectrum}}\\ \hline
\rm sector       &\hbox{$\cN=1$ multiplets}\\ \hline
\rm untw.        &\hbox{gravity, 1 lin., 6 chir.} \\ \hline
\hbox{order-two} &\hbox{48 chir.} \\ \hline
\rm remaining    &- \\ \hline
\end{array}$$
\caption{Spectrum of the $\bZ_2\times\bZ_2$ orientifold with discrete torsion
         and $(\mu_3,\mu_1)=(-1,-1)$. The signs $\alpha_i$ are the charges
         of the $D5_i$-branes. The open string states are in \N1 vector
         (gauge) and chiral (matter) multiplets. The spectrum in the 99 sector
         and the three $5_i5_i$ sectors is identical. Thus, in total, one has
         four copies of the $USp(16)$ gauge group. In the same way, there are
         six copies of bifundamentals connecting the four gauge group factors.
         \label{Z2Z2spec1}}
\end{table}

\begin{table}[t]
\renewcommand{\arraystretch}{1.25}
$$\begin{array}{|c|c|}
\hline
\multicolumn{2}{|c|}{\IZ_2\times\IZ_2,\ 
      (\alpha_1,\alpha_2,\alpha_3)=(+1,-1,-1),\ \eps=-1}
   \\ \hline\hline
\multicolumn{2}{|c|}{\hbox{open string spectrum}}\\ \hline
\rm sector &\hbox{gauge group / matter fields}\\ \hline
99,\ 5_15_1  &SO(16)\\
\sus   &2\,\Yasym,\ \Ysym\\
\hline
5_25_2,\ 5_35_3 &USp(16)\\
\nsus          &\spin 3\,\Yasym,\ \Ysym,\quad 
                \scal 2\,\Ysym,\ \Yasym\\
\hline
95_i,\ 5_i5_j    & (\Yfun,\Yfun)\\ 
\hline\hline
\multicolumn{2}{|c|}{\hbox{closed string spectrum}}\\ \hline
\rm sector       &\hbox{$\cN=1$ multiplets}\\ \hline
\rm untw.        &\hbox{gravity, 1 lin., 6 chir.} \\ \hline
\hbox{order-two} &\hbox{16 chir., 32 vec.} \\ \hline
\rm remaining    &- \\ \hline
\end{array}$$
\caption{Spectrum of the $\bZ_2\times\bZ_2$ orientifold with discrete torsion
         and $(\mu_3,\mu_1)=(+1,-1)$. The signs $\alpha_i$ are the charges
         of the $D5_i$-branes. In the susy open string sectors, the gauge 
         fields are in \N1 vector multiplets and the matter fields in \N1
         chiral multiplets. In the non-susy sectors, we display the spin 0
         (complex scalars) and spin 1/2 (Weyl or Majorana fermions) fields
         separately. In addition, there are spin 1 gauge bosons.
         The 99 and $5_15_1$ spectrum coincides and the $5_25_2$ and $5_35_3$
         spectrum coincides as well. Thus, in total, one has two copies of
         the gauge groups $SO(16)$ and $USp(16)$. In the same way, there are
         six copies of bifundamentals connecting the four gauge group factors.
         \label{Z2Z2spec2}}
\end{table}

The quiver diagrams for these two models are trivial. As there is a unique
projective irrep, they consist only of one node. For each of the sectors
99, $5_i5_i$, $i=1,2,3$, one finds 3 matter fields\footnote{In the
supersymmetric sectors, these are \N1 chiral multiplets. In the \nonsusy
sectors, these are pairs of fermions and bosons which may transform in
different representations.} transforming as second rank tensors under the
gauge group. To decide whether these are symmetric or antisymmetric tensors,
one needs to evaluate the index defined in eq.\ (A.9) of \cite{kr2}. For 
each of the six mixed sectors $95_i$, $5_i5_j$, one finds one matter field
in the bifundamental representation.

The complete spectrum of these two models is displayed in tables 
\ref{Z2Z2spec1}, \ref{Z2Z2spec2}. 

The two $\IZ_2\times\IZ_2$ orientifolds without discrete torsion
--- $(-1,-1,-1)$ and $(+1,+1,-1)$ --- are slightly more complicated.
In this case, there are four projective irreps, eq.\ (\ref{irrep_nodt}).
If $(\mu_3,\mu_1)=(+1,+1)$, then all of them are real, else they are
all complex. From (\ref{all_signs}) and (\ref{Gp_sym}), (\ref{Gp_asym}), 
one then finds that the $(-1,-1,-1)$ model has a gauge group of the form
$SO\times SO\times SO\times SO$ on the $D9$-branes and a gauge group of
the form $U\times U$ on the $D5_i$-branes, whereas the $(+1,+1,-1)$ model 
has a gauge group of the form $U\times U$ on the $D9$-branes and on the
$D5_j$-branes, $j=1,2$, and a gauge group of the form 
$USp\times USp\times USp\times USp$ on the $D5_3$-branes. 
The untwisted tadpole conditions (\ref{tad_untw}) fix the rank of the
gauge group to be 16 for each of the four sets of $D$-branes. Now, the
twisted tadpole conditions (\ref{tad_fixodd}) impose additional constraints
because, generically, the $\gamma$ matrices are not traceless. 
For the three order-two sectors $\bar k_1=(0,1)$, $\bar k_2=(1,1)$,
$\bar k_3=(1,0)$, these conditions are
\be  \label{tadpZ2Z2}
\Tr(\gamma_{\bar k_i,9})=\Tr(\gamma_{\bar k_i,5_j})=0,\quad i,j=1,2,3.
\ee
Here, we used the fact that eq.\ (\ref{tad_fixodd}) must be imposed at
each of the 16 $\bar k_i$ fixed points (more precisely, fixed tori
extended in the $i\th$ complex plane). As all $D5_i$-branes are located
at the origin, their contribution to (\ref{tad_fixodd}) vanishes at some
fixed points. In the $\bar k_i$ sector, the $D5_i$-branes only contribute
to the fixed point at the origin in the $j\th$ and $k\th$ complex plane,
where $(ijk)$ is a permutation of $(123)$. The $D5_j$-branes contribute
to the four fixed points at the origin in the $k\th$ complex plane.
Therefore, the tadpole conditions take the form given in (\ref{tadpZ2Z2}).
The complete spectrum of these two models is displayed in tables 
\ref{Z2Z2spec3}, \ref{Z2Z2spec4}.

\begin{table}[htp]
\renewcommand{\arraystretch}{1.25}
$$\begin{array}{|c|c|}
\hline
\multicolumn{2}{|c|}{\IZ_2\times\IZ_2,\ 
       (\alpha_1,\alpha_2,\alpha_3)=(-1,-1,-1),\ \eps=1}
   \\ \hline\hline
\multicolumn{2}{|c|}{\hbox{open string spectrum}}\\ \hline
\rm sector &\hbox{gauge group / matter fields}\\ \hline
99     &SO(8)_1\times SO(8)_2\times SO(8)_3\times SO(8)_4\\
\sus   &(\Yfun_1,\Yfun_2),\ (\Yfun_1,\Yfun_3),\ (\Yfun_1,\Yfun_4),\ 
        (\Yfun_2,\Yfun_3),\ (\Yfun_2,\Yfun_4),\ (\Yfun_3,\Yfun_4)\\
\hline
5_i5_i        &U(8)_1\times U(8)_2\\ 
i=1,2,3       &\smult (\Yfun_1,\Yfun_2),\ (\Yfun_1,\Yfunb_2),\ 
               (\Yfunb_1,\Yfun_2),\ (\Yfunb_1,\Yfunb_2)\\  
\nsus &\spin adj_1,\ adj_2,\ \Yasym_1,\ \Yasymb_1,\ \Yasym_2,\ \Yasymb_2\\ 
      &\scal \Ysym_1,\ \Ysymb_1,\ \Ysym_2,\ \Ysymb_2\\
\hline
95_1  &\spin (\Yfun_1,\Yfunb_1),\ (\Yfun_2,\Yfun_2),\ 
             (\Yfun_3,\Yfun_1),\ (\Yfun_4,\Yfunb_2)\\
\nsus &\scal (\Yfun_1,\Yfun_2),\ (\Yfun_2,\Yfunb_1),\ 
             (\Yfun_3,\Yfunb_2),\  (\Yfun_4,\Yfun_1)\\ 
\hline
95_2  &\spin (\Yfun_1,\Yfun_2),\  (\Yfun_2,\Yfun_1),\ 
             (\Yfun_3,\Yfunb_1),\ (\Yfun_4,\Yfunb_2)\\
\nsus &\scal (\Yfun_1,\Yfunb_1),\ (\Yfun_2,\Yfunb_2),\ 
             (\Yfun_3,\Yfun_2),\  (\Yfun_4,\Yfun_1)\\
\hline
95_3  &\spin (\Yfun_1,\Yfun_1),\ (\Yfun_2,\Yfunb_1),\ 
             (\Yfun_3,\Yfun_2),\ (\Yfun_4,\Yfunb_2)\\
\nsus &\scal (\Yfun_1,\Yfunb_2),\ (\Yfun_2,\Yfun_2),\ 
             (\Yfun_3,\Yfunb_1),\ (\Yfun_4,\Yfun_1)\\
\hline
5_i5_j\ \sus &(\Yfunb_1,\Yfun_1),\ (\Yfun_1,\Yfunb_2),\ 
               (\Yfunb_2,\Yfunb_1),\ (\Yfun_2,\Yfun_2)\\
\hline\hline
\multicolumn{2}{|c|}{\hbox{closed string spectrum}}\\ \hline
\rm sector       &\hbox{$\cN=1$ multiplets}\\ \hline
\rm untw.        &\hbox{gravity, 1 lin., 6 chir.} \\ \hline
\hbox{order-two} &\hbox{48 lin.} \\ \hline
\rm remaining    &- \\ \hline
\end{array}$$
\caption{Spectrum of the $\bZ_2\times\bZ_2$ orientifold without discrete
         torsion and $(\mu_3,\mu_1)=(+1,+1)$. The signs $\alpha_i$ are the 
         charges of the $D5_i$-branes. In the susy open string sectors, the 
         gauge  fields are in \N1 vector multiplets and the matter fields in 
         \N1 chiral multiplets. In the non-susy sectors, we display the spin 0
         (complex scalars) and spin 1/2 (Weyl or Majorana fermions) fields
         separately. In addition, there are spin 1 gauge bosons.
         The spectrum in the three $5_i5_i$ and in the three $5_i5_j$ sectors 
         is identical (our convention is that $(ijk)$ is a cyclic permutation
         of $(123)$). Thus, in total, one has three copies of the 
         $U(8)_1\times U(8)_2$ gauge group.
         \label{Z2Z2spec3}}
\end{table}

\begin{table}[htp]
\renewcommand{\arraystretch}{1.25}
$$\begin{array}{|c|c|}
\hline
\multicolumn{2}{|c|}{\IZ_2\times\IZ_2,\ 
       (\alpha_1,\alpha_2,\alpha_3)=(+1,+1,-1),\ \eps=1}
   \\ \hline\hline
\multicolumn{2}{|c|}{\hbox{open string spectrum}}\\ \hline
\rm sector &\hbox{gauge group / matter fields}\\ \hline
99,\ 5_15_1,\ 5_25_2 &U(8)_1\times U(8)_2\\
\sus    &(\Yfun_1,\Yfun_2),\ (\Yfun_1,\Yfunb_2),\ 
         (\Yfunb_1,\Yfun_2),\ (\Yfunb_1,\Yfunb_2),\\
        &\Yasym_1,\ \Yasymb_1,\ \Yasym_2,\ \Yasymb_2\\
\hline
5_35_3  &USp(8)_1\times USp(8)_2\times USp(8)_3\times USp(8)_4\\
\nsus   &\smult (\Yfun_1,\Yfun_2),\ (\Yfun_1,\Yfun_3),\ (\Yfun_1,\Yfun_4),\ 
         (\Yfun_2,\Yfun_3),\ (\Yfun_2,\Yfun_4),\ (\Yfun_3,\Yfun_4)\\
        &\spin \Yasym_1,\ \Yasym_2,\ \Yasym_3,\ \Yasym_4\\
\hline
95_1\ \sus &(\Yfunb_1,\Yfun_1),\ (\Yfun_1,\Yfun_2),\ 
             (\Yfun_2,\Yfunb_1),\ (\Yfunb_2,\Yfunb_2)\\
\hline
95_2\ \sus &(\Yfun_1,\Yfunb_1),\ (\Yfunb_1,\Yfunb_2),\ 
            (\Yfun_2,\Yfun_1),\ (\Yfunb_2,\Yfun_2)\\
\hline
95_3       &\spin (\Yfun_1,\Yfun_1),\ (\Yfunb_1,\Yfun_2),\ 
                  (\Yfun_2,\Yfun_3),\ (\Yfunb_2,\Yfun_4)\\  
\nsus      &\scal (\Yfunb_1,\Yfun_3),\ (\Yfun_1,\Yfun_4),\ 
                  (\Yfunb_2,\Yfun_1),\ (\Yfun_2,\Yfun_2)\\
\hline
5_25_3     &\spin (\Yfun_1,\Yfun_1),\ (\Yfunb_1,\Yfun_3),\ 
                  (\Yfun_2,\Yfun_2),\ (\Yfunb_2,\Yfun_4)\\
\nsus      &\scal (\Yfunb_1,\Yfun_2),\ (\Yfun_1,\Yfun_4),\ 
                  (\Yfunb_2,\Yfun_1),\ (\Yfun_2,\Yfun_3)\\
\hline
5_35_1     &\spin (\Yfun_1,\Yfunb_2),\ (\Yfun_2,\Yfunb_1),\ 
                  (\Yfun_3,\Yfun_1),\ (\Yfun_4,\Yfun_2)\\
\nsus      &\scal (\Yfun_1,\Yfunb_1),\ (\Yfun_2,\Yfunb_2),\ 
                  (\Yfun_3,\Yfun_2),\ (\Yfun_4,\Yfun_1)\\
\hline
5_15_2\ \sus &(\Yfun_1,\Yfunb_1),\ (\Yfunb_1,\Yfun_2),\ 
              (\Yfunb_2,\Yfun_1),\ (\Yfun_2,\Yfunb_2)\\
\hline\hline
\multicolumn{2}{|c|}{\hbox{closed string spectrum}}\\ \hline
\rm sector       &\hbox{$\cN=1$ multiplets}\\ \hline
\rm untw.        &\hbox{gravity, 1 lin., 6 chir.} \\ \hline
\hbox{order-two} &\hbox{16 lin., 32 chir.} \\ \hline
\rm remaining    &- \\ \hline
\end{array}$$
\caption{Spectrum of the $\bZ_2\times\bZ_2$ orientifold without discrete
         torsion and $(\mu_3,\mu_1)=(+1,-1)$. The notation is explained
         in the previous table.
         \label{Z2Z2spec4}}
\end{table}

\subsection{$\bZ_2\times\bZ_4$, $v=\frac12(1,-1,0)$, $w=\frac14(0,1,-1)$}
\label{Z2Z4}

This orientifold is not symmetric under a permutation of the three sets 
of $D5_i$-branes. The three order-two twists $\bar k_i$ that fix the 
world-volume of the $D5_i$-branes are $\bar k_1=(0,2)$, $\bar k_2=(1,2)$, 
$\bar k_3=(1,0)$. Only $\bar k_1$ is of even order. As a consequence, the 
tadpole conditions for the $D5_1$-branes differ from those for the 
$D5_2$- and $D5_3$-branes. In contrast to the $\IZ_2\times\IZ_2$ orientifold, 
we expect that six of the eight possible $\IZ_2\times\IZ_4$ models are 
inequivalent. However, only for the models with vector structure in the
$\IZ_4$ factor, \ie $\mu_1=+1$, does a consistent solution to the
tadpole equations exist, at least in the absence of Wilson lines.

From (\ref{all_signs}), we find that $\alpha_1\alpha_2\alpha_3=-1$. 
Thus, only an odd number of negative $\alpha_i$'s is allowed. We found
a solution for $(-1,-1,-1)$ and $(-1,+1,+1)$ both with and without discrete
torsion.

Let us first analyse the cases with discrete torsion. Most of the twisted
tadpole conditions are trivially satisfied because the matrices
$\gamma_{(1,0),p}$ and $\gamma_{(0,1),p}$ are traceless, eq.\ 
(\ref{irrep_dt}). The only \nontriv tadpole conditions are (\ref{tad_untw})
and (\ref{tad_fixeven1}), corresponding to the untwisted sector and the
$(0,2)$ sector. The former fixes the rank of the gauge group to be 8 for
each of the four sectors 99, $5_i5_i$. If $\alpha_1=-1$, then the latter reads
\be  \label{tadpZ2Z4}
\Tr(\gamma_{(0,2),9})=0,\quad \Tr(\gamma_{(0,2),5_1,n})=8,\quad
\Tr(\gamma_{(0,2),5_2})=\Tr(\gamma_{(0,2),5_3})=0,
\ee
where $n=0,1,2,3$ denotes the $\IZ_4$ fixed points in the second and
third complex plane. We assumed that there are no $D5_1$-branes at the
12 remaining $\IZ_2$ fixed points and we located all $D5_2$- and
$D5_3$-branes at the origin. By a similar reasoning to the one used
at the end of section \ref{Z2Z2}, the tadpole equations (\ref{tad_fixeven1})
simplify to the form given in (\ref{tadpZ2Z4}). 
One can verify that the $D5_2$- and 
$D5_3$-branes have $\mu_1=-1$ if $\alpha_1=-1$, which implies, using 
(\ref{irrep_dt}), that the matrices $\gamma_{(0,2),5_{2/3}}$ are 
traceless. Therefore the last two conditions in (\ref{tadpZ2Z4}) are trivially 
satisfied. The second condition in (\ref{tadpZ2Z4}) forces us to put some 
$D5_1$-branes at each of the four $\IZ_4$ fixed points. In the following,
we will consider the most symmetric situation where eight $D5_1$-branes
sit at each of the $\IZ_4$ fixed points. It is straightforward to obtain
the spectrum of this orientifold using the methods described in section
\ref{construct}. 

The two cases without discrete torsion are slightly more complicated.
But it is easy to see that again some $D5_1$-branes are needed at the
four $\IZ_4$ fixed points. The simplest consistent brane configuration
is the one described in the previous paragraph. The tadpole conditions
of the $(0,2)$ sector are again given by (\ref{tadpZ2Z4}) and the last two 
of these conditions are again trivially satisfied. The untwisted tadpole
conditions now imply that the rank of the gauge group is 16 for each
of the four sectors 99, $5_i5_i$. The complete spectrum was computed
with the help of a computer algebra program.
The results are displayed in tables \ref{Z2Z4spec1} -- \ref{Z2Z4spec4b}. 

\begin{table}[htp]
\renewcommand{\arraystretch}{1.25}
$$\begin{array}{|c|c|}
\hline
\multicolumn{2}{|c|}{\IZ_2\times\IZ_4,\ 
       (\alpha_1,\alpha_2,\alpha_3)=(-1,-1,-1),\ \eps=-1}
   \\ \hline\hline
\multicolumn{2}{|c|}{\hbox{open string spectrum}}\\ \hline
\rm sector &\hbox{gauge group / matter fields}\\ \hline
99    &SO(8)_1\times SO(8)_2\\
\sus  &2\,(\Yfun_1,\Yfun_2),\ \Yasym_1,\ \Ysym_2\\
\hline
5_{1,n}5_{1,n},\ n=0,1,2,3 &USp(4)\\
\nsus     &\spin 2\,\Yasym,\quad\scal \Ysym\\
\hline
5_25_2,\ 5_35_3    &U(8)\\
\nsus     &\spin 2\,adj,\ \Yasym,\ 2\,\Yasymb, \Ysym\\  
          &\scal adj,\ \Ysym,\ 2\,\Ysymb,\ \Yasym\\
\hline
95_{1,n}\ \nsus  &\spin (\Yfun_1,\Yfun),\quad\scal (\Yfun_2,\Yfun)\\
\hline
95_2\ \nsus &\spin (\Yfun_1,\Yfunb),\ (\Yfun_2,\Yfun),\quad  
             \scal (\Yfun_1,\Yfunb),\ (\Yfun_2,\Yfun)\\
\hline
95_3\ \nsus &\spin (\Yfun_1,\Yfunb),\ (\Yfun_2,\Yfun),\quad  
             \scal (\Yfun_1,\Yfun),\ (\Yfun_2,\Yfunb)\\
\hline
5_25_3\ \sus &(\Yfun,\Yfun),\ (\Yfunb,\Yfunb)\\
\hline
5_{1,n}5_3,\ 5_{1,n}5_2\ \sus &(\Yfun,\Yfunb)\\
\hline\hline
\multicolumn{2}{|c|}{\hbox{closed string spectrum}}\\ \hline
\rm sector       &\hbox{$\cN=1$ multiplets}\\ \hline
\rm untw.        &\hbox{gravity, 1 lin., 4 chir.} \\ \hline
\hbox{order-two} &\hbox{18 lin.} \\ \hline
\rm remaining    &\hbox{4 chir., 4 vec.} \\ \hline
\end{array}$$
\caption{Spectrum of the $\bZ_2\times\bZ_4$ orientifold with discrete
         torsion and $(\mu_3,\mu_1)=(+1,+1)$. The four sets of 
         $D5_{1,n}$-branes are located at the four $\bZ_4$ fixed points
         in the second and third torus. The notation is explained
         in the tables of section \ref{Z2Z2}.
         \label{Z2Z4spec1}}
\end{table}

Surprisingly, we found no consistent solution for the $(+1,+1,-1)$
model with or without discrete torsion and without Wilson lines.%
\footnote{The $(+1,+1,-1)$ and $(+1,-1,+1)$ models are equivalent 
up to a permutation of $D5_2$- and $D5_3$-branes.} 
This is due to the impossibility
of cancelling the tadpoles in the $(0,2)$ sector. To write down the
tadpole conditions, let us label the four $\IZ_2$ fixed points in the
second and in the third complex plane by $n_2$ and $n_3$. From
(\ref{tad_fixeven1}) we find
\bea  \label{tadpoleZ2Z4}
&&\Tr(\gamma_{(0,2),9})=Tr(\gamma_{(0,2),5_1})=0,\\
&&\Tr(\gamma_{(0,2),5_2,n_3})+\Tr(\gamma_{(0,2),5_3,n_2})=16\,\eps
  \qquad\hbox{if $(n_2,n_3)$ is fixed under }\IZ_4,\nonumber\\
&&\Tr(\gamma_{(0,2),5_2,n_3})+\Tr(\gamma_{(0,2),5_3,n_2})=0
  \qquad\hbox{if $(n_2,n_3)$ is not fixed under }\IZ_4.\nonumber
\eea
The two equations in the first line of (\ref{tadpoleZ2Z4}) are trivially 
satisfied, because both $D9$- and $D5_1$-branes have no vector structure 
in the $\IZ_4$ factor. 
However, the last two equations are incompatible, \ie there is no possible 
brane configuration that satisfies both of them. To see this, denote the
$\IZ_4$ fixed points by $n_{2/3}=0,2$ and the remaining two $\IZ_2$ fixed
points by $n_{2/3}=1,3$. Summing the two equations for the fixed points
$(0,1)$ and $(1,0)$ and subtracting the equation for the fixed point $(1,1)$,
one finds $\Tr(\gamma_{5_2,0})+\Tr(\gamma_{5_3,0})=0$ which contradicts
the second line in (\ref{tadpoleZ2Z4}).
To derive this result, it was crucial that $\Tr(\gamma_{(0,2),5_2})$
does only depend on $n_3$ but not on $n_2$. This is no longer true if
Wilson lines in the second complex plane are added. Probably, there is
a solution to the tadpole conditions if appropriate Wilson lines in
the second and third complex plane are added.

As we will see, a similar problem arises in the $\IZ_4\times\IZ_4$
orientifold. Indeed, it is easy to see that a \ZNM orientifold with $N=4$ 
and/or $M=4$ is only consistent if its $\IZ_4$ has vector structure.

\begin{table}[t]
\renewcommand{\arraystretch}{1.25}
$$\begin{array}{|c|c|}
\hline
\multicolumn{2}{|c|}{\IZ_2\times\IZ_4,\ 
       (\alpha_1,\alpha_2,\alpha_3)=(-1,+1,+1),\ \eps=-1}
   \\ \hline\hline
\multicolumn{2}{|c|}{\hbox{open string spectrum}}\\ \hline
\rm sector &\hbox{gauge group / matter fields}\\ \hline
99    &SO(8)_1\times USp(8)_2\\
\sus  &2\,(\Yfun_1,\Yfun_2),\ \Yasym_1,\ \Yasym_2\\
\hline
5_{1,n}5_{1,n},\ n=0,1,2,3 &USp(4)\\
\nsus     &\spin 2\,\Yasym,\quad\scal \Ysym\\
\hline
5_25_2,\ 5_35_3  &U(8)\\
\sus     &adj,\ 2\,\Yasym,\ \Yasymb,\ \Ysymb\\  
\hline
95_{1,n}\ \nsus  &\spin (\Yfun_1,\Yfun),\quad\scal (\Yfun_2,\Yfun)\\
\hline
95_2,\ 95_3\ \sus & (\Yfun_1,\Yfun),\ (\Yfun_2,\Yfunb)\\
\hline
5_25_3\ \sus &(\Yfun,\Yfun),\ (\Yfunb,\Yfunb)\\
\hline
5_35_{1,n}\ \nsus &\spin (\Yfun,\Yfun),\quad\scal (\Yfun,\Yfun)\\
\hline
5_{1,n}5_2\ \nsus &\spin (\Yfun,\Yfun),\quad\scal (\Yfun,\Yfunb)\\
\hline\hline
\multicolumn{2}{|c|}{\hbox{closed string spectrum}}\\ \hline
\rm sector       &\hbox{$\cN=1$ multiplets}\\ \hline
\rm untw.        &\hbox{gravity, 1 lin., 4 chir.} \\ \hline
\hbox{order-two} &\hbox{10 lin., 8 chir.} \\ \hline
\rm remaining    &\hbox{4 chir., 4 vec.} \\ \hline
\end{array}$$
\caption{Spectrum of the $\bZ_2\times\bZ_4$ orientifold with discrete
         torsion and $(\mu_3,\mu_1)=(-1,+1)$. The four sets of 
         $D5_{1,n}$-branes are located at the four $\bZ_4$ fixed points
         in the second and third torus. The notation is explained
         in the tables of section \ref{Z2Z2}.
         \label{Z2Z4spec2}}
\end{table}

\begin{table}[htp]
\renewcommand{\arraystretch}{1.2}
$$\begin{array}{|c|c|}
\hline
\multicolumn{2}{|c|}{\IZ_2\times\IZ_4,\ 
       (\alpha_1,\alpha_2,\alpha_3)=(-1,-1,-1),\ \eps=1}
   \\ \hline\hline
\multicolumn{2}{|c|}{\hbox{open string spectrum}}\\ \hline
\rm sector &\hbox{gauge group / matter fields}\\ \hline
99    &SO(4)_1\times U(4)_2\times SO(4)_3\times SO(4)_4
       \times U(4)_5\times SO(4)_6\\
\sus  &(\Yfun_1,\Yfun_2),\ (\Yfun_1,\Yfun_4),\ (\Yfun_1,\Yfunb_5),\ 
       (\Yfunb_2,\Yfun_3),\ (\Yfunb_2,\Yfun_4),\ (\Yfun_2,\Yfunb_5),\\ 
      &(\Yfun_2,\Yfun_6),\ (\Yfun_3,\Yfun_5),\ (\Yfun_3,\Yfun_6),\ 
       (\Yfun_4,\Yfun_5),\ (\Yfun_5,\Yfunb_2),\ (\Yfunb_5,\Yfun_6)\\
\hline
5_{1,n}5_{1,n} &U(2)_1\times U(2)_2\\
n=0,1,2,3 &\spin adj_1,\ adj_2,\ \Yasym_1,\ \Yasymb_1,\ \Yasym_2,\ \Yasymb_2\\
\nsus   &\scal \Ysym_1,\ \Ysymb_1,\ \Ysym_2,\ \Ysymb_2\\ 
\hline
5_25_2,\ 5_35_3 &U(4)_1\times U(4)_2\times U(4)_3\times U(4)_4\\
\nsus  &\smult (\Yfun_1,\Yfun_3),\ (\Yfun_1,\Yfun_4),\ (\Yfun_1,\Yfunb_4),\ 
               (\Yfunb_1,\Yfunb_4),\ (\Yfun_2,\Yfunb_1),\\
       &\qquad (\Yfun_2,\Yfun_3),\ (\Yfunb_2,\Yfunb_3),\ (\Yfunb_2,\Yfunb_4),\
               (\Yfun_3,\Yfunb_2),\ (\Yfun_4,\Yfunb_3)\\
       &\spin adj_1,\ adj_2,\ adj_3,\ adj_4,\ 
              \Yasymb_1,\ \Yasym_2,\ \Yasymb_3,\ \Yasym_4\\
       &\scal \Ysymb_1,\ \Ysym_2,\ \Ysymb_3,\ \Ysym_4\\  
\hline
95_{1,n}  &\spin (\Yfun_1,\Yfunb_1),\ (\Yfun_3,\Yfunb_2),\ 
                 (\Yfun_4,\Yfun_1),\ (\Yfun_6,\Yfun_2)\\
\nsus     &\scal (\Yfun_2,\Yfunb_1),\ (\Yfunb_2,\Yfunb_2),\ 
                 (\Yfun_5,\Yfun_1),\ (\Yfunb_5,\Yfun_2)\\
\hline
95_2    &\spin (\Yfun_1,\Yfun_4),\ (\Yfunb_2,\Yfun_1),\ 
               (\Yfun_2,\Yfun_3),\ (\Yfun_3,\Yfun_2),\\ 
        &\qquad (\Yfun_4,\Yfunb_1),\ (\Yfun_5,\Yfunb_2),\ 
                (\Yfunb_5,\Yfunb_4),\ (\Yfun_6,\Yfunb_3)\\
\nsus   &\scal (\Yfun_1,\Yfunb_1),\ (\Yfun_2,\Yfunb_2),\ 
               (\Yfunb_2,\Yfunb_4),\ (\Yfun_3,\Yfunb_3),\\
        &\quad (\Yfun_4,\Yfun_4),\ (\Yfunb_5,\Yfun_1),\ 
               (\Yfun_5,\Yfun_3),\ (\Yfun_6,\Yfun_2)\\
\hline
95_3    &\spin (\Yfun_1,\Yfunb_1),\ (\Yfun_2,\Yfun_1),\ 
               (\Yfunb_2,\Yfun_3),\ (\Yfun_3,\Yfunb_3),\\
        &\qquad (\Yfun_4,\Yfun_4),\ (\Yfun_5,\Yfunb_4),\ 
                (\Yfunb_5,\Yfunb_2),\ (\Yfun_6,\Yfun_2)\\
\nsus   &\scal (\Yfun_1,\Yfunb_4),\ (\Yfunb_2,\Yfun_4),\ 
               (\Yfun_2,\Yfun_2),\ (\Yfun_3,\Yfunb_2),\\
        &\qquad (\Yfun_4,\Yfun_1),\ (\Yfunb_6,\Yfunb_1),\ 
                (\Yfun_6,\Yfunb_3),\ (\Yfun_6,\Yfun_3)\\
\hline
5_25_3  &(\Yfun_1,\Yfun_1),\ (\Yfunb_1,\Yfun_4),\
         (\Yfun_2,\Yfunb_3),\ (\Yfunb_2,\Yfunb_2),\\
\sus    &(\Yfun_3,\Yfun_3),\ (\Yfunb_3,\Yfun_2),\ 
         (\Yfun_4,\Yfunb_1),\ (\Yfunb_4,\Yfunb_4)\\
\hline
5_35_{1,n}\ \sus &(\Yfunb_1,\Yfunb_1),\ (\Yfunb_3,\Yfunb_2),\ 
                  (\Yfun_4,\Yfun_1),\ (\Yfun_2,\Yfun_2)\\
\hline
5_{1,n}5_2\ \sus &(\Yfun_1,\Yfunb_1),\ (\Yfunb_1,\Yfun_4),\ 
                  (\Yfunb_3,\Yfun_2),\ (\Yfun_3,\Yfunb_3),\\
\hline
\end{array}$$
\caption{Spectrum of the $\bZ_2\times\bZ_4$ orientifold without discrete
         torsion and $(\mu_3,\mu_1)=(+1,+1)$. The four sets of 
         $D5_{1,n}$-branes are located at the four $\bZ_4$ fixed points
         in the second and third torus. The notation is explained
         in the tables of section \ref{Z2Z2}.
         \label{Z2Z4spec3}}
\end{table}

\begin{table}[htp]
\renewcommand{\arraystretch}{1.2}
$$\begin{array}{|c|c|}
\hline
\multicolumn{2}{|c|}{\IZ_2\times\IZ_4,\ 
       (\alpha_1,\alpha_2,\alpha_3)=(-1,+1,+1),\ \eps=1}
   \\ \hline\hline
\multicolumn{2}{|c|}{\hbox{open string spectrum}}\\ \hline
\rm sector &\hbox{gauge group / matter fields}\\ \hline
99    &U(4)_1\times U(4)_2\times U(4)_3\times U(4)_4\\
\sus  &(\Yfunb_1,\Yfunb_2),\  (\Yfun_1,\Yfun_4),\ (\Yfun_1,\Yfunb_4),\ 
       (\Yfun_2,\Yfunb_1),\ (\Yfun_2,\Yfun_3), (\Yfun_2,\Yfun_4),\\ 
      &(\Yfunb_2,\Yfunb_4),\ (\Yfun_3,\Yfunb_2),\ (\Yfunb_3,\Yfunb_4),\ 
       (\Yfun_4,\Yfunb_3),\ \Yasym_1,\ \Yasymb_1,\ \Yasym_3,\ \Yasymb_3\\
\hline
5_{1,n}5_{1,n} &USp(2)_1\times USp(2)_2\times USp(2)_3\times USp(2)_4\\
n=0,1,2,3 &\smult (\Yfun_1,\Yfun_3),\ (\Yfun_2,\Yfun_4)\\
\nsus     &\spin \Yasym_1,\ \Yasym_2,\ \Yasym_3,\ \Yasym_4\\ 
\hline
5_25_2,\ 5_35_3 &U(4)_1\times U(4)_2\times U(4)_3\times U(4)_4\\
\sus     &(\Yfun_1,\Yfunb_3),\ (\Yfunb_1,\Yfunb_3),\ (\Yfun_1,\Yfunb_4),\ 
          (\Yfun_2,\Yfunb_1),\ (\Yfun_2,\Yfun_4),\ (\Yfun_2,\Yfunb_4),\\
         &(\Yfun_3,\Yfunb_1),\ (\Yfun_3,\Yfunb_2),\ (\Yfun_4,\Yfunb_2),\ 
          (\Yfun_4,\Yfunb_3),\ \Yasym_1,\ \Yasymb_2,\ \Yasym_3,\ \Yasymb_4\\  
\hline
95_{1,n}  &\spin (\Yfunb_1,\Yfun_1),\ (\Yfun_1,\Yfun_3),\ 
                 (\Yfunb_3,\Yfun_2),\ (\Yfun_3,\Yfun_4)\\
\nsus     &\scal (\Yfunb_2,\Yfun_2),\ (\Yfun_2,\Yfun_3),\ 
                 (\Yfunb_4,\Yfun_1),\ (\Yfun_4,\Yfun_4)\\
\hline
95_2      &(\Yfunb_1,\Yfun_1),\ (\Yfun_1,\Yfun_3),\ (\Yfunb_2,\Yfun_2),\ 
           (\Yfun_2,\Yfunb_3),\\ 
\sus      &(\Yfunb_3,\Yfunb_2),\ (\Yfun_3,\Yfunb_4),\ (\Yfunb_4,\Yfunb_1),\ 
           (\Yfun_4,\Yfun_4)\\
\hline
95_3      &(\Yfun_1,\Yfun_1),\ (\Yfunb_1,\Yfun_3),\ (\Yfunb_2,\Yfunb_1),\ 
            (\Yfun_2,\Yfun_4),\\
\sus      &(\Yfunb_3,\Yfunb_4),\ (\Yfun_3,\Yfunb_2),\ (\Yfunb_4,\Yfun_2),\ 
           (\Yfun_4,\Yfunb_3)\\
\hline
5_25_3    &(\Yfunb_1,\Yfunb_1),\ (\Yfun_1,\Yfun_3),\ (\Yfunb_2,\Yfunb_4),\ 
           (\Yfun_2,\Yfun_2),\\
\sus      &(\Yfun_3,\Yfun_1),\ (\Yfunb_3,\Yfunb_3),\ (\Yfun_4,\Yfun_4),\ 
           (\Yfunb_4,\Yfunb_2)\\
\hline
5_35_{1,n}  &\spin (\Yfun_1,\Yfun_3),\ (\Yfunb_4,\Yfun_2),\ 
                   (\Yfunb_2,\Yfun_4),\ (\Yfun_3,\Yfun_1)\\
\nsus       &\scal (\Yfun_1,\Yfun_1),\ (\Yfunb_4,\Yfun_4),\ 
                   (\Yfunb_2,\Yfun_2),\ (\Yfun_3,\Yfun_3)\\
\hline
5_{1,n}5_2  &\spin (\Yfun_1,\Yfun_1),\ (\Yfun_2,\Yfunb_2),\ 
                   (\Yfun_3,\Yfun_3),\ (\Yfun_4,\Yfunb_4)\\
\nsus       &\scal (\Yfun_1,\Yfunb_3),\ (\Yfun_2,\Yfun_4),\ 
                   (\Yfun_3,\Yfunb_1),\ (\Yfun_4,\Yfun_2)\\
\hline
\end{array}$$
\caption{Spectrum of the $\bZ_2\times\bZ_4$ orientifold without discrete
         torsion and $(\mu_3,\mu_1)=(-1,+1)$. The four sets of 
         $D5_{1,n}$-branes are located at the four $\bZ_4$ fixed points
         in the second and third torus. 
         The notation is explained in the tables of section \ref{Z2Z2}.
         \label{Z2Z4spec4}}
\end{table}

\begin{table}[htp]
\renewcommand{\arraystretch}{1.25}
$$\begin{array}{|c|c|}
\hline
\multicolumn{2}{|c|}{\IZ_2\times\IZ_4,\ 
       (\alpha_1,\alpha_2,\alpha_3)=(-1,-1,-1),\ \eps=1}
   \\ \hline\hline
\multicolumn{2}{|c|}{\hbox{closed string spectrum}}\\ \hline
\rm sector       &\hbox{$\cN=1$ multiplets}\\ \hline
\rm untw.        &\hbox{gravity, 1 lin., 4 chir.} \\ \hline
\hbox{order-two} &\hbox{34 lin.} \\ \hline
\rm remaining    &\hbox{20 lin., 4 chir.} \\ \hline
\end{array}$$
\caption{Closed string spectrum of the $\bZ_2\times\bZ_4$ orientifold 
         without discrete torsion and $(\mu_3,\mu_1)=(+1,+1)$. 
         \label{Z2Z4spec3b}}
\end{table}

\begin{table}[htp]
\renewcommand{\arraystretch}{1.25}
$$\begin{array}{|c|c|}
\hline
\multicolumn{2}{|c|}{\IZ_2\times\IZ_4,\ 
       (\alpha_1,\alpha_2,\alpha_3)=(-1,+1,+1),\ \eps=1}
   \\ \hline\hline
\multicolumn{2}{|c|}{\hbox{closed string spectrum}}\\ \hline
\rm sector       &\hbox{$\cN=1$ multiplets}\\ \hline
\rm untw.        &\hbox{gravity, 1 lin., 4 chir.} \\ \hline
\hbox{order-two} &\hbox{10 lin., 24 chir.} \\ \hline
\rm remaining    &\hbox{20 lin., 4 chir.} \\ \hline
\end{array}$$
\caption{Closed string spectrum of the $\bZ_2\times\bZ_4$ orientifold 
         without discrete torsion and $(\mu_3,\mu_1)=(-1,+1)$. 
         \label{Z2Z4spec4b}}
\end{table}

\subsection{$\bZ_4\times\bZ_4$, $v=\frac14(1,-1,0)$, $w=\frac14(0,1,-1)$}
\label{Z4Z4}

This orientifold is symmetric under a permutation of the three sets 
of $D5_i$-branes. Thus, only four of the eight possible $\IZ_4\times\IZ_4$ 
models are inequivalent. However, only for the models with vector structure 
in each of the two $\IZ_4$ factors, \ie $\mu_3=\mu_1=+1$, does a consistent 
solution to the tadpole equations exist, at least in the absence of
Wilson lines. From (\ref{all_signs}), we find that 
$\alpha_1\alpha_2\alpha_3=-1$. Thus, only an odd number of negative 
$\alpha_i$'s is allowed. We found a solution for $(-1,-1,-1)$ with and
without discrete torsion.

The only tadpole conditions with \nonvan Klein bottle contribution are
(\ref{tad_untw}) and (\ref{tad_fixeven1}), corresponding to the untwisted 
sector and the the three order-two sectors $\bar k_1=(0,2)$, $\bar k_2=(2,2)$,
$\bar k_3=(2,0)$. The untwisted tadpoles fix the rank of the gauge group to 
be 8 (16) for each of the four sectors 99, $5_i5_i$ in the case with
(without) discrete torsion. For the $(-1,-1,-1)$ model, the tadpole 
conditions of the $\bar k_i$ sector read
\be  \label{tadpZ4Z4}
\Tr(\gamma_{\bar k_i,9})=0,\quad \Tr(\gamma_{\bar k_i,5_i,n_i})=
 \eps^{-k_{i,1}k_{i,2}/4}\,8,\quad
\Tr(\gamma_{\bar k_i,5_j,n_i})=\Tr(\gamma_{\bar k_i,5_k,n_i})=0,
\ee
where $k_{i,1}$, $k_{i,2}$ are the components of the two-vector $\bar k_i$
and $n_i=0,1,2,3$ denotes the $\IZ_4$ fixed points in the $j\th$ and
$k\th$ complex plane.\footnote{Here, $(ijk)$ is a permutation of $(123)$.}
We assumed that there are no $D5_i$-branes at the 12 remaining $\IZ_2$ fixed 
points. The sign $\eps^{-k_{i,1}k_{i,2}/4}$ is $-1$ for the $(2,2)$ sector 
in the model with discrete torsion and $+1$ else.
The conditions (\ref{tadpZ4Z4}) were obtained from (\ref{tad_fixeven1}) by
analysing the contributions of the $D5_i$-branes to the 16 fixed points 
and using the arguments given at the end of section \ref{Z2Z2}.
Generalising the argument showing that in the $\IZ_2\times\IZ_4$ orientifold
$\gamma_{(0,2),5_2}$ is traceless, one can verify that in the present case 
the matrices $\gamma_{\bar k_i,5_j}$ are traceless for $i\neq j$. Therefore 
the last two conditions in (\ref{tadpZ4Z4}) are trivially satisfied. The 
second condition in (\ref{tadpZ4Z4}) forces us to put some $D5_i$-branes 
at each of the four $\IZ_4$ fixed points in the $j\th$ and $k\th$ complex 
plane. In the following, we will consider the most symmetric situation 
where eight $D5_i$-branes sit at each $\IZ_4$ fixed point. 
This leads to a unique solution of the tadpole equations. The complete
spectrum is displayed in tables \ref{Z4Z4spec1} -- \ref{Z4Z4spec2b}.

\begin{table}[htp]
\renewcommand{\arraystretch}{1.25}
$$\begin{array}{|c|c|}
\hline
\multicolumn{2}{|c|}{\IZ_4\times\IZ_4,\ 
       (\alpha_1,\alpha_2,\alpha_3)=(-1,-1,-1),\ \eps=-1}
   \\ \hline\hline
\multicolumn{2}{|c|}{\hbox{open string spectrum}}\\ \hline
\rm sector &\hbox{gauge group / matter fields}\\ \hline
99    &SO(4)_1\times SO(4)_2\times SO(4)_3\times USp(4)_4\\
\sus  &(\Yfun_1,\Yfun_2),\ (\Yfun_1,\Yfun_3),\ (\Yfun_1,\Yfun_4),\ 
       (\Yfun_2,\Yfun_3),\ (\Yfun_2,\Yfun_4),\ (\Yfun_3,\Yfun_4)\\
\hline
5_{i,n}5_{i,n} &U(2)\\
\nsus          &\spin adj,\ \Yasym,\ \Yasymb,\quad\scal \Ysym,\ \Ysymb\\
\hline
95_{1,n} &\spin (\Yfun_1,\Yfunb),\ (\Yfun_3,\Yfun)\\
\nsus    &\scal (\Yfun_2,\Yfunb),\ (\Yfun_4,\Yfun)\\
\hline
95_{2,n} &\spin (\Yfun_2,\Yfun),\ (\Yfun_3,\Yfunb)\\
\nsus    &\scal (\Yfun_1,\Yfunb),\ (\Yfun_4,\Yfun)\\
\hline
95_{3,n} &\spin (\Yfun_1,\Yfun),\ (\Yfun_2,\Yfunb)\\
\nsus    &\scal (\Yfun_3,\Yfunb),\ (\Yfun_4,\Yfun)\\
\hline
5_{i,n}5_{j,m}\ \sus &(\Yfunb,\Yfun)\\
\hline\hline
\multicolumn{2}{|c|}{\hbox{closed string spectrum}}\\ \hline
\rm sector       &\hbox{$\cN=1$ multiplets}\\ \hline
\rm untw.        &\hbox{gravity, 1 lin., 3 chir.} \\ \hline
\hbox{order-two} &\hbox{27 lin.} \\ \hline
\rm remaining    &\hbox{12 lin.} \\ \hline
\end{array}$$
\caption{Spectrum of the $\bZ_4\times\bZ_4$ orientifold with discrete 
        torsion and $(\mu_3,\mu_1)=(+1,+1)$. The 12 sets of 
        $D5_{i,n}$-branes, $i=1,2,3$, $n=0,1,2,3$, are located at 
        the four $\bZ_4$ fixed points in the $j\th$ and $k\th$ torus, 
        where $(ijk)$ is a permutation of $(123)$. The matter in the 
        $5_{i,n}5_{j,m}$ sectors is only present if the fixed points $n$
        and $m$ are located at the same point in the $k\th$ torus. 
        The notation is explained in the tables of section \ref{Z2Z2}. 
         \label{Z4Z4spec1}}
\end{table}

\begin{table}[htp]
\renewcommand{\arraystretch}{1.18}
$$\begin{array}{|c|c|}
\hline
\multicolumn{2}{|c|}{\IZ_4\times\IZ_4,\ 
       (\alpha_1,\alpha_2,\alpha_3)=(-1,-1,-1),\ \eps=1}
   \\ \hline\hline
\multicolumn{2}{|c|}{\hbox{open string spectrum}}\\ \hline
\rm sector &\hbox{gauge group / matter fields}\\ \hline
99    &SO(2)_1\times U(2)_2\times SO(2)_3\times U(2)_4\times U(2)_5\times 
       U(2)_6\\
      &\qquad\times U(2)_7\times SO(2)_8\times U(2)_9\times SO(2)_{10}\\
\sus  &(\Yfun_1,\Yfun_2),\ (\Yfun_1,\Yfunb_4),\ (\Yfun_1,\Yfun_7),\ 
       (\Yfunb_2,\Yfun_3),\ (\Yfun_2,\Yfunb_5),\ (\Yfun_2,\Yfun_6),\\
      &(\Yfunb_2,\Yfunb_7),\ (\Yfun_3,\Yfun_5),\ (\Yfun_3,\Yfunb_6),\ 
       (\Yfun_4,\Yfunb_2),\ (\Yfun_4,\Yfunb_7),\ (\Yfun_4,\Yfun_8),\\
      &(\Yfunb_4,\Yfunb_9),\ (\Yfun_5,\Yfunb_4),\ (\Yfunb_5,\Yfun_8),\ 
       (\Yfun_5,\Yfunb_9),\ (\Yfun_6,\Yfunb_5),\ (\Yfun_6,\Yfun_{10}),\\
      &(\Yfun_7,\Yfunb_6),\ (\Yfun_7,\Yfun_9),\ (\Yfunb_7,\Yfun_{10}),\ 
       (\Yfun_8,\Yfun_9),\ (\Yfun_9,\Yfunb_6),\ (\Yfunb_9,\Yfun_{10})\\
\hline
5_{i,n}5_{i,n}  &U(1)_1\times U(1)_2\times U(1)_3\times U(1)_4\\
\nsus     &\smult (\Yfun_1,\Yfunb_3),\ (\Yfun_2,\Yfunb_4)\\ 
          &\spin 4\ {\rm singlets},\quad
           \scal \Ysymb_1,\ \Ysymb_2,\ \Ysym_3,\ \Ysym_4\\
\hline
95_{1,n} &\spin (\Yfun_1,\Yfunb_1),\ (\Yfun_3,\Yfunb_2),\ 
                (\Yfunb_4,\Yfun_1),\ (\Yfun_4,\Yfunb_3),\\ 
\nsus    &\qquad (\Yfunb_6,\Yfun_2),\ (\Yfun_6,\Yfunb_4),\ 
                 (\Yfun_8,\Yfun_3),\ (\Yfun_{10},\Yfun_4)\\
         &\scal (\Yfun_2,\Yfunb_1),\ (\Yfunb_2,\Yfunb_2),\ 
                (\Yfunb_5,\Yfun_2),\ (\Yfun_5,\Yfunb_3),\\
         &\qquad (\Yfunb_7,\Yfun_1),\ (\Yfun_7,\Yfunb_4),\ 
                 (\Yfun_9,\Yfun_3),\ (\Yfunb_9,\Yfun_4)\\
\hline
95_{2,n} &\spin (\Yfun_1,\Yfun_4),\ (\Yfun_3,\Yfun_3),\ 
                (\Yfun_5,\Yfunb_3),\ (\Yfunb_5,\Yfun_1),\\
\nsus    &\qquad (\Yfun_7,\Yfunb_4),\ (\Yfunb_7,\Yfun_2),\ 
                 (\Yfun_8,\Yfunb_1),\ (\Yfun_{10},\Yfunb_2)\\
         &\scal (\Yfun_2,\Yfunb_3),\ (\Yfunb_2,\Yfunb_4),\ 
                (\Yfunb_4,\Yfun_4),\ (\Yfun_4,\Yfunb_1),\\
         &\qquad (\Yfunb_6,\Yfun_3),\ (\Yfun_6,\Yfunb_2),\ 
                 (\Yfunb_9,\Yfun_1),\ (\Yfun_9,\Yfun_2)\\
\hline
95_{3,n} &\spin (\Yfun_1,\Yfun_4),\ (\Yfun_2,\Yfunb_4),\ 
                (\Yfunb_2,\Yfun_2),\ (\Yfun_3,\Yfunb_2),\\
\nsus    &\qquad (\Yfun_8,\Yfun_3),\ (\Yfun_9,\Yfunb_3),\ 
                 (\Yfunb_9,\Yfun_1),\ (\Yfun_{10},\Yfunb_1)\\ 
         &\scal (\Yfun_4,\Yfunb_4),\ (\Yfunb_4,\Yfunb_3),\ 
                (\Yfun_5,\Yfunb_2),\ (\Yfunb_5,\Yfun_3),\\
         &\qquad (\Yfun_6,\Yfun_2),\ (\Yfunb_6,\Yfun_1),\ 
                 (\Yfun_7,\Yfun_4),\ (\Yfunb_7,\Yfunb_1)\\
\hline
5_{2,n}5_{3,m}\ \sus &(\Yfun_3,\Yfunb_2),\ (\Yfun_4,\Yfun_4),\ 
                      (\Yfunb_1,\Yfun_3),\ (\Yfunb_2,\Yfunb_1)\\
\hline
5_{3,n}5_{1,m}\ \sus &(\Yfun_4,\Yfunb_1),\ (\Yfunb_2,\Yfunb_2),\ 
                      (\Yfun_3,\Yfun_3),\ (\Yfunb_1,\Yfun_4)\\
\hline
5_{1,n}5_{2,m}\ \sus &(\Yfunb_1,\Yfun_4),\ (\Yfunb_2,\Yfun_3),\ 
                      (\Yfun_3,\Yfunb_1),\ (\Yfun_4,\Yfunb_2)\\
\hline
\end{array}$$
\caption{Spectrum of the $\bZ_4\times\bZ_4$ orientifold without discrete 
        torsion and $(\mu_3,\mu_1)=(+1,+1)$. There are 12 sets of 
        $D5_{i,n}$-branes, $i=1,2,3$, $n=0,1,2,3$, as in the model of
        table \ref{Z4Z4spec1}. The matter in the $5_{i,n}5_{j,m}$ sectors 
        is only present if the fixed points $n$ and $m$ are located at the 
        same point in the $k\th$ torus. The symmetric tensors in the
        $5_{i,n}5_{i,n}$ sectors are fields that carry double $U(1)$ charge.
         \label{Z4Z4spec2}}
\end{table}

\begin{table}[htp]
\renewcommand{\arraystretch}{1.25}
$$\begin{array}{|c|c|}
\hline
\multicolumn{2}{|c|}{\IZ_4\times\IZ_4,\ 
       (\alpha_1,\alpha_2,\alpha_3)=(-1,-1,-1),\ \eps=1}
   \\ \hline\hline
\multicolumn{2}{|c|}{\hbox{closed string spectrum}}\\ \hline
\rm sector       &\hbox{$\cN=1$ multiplets}\\ \hline
\rm untw.        &\hbox{gravity, 1 lin., 3 chir.} \\ \hline
\hbox{order-two} &\hbox{27 lin.} \\ \hline
\rm remaining    &\hbox{48 lin., 12 chir.} \\ \hline
\end{array}$$
\caption{Closed string spectrum of the $\bZ_4\times\bZ_4$ orientifold without 
        discrete torsion and $(\mu_3,\mu_1)=(+1,+1)$.          
        \label{Z4Z4spec2b}}
\end{table}

It is straightforward to generalise the argument for the impossibility of 
cancelling the RR twisted tadpoles in the $\IZ_2\times\IZ_4$ $(+1,+1,-1)$
model. In the $\IZ_4\times\IZ_4$ $(+1,+1,-1)$ model\footnote{The three models
$(+1,+1,-1)$, $(+1,-1,+1)$  and $(-1,+1,+1)$ are equivalent up to a 
permutation of the $D5_i$-branes.}, the incompatibility between the 
tadpole equations appears in in the sectors $(0,2)$ and $(2,2)$, 
independently of the value of discrete torsion. The problem appears in 
order-two sectors that fix a set of branes without vector structure. 
Only the $(-1,-1,-1)$ model with and without discrete torsion is found 
to be consistent.

\subsection{$\bZ_2\times\bZ_6$, $v=\frac12(1,-1,0)$, $w=\frac16(0,1,-1)$}
\label{Z2Z6}
This orientifold is not symmetric under an arbitrary permutation of the 
three sets of $D5_i$-branes. The only even twist, $\bar k=(0,2)$, fixes
the world-volume of the $D5_1$-branes. As in the $\IZ_2\times\IZ_4$ case, 
only the $D5_2$- and $D5_3$-branes appear on the same footing. Therefore, 
we expect that six of the eight possible $\IZ_2\times\IZ_6$ models are 
inequivalent.
The $(+1,+1,+1)$ model has already been constructed in \cite{z}.

Let us first concentrate on the cases with discrete torsion. 
The only \nontriv tadpole conditions are (\ref{tad_untw}) and 
(\ref{tad_fixeven1}), corresponding to the untwisted sector and the
$(0,2)$ sector. The former fixes the rank of the gauge group to be 8 for
each of the four sectors 99, $5_i5_i$. The latter read
\be  \label{tadpZ2Z6}
\Tr(\gamma_{(0,2),9})=\Tr(\gamma_{(0,2),5_1})=-\alpha_1\,8,\quad
\Tr(\gamma_{(0,2),5_2})=\Tr(\gamma_{(0,2),5_3})=\alpha_1\,8.
\ee
We assumed that all $D5_i$-branes are located at the origin. To see that
the tadpole conditions (\ref{tad_fixeven1}) indeed simplify to the form
(\ref{tadpZ2Z6}), we need to consider the Klein bottle contributions to
the various fixed points of the $(0,2)$ sector, using the method described
at the end of section \ref{tadpoles}. Label the three $\IZ_3$
fixed points in the second plane by $n_2=0,1,2$ and the three $\IZ_3$ 
fixed points in the third plane $n_3=0,1,2$, where $0$ denotes the origin.
The Klein bottle contribution $\cK_0$ (the term proportional to $c_2c_3$ in
(\ref{tad_fixeven1})) is only present at the origin $(0,0)$, $\cK_1$ (the 
term proportional to $\eps_1\,\beta_1\,s_2s_3$) is present at all nine 
$\IZ_3$ fixed points $(n_2,n_3)$, $\cK_2$ (the term proportional to 
$\eps_2\,\beta_2$) is present at the three fixed points $(0,n_3)$ and
$\cK_3$ (the term proportional to $\eps_3\,\beta_3$) is present at the 
three fixed points $(n_2,0)$. Using this and $\alpha_1=\alpha_2\alpha_3$,
it is easy to see that (\ref{tad_fixeven1}) reduces to (\ref{tadpZ2Z6}).
These equations have a unique solution. The results are displayed in tables
\ref{Z2Z6spec1} -- \ref{Z2Z6spec3}.

In the cases without discrete torsion, the tadpole conditions for the $(0,2)$
sector are identical to eq.\ (\ref{tadpZ2Z6}) if all $D5_i$-branes sit at
the origin. Now, the remaining twisted sectors give additional conditions
because, in general, the $\gamma$ matrices are not traceless. The tadpole
equations have 25 distinct solutions for each of the three possible models. 
We display the complete spectrum of the $(-1,-1,-1)$ model in tables
\ref{Z2Z6spec4} and \ref{Z2Z6spec4b} and restrict ourselves to give the 
99 and $5_i5_i$ spectrum of the two remaining models in tables 
\ref{Z2Z6spec5} and \ref{Z2Z6spec6}. We determined also the $95_i$
and $5_i5_j$ spectrum of the last two models and verified that the
complete spectrum is free of \nonab gauge anomalies.

\begin{table}[htp]
\renewcommand{\arraystretch}{1.25}
$$\begin{array}{|c|c|}
\hline
\multicolumn{2}{|c|}{\IZ_2\times\IZ_6,\ 
       (\alpha_1,\alpha_2,\alpha_3)=(+1,+1,+1),\ \eps=-1}
   \\ \hline\hline
\multicolumn{2}{|c|}{\hbox{open string spectrum}}\\ \hline
\rm sector &\hbox{gauge group / matter fields}\\ \hline
99,\ 5_i5_i    &U(4)_1\times USp(8)_2\\
\sus  &(\Yfun_1,\Yfun_2),\  (\Yfunb_1,\Yfun_2),\ 
       adj_1,\ \Yasym_1,\ \Yasymb_1,\ \Yasym_2\\
\hline
95_1,\ 5_25_3\ \sus 
     &(\Yfun_1,\Yfunb_1),\ (\Yfunb_1,\Yfun_1),\ (\Yfun_2,\Yfun_2)\\
\hline
95_2,\ 5_15_3\ \sus 
     &(\Yfun_1,\Yfun_2),\ (\Yfunb_1,\Yfun_1),\ (\Yfun_2,\Yfunb_1)\\
\hline
95_3,\ 5_15_2\ \sus 
     &(\Yfunb_1,\Yfun_2),\ (\Yfun_1,\Yfunb_1),\ (\Yfun_2,\Yfun_1)\\
\hline\hline
\multicolumn{2}{|c|}{\hbox{closed string spectrum}}\\ \hline
\rm sector       &\hbox{$\cN=1$ multiplets}\\ \hline
\rm untw.        &\hbox{gravity, 1 lin., 4 chir.} \\ \hline
\hbox{order-two} &\hbox{14 chir.} \\ \hline
\rm remaining    &\hbox{12 lin., 6 chir., 2 vec.} \\ \hline
\end{array}$$
\caption{Spectrum of the $\bZ_2\times\bZ_6$ orientifold with discrete
         torsion and $(\mu_3,\mu_1)=(-1,-1)$. The notation is explained
         in the tables of section \ref{Z2Z2}.
         \label{Z2Z6spec1}}
\end{table}

\begin{table}[htp]
\renewcommand{\arraystretch}{1.2}
$$\begin{array}{|c|c|}
\hline
\multicolumn{2}{|c|}{\IZ_2\times\IZ_6,\ 
       (\alpha_1,\alpha_2,\alpha_3)=(-1,+1,-1),\ \eps=-1}
   \\ \hline\hline
\multicolumn{2}{|c|}{\hbox{open string spectrum}}\\ \hline
\rm sector &\hbox{gauge group / matter fields}\\ \hline
99,\ 5_25_2  &SO(8)_1\times U(4)_2\\
\sus  &(\Yfun_1,\Yfun_2),\ (\Yfun_1,\Yfunb_2),\ 
       \Yasym_1,\ \Yasymb_2,\ \Ysym_2,\ adj_2\\
\hline
5_15_1,\ 5_35_3  &USp(8)_1\times U(4)_2\\
\nsus  &\smult (\Yfun_1,\Yfun_2),\ (\Yfun_1,\Yfunb_2)\\
       &\spin 2\,\Yasym_1,\ \Yasymb_2,\ \Ysym_2,\ 2\,adj_2,\quad
        \scal \Ysym_1,\ \Ysymb_2,\ \Yasym_2,\ adj_2,\\
\hline
95_1,\ 5_25_3   
       &\spin (\Yfun_1,\Yfun_1),\ (\Yfun_2,\Yfunb_2),\ (\Yfunb_2,\Yfun_2)\\  
\nsus  &\scal (\Yfun_1,\Yfun_2),\ (\Yfun_2,\Yfun_1),\ (\Yfunb_2,\Yfunb_2)\\
\hline
95_2,\ 5_35_1\ \sus 
       &(\Yfun_1,\Yfunb_2),\ (\Yfun_2,\Yfun_2),\ (\Yfunb_2,\Yfun_1)\\
\hline
95_3,\ 5_15_2  
       &\spin (\Yfun_1,\Yfunb_2),\ (\Yfun_2,\Yfun_2),\ (\Yfunb_2,\Yfun_1)\\
\nsus  &\scal (\Yfun_1,\Yfun_2),\ (\Yfunb_2,\Yfunb_2),\ (\Yfun_2,\Yfun_1)\\
\hline\hline
\multicolumn{2}{|c|}{\hbox{closed string spectrum}}\\ \hline
\rm sector       &\hbox{$\cN=1$ multiplets}\\ \hline
\rm untw.        &\hbox{gravity, 1 lin., 4 chir.} \\ \hline
\hbox{order-two} &\hbox{4 chir., 10 vec.} \\ \hline
\rm remaining    &\hbox{12 lin., 6 chir., 2 vec.} \\ \hline
\end{array}$$
\caption{Spectrum of the $\bZ_2\times\bZ_6$ orientifold with discrete
         torsion and $(\mu_3,\mu_1)=(+1,+1)$. The notation is explained
         in the tables of section \ref{Z2Z2}.
         \label{Z2Z6spec2}}
\end{table}

\begin{table}[htp]
\renewcommand{\arraystretch}{1.2}
$$\begin{array}{|c|c|}
\hline
\multicolumn{2}{|c|}{\IZ_2\times\IZ_6,\ 
       (\alpha_1,\alpha_2,\alpha_3)=(+1,-1,-1),\ \eps=-1}
   \\ \hline\hline
\multicolumn{2}{|c|}{\hbox{open string spectrum}}\\ \hline
\rm sector &\hbox{gauge group / matter fields}\\ \hline
99,\ 5_15_1  &SO(8)_1\times U(4)_2\\
\sus  &(\Yfun_1,\Yfun_2),\ (\Yfun_1,\Yfunb_2),\ 
       \Ysym_1,\ \Yasym_2,\ \Yasymb_2,\ adj_2\\
\hline
5_25_2,\ 5_35_3  &USp(8)_1\times U(4)_2\\
\nsus  &\smult (\Yfun_1,\Yfun_2),\ (\Yfun_1,\Yfunb_2)\\
       &\spin \Yasym_1,\ \Ysym_1,\ \Yasym_2,\ \Yasymb_2,\ 2\,adj_2\quad
        \scal \Yasym_1,\ \Ysym_2,\ \Ysymb_2,\ adj_2\\
\hline
95_1,\ 5_25_3\ \sus
       &(\Yfun_1,\Yfun_1),\ (\Yfun_2,\Yfunb_2),\ (\Yfunb_2,\Yfun_2)\\ 
\hline
95_2,\ 5_15_3
       &\spin (\Yfun_1,\Yfun_2),\ (\Yfunb_2,\Yfun_1),\ (\Yfun_2,\Yfunb_2)\\
\nsus  &\scal (\Yfun_1,\Yfun_2),\ (\Yfunb_2,\Yfun_1),\ (\Yfun_2,\Yfunb_2)\\
\hline
95_3,\ 5_15_2  
       &\spin (\Yfun_1,\Yfunb_2),\ (\Yfun_2,\Yfun_1),\ (\Yfunb_2,\Yfun_2)\\  
\nsus  &\scal (\Yfun_1,\Yfun_2),\ (\Yfunb_2,\Yfun_1),\ (\Yfun_2,\Yfunb_2)\\
\hline\hline
\multicolumn{2}{|c|}{\hbox{closed string spectrum}}\\ \hline
\rm sector       &\hbox{$\cN=1$ multiplets}\\ \hline
\rm untw.        &\hbox{gravity, 1 lin., 4 chir.} \\ \hline
\hbox{order-two} &\hbox{6 chir., 8 vec.} \\ \hline
\rm remaining    &\hbox{12 lin., 6 chir., 2 vec.} \\ \hline
\end{array}$$
\caption{Spectrum of the $\bZ_2\times\bZ_6$ orientifold with discrete
         torsion and $(\mu_3,\mu_1)=(+1,-1)$. The notation is explained
         in the tables of section \ref{Z2Z2}.
         \label{Z2Z6spec3}}
\end{table}

\begin{table}[htp]
\renewcommand{\arraystretch}{1.25}
$$\begin{array}{|c|c|}
\hline
\multicolumn{2}{|c|}{\IZ_2\times\IZ_6,\ 
       (\alpha_1,\alpha_2,\alpha_3)=(-1,-1,-1),\ \eps=1}
   \\ \hline\hline
\multicolumn{2}{|c|}{\hbox{open string spectrum}}\\ \hline
\rm sector &\hbox{gauge group / matter fields}\\ \hline
99   &SO(8-2t_1)_1\times U(4-t_1)_2\times U(t_1)_3\times SO(2t_1)_4\\
     &\qquad \times SO(8-2t_1)_5\times U(4-t_1)_6\times U(t_1)_7\times 
       SO(2t_1)_8\\
\sus  &(\Yfun_1,\Yfun_2),\ (\Yfun_1,\Yfun_5),\ (\Yfun_1,\Yfunb_6),\ 
       (\Yfunb_2,\Yfun_5),\ (\Yfun_2,\Yfunb_6),\ (\Yfun_2,\Yfunb_7),\\
      &(\Yfun_3,\Yfunb_2),\ (\Yfunb_3,\Yfun_4),\ (\Yfun_3,\Yfunb_7),\ 
       (\Yfun_3,\Yfun_8),\ (\Yfun_4,\Yfun_7),\ (\Yfun_4,\Yfun_8),\\
      &(\Yfun_5,\Yfun_6),\ (\Yfun_6,\Yfunb_2),\ (\Yfun_6,\Yfunb_3),\ 
       (\Yfun_7,\Yfunb_3),\ (\Yfun_7,\Yfunb_6),\ (\Yfunb_7,\Yfun_8)\\
\hline
5_15_1 &U(2t_1)_1\times U(t_1)_2\times U(4-t_1)_3\times U(8-2t_1)_4
        \times U(4-t_1)_5\times U(t_1)_6\\
\nsus  &\smult (\Yfunb_1,\Yfunb_2),\ (\Yfun_1,\Yfun_6),\ (\Yfun_1,\Yfunb_6),\ 
        (\Yfun_2,\Yfunb_1),\ (\Yfun_2,\Yfun_5),\ (\Yfun_2,\Yfun_6),\\
       &\qquad (\Yfunb_2,\Yfunb_6),\ (\Yfun_3,\Yfunb_2),\ (\Yfun_3,\Yfun_4),\ 
        (\Yfun_3,\Yfun_5),\ (\Yfunb_3,\Yfunb_5),\ (\Yfunb_3,\Yfunb_6),\\
       &\qquad (\Yfun_4,\Yfunb_3),\ (\Yfunb_4,\Yfunb_5),\ (\Yfun_5,\Yfunb_4),\
        (\Yfun_6,\Yfunb_5)\\
       &\spin adj_1,\ \ldots,\  adj_6,\ 
              \Yasym_1,\ \Yasymb_1,\ \Yasym_4,\ \Yasymb_4\\ 
       &\scal \Ysym_1,\ \Ysymb_1,\ \Ysym_4,\ \Ysymb_4\\ 
\hline
5_25_2,\ 5_35_3 &U(4-t_2)_1\times U(4)_2\times U(t_2)_3\times U(t_2)_4
        \times U(4)_5\times U(4-t_2)_6\\
\nsus  &(\Yfun_1,\Yfun_5),\ (\Yfun_1,\Yfun_6),\ (\Yfun_1,\Yfunb_6),\ 
        (\Yfunb_1,\Yfunb_6),\ (\Yfun_2,\Yfunb_1),\ (\Yfun_2,\Yfun_4),\\
       &(\Yfun_2,\Yfun_5),\ (\Yfunb_2,\Yfunb_5),\ (\Yfunb_2,\Yfunb_6),\ 
        (\Yfun_3,\Yfunb_2),\ (\Yfun_3,\Yfun_4),\ (\Yfunb_3,\Yfunb_4),\\
       &(\Yfunb_3,\Yfunb_5),\ (\Yfun_4,\Yfunb_3),\ (\Yfun_5,\Yfunb_4),\ 
        (\Yfun_6,\Yfunb_5)\\
       &\spin adj_1,\ \ldots,\ adj_6,\ 
              \Yasymb_1,\ \Yasym_3,\ \Yasymb_4,\ \Yasym_6\\
       &\scal \Ysymb_1,\ \Ysym_3,\ \Ysymb_4,\ \Ysym_6\\
\hline
95_1  &\spin (\Yfun_1,\Yfunb_1),\ (\Yfun_2,\Yfunb_2),\ (\Yfunb_2,\Yfunb_6),\ 
      (\Yfun_3,\Yfunb_3),\ (\Yfunb_3,\Yfunb_5),\ (\Yfun_4,\Yfunb_4),\\
\nsus &\qquad (\Yfun_5,\Yfun_1),\ (\Yfunb_6,\Yfun_2),\ (\Yfun_6,\Yfun_6),\ 
       (\Yfunb_7,\Yfun_3),\ (\Yfun_7,\Yfun_5),\ (\Yfun_8,\Yfun_4)\\
      &\scal (\Yfun_1,\Yfunb_6),\ (\Yfun_2,\Yfunb_1),\ (\Yfunb_2,\Yfunb_5),\ 
       (\Yfun_3,\Yfunb_2),\ (\Yfunb_3,\Yfunb_4),\ (\Yfun_4,\Yfunb_3),\\ 
      &\qquad (\Yfun_5,\Yfun_2),\ (\Yfun_6,\Yfun_1),\ (\Yfunb_6,\Yfun_3),\ 
       (\Yfunb_7,\Yfun_4),\ (\Yfun_7,\Yfun_6),\ (\Yfun_8,\Yfun_5)\\
\hline
\end{array}$$
\caption{Spectrum of the $\bZ_2\times\bZ_6$ orientifold without discrete
         torsion and $(\mu_3,\mu_1)=(+1,+1)$. There are 25 solutions to
         the tadpole equations, parametrised by $t_1,t_2=0,\ldots,4$.
         The notation is explained in the tables of section \ref{Z2Z2}.
         \label{Z2Z6spec4}}
\end{table}

\begin{table}[htp]
\renewcommand{\arraystretch}{1.25}
$$\begin{array}{|c|c|}
\hline
\multicolumn{2}{|c|}{\IZ_2\times\IZ_6,\ 
       (\alpha_1,\alpha_2,\alpha_3)=(-1,-1,-1),\ \eps=1}
   \\ \hline\hline
\multicolumn{2}{|c|}{\hbox{open string spectrum}}\\ \hline
\rm sector &\hbox{gauge group / matter fields}\\ \hline
95_2  &\spin (\Yfun_1,\Yfun_6),\ (\Yfunb_2,\Yfun_1),\ (\Yfun_2,\Yfun_5),\ 
       (\Yfunb_3,\Yfun_2),\ (\Yfun_3,\Yfun_4),\ (\Yfun_4,\Yfun_3),\\
\nsus &\qquad (\Yfun_5,\Yfunb_1),\ (\Yfun_6,\Yfunb_2),\ (\Yfunb_6,\Yfunb_6),\ 
       (\Yfun_7,\Yfunb_3),\ (\Yfunb_7,\Yfunb_5),\ (\Yfun_8,\Yfunb_4)\\
      &\scal (\Yfun_1,\Yfunb_1),\ (\Yfun_2,\Yfunb_2),\ (\Yfunb_2,\Yfunb_6),\ 
       (\Yfun_3,\Yfunb_3),\ (\Yfunb_3,\Yfunb_5),\ (\Yfun_4,\Yfunb_4),\\
      &\qquad (\Yfun_5,\Yfun_6),\ (\Yfunb_6,\Yfun_1),\ (\Yfun_6,\Yfun_5),\ 
       (\Yfunb_7,\Yfun_2),\ (\Yfun_7,\Yfun_4),\ (\Yfun_8,\Yfun_3)\\
\hline
95_3  &\spin (\Yfun_1,\Yfunb_1),\ (\Yfun_2,\Yfun_1),\ (\Yfunb_2,\Yfun_5),\ 
       (\Yfun_3,\Yfunb_5),\ (\Yfunb_3,\Yfunb_3),\ (\Yfun_4,\Yfun_3),\\
\nsus &\qquad (\Yfun_5,\Yfun_6),\ (\Yfun_6,\Yfunb_6), (\Yfunb_6,\Yfunb_2),\ 
       (\Yfun_7,\Yfun_2),\ (\Yfunb_7,\Yfun_4),\ (\Yfun_8,\Yfunb_4)\\  
      &\scal (\Yfun_1,\Yfunb_6),\ (\Yfunb_2,\Yfun_6),\ (\Yfun_2,\Yfun_2),\ 
       (\Yfunb_3,\Yfunb_2),\ (\Yfun_3,\Yfunb_4), (\Yfun_4,\Yfun_4),\\
      &\qquad (\Yfun_5,\Yfun_1),\ (\Yfunb_6,\Yfunb_1),\ (\Yfun_6,\Yfunb_5),\ 
       (\Yfunb_7,\Yfun_5),\ (\Yfun_7,\Yfun_3),\ (\Yfun_8,\Yfunb_3)\\
\hline
5_25_3  &(\Yfun_1,\Yfun_1),\ (\Yfunb_1,\Yfun_6),\ (\Yfun_2,\Yfunb_5),\ 
         (\Yfunb_2,\Yfunb_2),\ (\Yfun_3,\Yfun_3),\ (\Yfunb_3,\Yfun_4),\\
\sus    &(\Yfun_4,\Yfunb_3),\ (\Yfunb_4,\Yfunb_4),\ (\Yfun_5,\Yfun_5),\ 
         (\Yfunb_5,\Yfun_2),\ (\Yfun_6,\Yfunb_1),\ (\Yfunb_6,\Yfunb_6)\\
\hline
5_35_1  &(\Yfunb_1,\Yfunb_1),\ (\Yfun_1,\Yfunb_6),\ (\Yfun_5,\Yfunb_2),\ 
         (\Yfunb_5,\Yfunb_5),\ (\Yfunb_3,\Yfunb_3),\ (\Yfun_3,\Yfunb_4),\\
\sus    &(\Yfun_6,\Yfun_1),\ (\Yfunb_6,\Yfun_2),\ (\Yfun_2,\Yfun_3),\
         (\Yfunb_2,\Yfun_6),\ (\Yfunb_4,\Yfun_4),\ (\Yfun_4,\Yfun_5)\\
\hline
5_15_2  &(\Yfun_1,\Yfunb_1),\ (\Yfunb_1,\Yfun_6),\ (\Yfunb_2,\Yfun_1),\
         (\Yfun_2,\Yfunb_2),\ (\Yfunb_3,\Yfun_2),\ (\Yfun_3,\Yfunb_3),\\
\sus    &(\Yfunb_4,\Yfun_3),\ (\Yfun_4,\Yfunb_4),\ (\Yfunb_5,\Yfun_4),\ 
         (\Yfun_5,\Yfunb_5),\ (\Yfunb_6,\Yfun_5),\ (\Yfun_6,\Yfunb_6)\\
\hline\hline
\multicolumn{2}{|c|}{\hbox{closed string spectrum}}\\ \hline
\rm sector       &\hbox{$\cN=1$ multiplets}\\ \hline
\rm untw.        &\hbox{gravity, 1 lin., 4 chir.} \\ \hline
\hbox{order-two} &\hbox{22 lin.} \\ \hline
\rm remaining    &\hbox{21 lin., 6 chir., 1 vec.} \\ \hline
\end{array}$$
\caption{Spectrum of the $\bZ_2\times\bZ_6$ orientifold without discrete
         torsion and $(\mu_3,\mu_1)=(+1,+1)$. (continued)
         \label{Z2Z6spec4b}}
\end{table}

\begin{table}[htp]
\renewcommand{\arraystretch}{1.25}
$$\begin{array}{|c|c|}
\hline
\multicolumn{2}{|c|}{\IZ_2\times\IZ_6,\ 
       (\alpha_1,\alpha_2,\alpha_3)=(+1,+1,-1),\ \eps=1}
   \\ \hline\hline
\multicolumn{2}{|c|}{\hbox{open string spectrum}}\\ \hline
\rm sector &\hbox{gauge group / matter fields}\\ \hline
99,\ 5_15_1 &U(4-t_1)_1\times U(4)_2\times U(t_1)_3\times 
               U(t_1)_4\times U(4)_5\times U(4-t_1)_6\\
\sus    &(\Yfun_1,\Yfun_5),\ (\Yfun_1,\Yfun_6),\ (\Yfun_1,\Yfunb_6),\ 
         (\Yfunb_1,\Yfunb_6),\ (\Yfun_2,\Yfunb_1),\ (\Yfun_2,\Yfun_4),\\
        &(\Yfun_2,\Yfun_5),\ (\Yfunb_2,\Yfunb_5),\ (\Yfunb_2,\Yfunb_6),\ 
         (\Yfun_3,\Yfunb_2),\ (\Yfun_3,\Yfun_4),\ (\Yfunb_3,\Yfunb_4),\\
        &(\Yfunb_3,\Yfunb_5),\ (\Yfun_4,\Yfunb_3),\ (\Yfun_5,\Yfunb_4),\
         (\Yfun_6,\Yfunb_5),\ \Yasymb_1,\ \Yasym_3,\ \Yasymb_4,\ \Yasym_6\\
\hline
5_25_2  &U(8-2t_2)_1\times U(4-t_2)_2\times U(t_2)_3\times 
           U(2t_2)_4\times U(t_2)_5\times U(4-t_2)_6\\
\sus    &(\Yfunb_1,\Yfunb_2),\ (\Yfun_1,\Yfun_6),\ (\Yfun_1,\Yfunb_6),\ 
         (\Yfun_2,\Yfunb_1),\ (\Yfun_2,\Yfun_5),\ (\Yfun_2,\Yfun_6),\\
        &(\Yfunb_2,\Yfunb_6),\ (\Yfun_3,\Yfunb_2),\ (\Yfun_3,\Yfun_4),\
         (\Yfun_3,\Yfun_5),\ (\Yfunb_3,\Yfunb_5),\ (\Yfunb_3,\Yfunb_6),\\
        &(\Yfun_4,\Yfunb_3),\ (\Yfunb_4,\Yfunb_5),\ (\Yfun_5,\Yfunb_4),\ 
         (\Yfun_6,\Yfunb_5),\ \Yasym_1,\ \Yasymb_1,\ \Yasym_4,\ \Yasymb_4\\
\hline
5_35_3  &USp(2t_2)_1\times U(t_2)_2\times U(4-t_2)_3\times USp(8-2t_2)_4\\
        &\qquad\times USp(2t_2)_5\times U(t_2)_6\times U(4-t_2)_7
                                                     \times USp(8-2t_2)_8\\
\nsus   &\smult (\Yfun_1,\Yfun_2),\ (\Yfun_1,\Yfun_5),\ (\Yfun_1,\Yfunb_6),\ 
         (\Yfunb_2,\Yfun_5),\ (\Yfun_2,\Yfunb_6),\ (\Yfun_2,\Yfunb_7),\\
        &\qquad (\Yfun_3,\Yfunb_2),\ (\Yfunb_3,\Yfun_4),\ (\Yfun_3,\Yfunb_7),\ 
         (\Yfun_3,\Yfun_8),\ (\Yfun_4,\Yfun_7),\ (\Yfun_4,\Yfun_8),\\ 
        &(\Yfun_5,\Yfun_6),\ (\Yfun_6,\Yfunb_2),\ (\Yfun_6,\Yfunb_3),\ 
         (\Yfun_7,\Yfunb_3),\ (\Yfun_7,\Yfunb_8),\ (\Yfunb_7,\Yfun_8)\\
        &\spin \Yasym_1,\  adj_2,\  adj_3,\  \Yasym_4,\  \Yasym_5,\  
               adj_6,\ adj_7,\ \Yasym_8\\
\hline
95_i,\ 5_i5_j &\hbox{many bifundamentals}\\
\hline\hline
\multicolumn{2}{|c|}{\hbox{closed string spectrum}}\\ \hline
\rm sector       &\hbox{$\cN=1$ multiplets}\\ \hline
\rm untw.        &\hbox{gravity, 1 lin., 4 chir.} \\ \hline
\hbox{order-two} &\hbox{8 lin., 14 chir.} \\ \hline
\rm remaining    &\hbox{21 lin., 6 chir., 1 vec.} \\ \hline
\end{array}$$
\caption{Spectrum of the $\bZ_2\times\bZ_6$ orientifold without discrete
         torsion and $(\mu_3,\mu_1)=(+1,-1)$. There are 25 solutions to
         the tadpole equations, parametrised by $t_1,t_2=0,\ldots,4$.
         The notation is explained in the tables of section \ref{Z2Z2}.
         \label{Z2Z6spec5}}
\end{table}

\begin{table}[htp]
\renewcommand{\arraystretch}{1.25}
$$\begin{array}{|c|c|}
\hline
\multicolumn{2}{|c|}{\IZ_2\times\IZ_6,\ 
       (\alpha_1,\alpha_2,\alpha_3)=(-1,+1,+1),\ \eps=1}
   \\ \hline\hline
\multicolumn{2}{|c|}{\hbox{open string spectrum}}\\ \hline
\rm sector &\hbox{gauge group / matter fields}\\ \hline
99      &U(8-2t_1)_1\times U(4-t_1)_2\times U(t_1)_3\times 
         U(2t_1)_4\times U(t_1)_5\times U(4-t_1)_6\\
\sus    &(\Yfunb_1,\Yfunb_2),\ (\Yfun_1,\Yfun_6),\ (\Yfun_1,\Yfunb_6),\ 
         (\Yfun_2,\Yfunb_1),\ (\Yfun_2,\Yfun_5),\ (\Yfun_2,\Yfun_6),\\
        &(\Yfunb_2,\Yfunb_6),\ (\Yfun_3,\Yfunb_2),\ (\Yfun_3,\Yfun_4),\ 
         (\Yfun_3,\Yfun_5),\ (\Yfunb_3,\Yfunb_5),\ (\Yfunb_3,\Yfunb_6),\\
        &(\Yfun_4,\Yfunb_3),\ (\Yfunb_4,\Yfunb_5),\ (\Yfun_5,\Yfunb_4),\ 
         (\Yfun_6,\Yfunb_5),\ \Yasym_1,\ \Yasymb_1,\ \Yasym_4,\ \Yasymb_4\\
\hline
5_15_1  &USp(2t_1)_1\times U(t_1)_2\times U(4-t_1)_3\times USp(8-2t_1)_4\\
        &\qquad \times USp(2t_1)_5\times U(t_1)_6\times U(4-t_1)_7
                                                     \times USp(8-2t_1)_8\\
\nsus   &\smult (\Yfun_1,\Yfun_2),\ (\Yfun_1,\Yfun_5),\ (\Yfun_1,\Yfunb_6),\ 
         (\Yfunb_2,\Yfun_5),\ (\Yfun_2,\Yfunb_6),\ (\Yfun_2,\Yfunb_7),\\
        &\qquad (\Yfun_3,\Yfunb_2),\ (\Yfunb_3,\Yfun_4),\ (\Yfun_3,\Yfunb_7),\ 
         (\Yfun_3,\Yfun_8),\ (\Yfun_4,\Yfun_7),\ (\Yfun_4,\Yfun_8),\\
        &\qquad (\Yfun_5,\Yfun_6),\ (\Yfun_6,\Yfunb_2),\ (\Yfun_6,\Yfunb_3),\ 
         (\Yfun_7,\Yfunb_3),\ (\Yfun_7,\Yfunb_6),\ (\Yfunb_7,\Yfun_8)\\
        &\spin \Yasym_1,\  adj_2,\  adj_3,\  \Yasym_4,\  \Yasym_5,\  
               adj_6,\ adj_7,\ \Yasym_8\\
\hline
5_25_2,\ 5_35_3 &U(t_2)_1\times U(4)_2\times U(4-t_2)_3\times 
                 U(4-t_2)_4\times U(4)_5\times U(t_2)_6\\
\sus    &(\Yfun_1,\Yfun_5),\ (\Yfun_1,\Yfun_6),\ (\Yfun_1,\Yfunb_6),\ 
         (\Yfunb_1,\Yfunb_6),\ (\Yfun_2,\Yfunb_1),\ (\Yfun_2,\Yfun_4),\\
        &(\Yfun_2,\Yfun_5),\ (\Yfunb_2,\Yfunb_5),\ (\Yfunb_2,\Yfunb_6),\ 
         (\Yfun_3,\Yfunb_2),\ (\Yfun_3,\Yfun_4),\ (\Yfunb_3,\Yfunb_4),\\
        &(\Yfunb_3,\Yfunb_5),\ (\Yfun_4,\Yfunb_3),\ (\Yfun_5,\Yfunb_4),\ 
         (\Yfun_6,\Yfunb_5),\ \Yasymb_1,\ \Yasym_3,\ \Yasymb_4,\ \Yasym_6\\
\hline
95_i,\ 5_i5_j &\hbox{many bifundamentals}\\
\hline\hline
\multicolumn{2}{|c|}{\hbox{closed string spectrum}}\\ \hline
\rm sector       &\hbox{$\cN=1$ multiplets}\\ \hline
\rm untw.        &\hbox{gravity, 1 lin., 4 chir.} \\ \hline
\hbox{order-two} &\hbox{6 lin., 16 chir.} \\ \hline
\rm remaining    &\hbox{21 lin., 6 chir., 1 vec.} \\ \hline
\end{array}$$
\caption{Spectrum of the $\bZ_2\times\bZ_6$ orientifold without discrete
         torsion and $(\mu_3,\mu_1)=(-1,+1)$. There are 25 solutions to
         the tadpole equations, parametrised by $t_1,t_2=0,\ldots,4$.
         The notation is explained in the tables of section \ref{Z2Z2}.
         \label{Z2Z6spec6}}
\end{table}

\subsection{$\bZ_2\times\bZ_6'$, $v=\frac12(1,-1,0)$, $w=\frac16(-2,1,1)$}
\label{Z2Z6'}

This orientifold is similar to the previous one. However, there are 
only four inequivalent models. This is because the only even twist,
$\bar k=(0,2)$ has no fixed planes. Therefore, all three sets of 
$D5_i$-branes appear on the same footing. The $(+1,+1,+1)$ model has 
already been constructed in \cite{k}. It is of some phenomenological 
interest because it contains a gauge group $SU(6)$ with three generations 
of matter fields in the antisymmetric tensor representation.

The only \nontriv tadpole conditions for the models with discrete
torsion are (\ref{tad_untw}) and (\ref{tad_evenk}), corresponding 
to the untwisted sector and the $(0,2)$ sector. The former fixes the 
rank of the gauge group to be 8 for each of the four sectors 99, 
$5_i5_i$. The latter reads
\be  \label{tadpZ2Z6'}
\Tr(\gamma_{(0,2),9})=\Tr(\gamma_{(0,2),5_1})=\alpha_1\,4,\quad
\Tr(\gamma_{(0,2),5_2})=\Tr(\gamma_{(0,2),5_3})=-\alpha_1\,4.
\ee
We assumed that all $D5_i$-branes are located at the origin. The 27 
fixed points of the $(0,2)$ sector can be labelled by the triples
$(n_1,n_2,n_3)$, where $n_i=0,1,2$ denotes the fixed points in the
$i\th$ complex plane and $0$ is the origin. One finds that the
Klein bottle contribution $\cK_0$ is present at the three fixed points
$(n_1,0,0)$, the contribution $\cK_1$ is present at all 27 fixed points,
the contribution $\cK_2$ is present at $(0,0,n_3)$ and the contribution
$\cK_3$ is present at $(0,n_2,0)$. This leads to the tadpole equations
\ref{tadpZ2Z6'}. They have a unique solution. The results are displayed 
in tables \ref{Z2Z6'spec1} and \ref{Z2Z6'spec2}.

\begin{table}[htp]
\renewcommand{\arraystretch}{1.25}
$$\begin{array}{|c|c|}
\hline
\multicolumn{2}{|c|}{\IZ_2\times\IZ_6',\ 
       (\alpha_1,\alpha_2,\alpha_3)=(+1,+1,+1),\ \eps=-1}
   \\ \hline\hline
\multicolumn{2}{|c|}{\hbox{open string spectrum}}\\ \hline
\rm sector &\hbox{gauge group / matter fields}\\ \hline
99,\ 5_i5_i    &U(6)_1\times USp(4)_2\\
\sus  &3\,(\Yfun_1,\Yfun_2),\ 3\,\Yasymb_1\\
\hline
95_1,\ 5_25_3\ \sus 
     &(\Yfunb_1,\Yfunb_1),\ (\Yfun_1,\Yfun_2),\ (\Yfun_2,\Yfun_1)\\
\hline
95_2,\ 5_35_1\ \sus 
     &(\Yfun_1,\Yfun_2),\ (\Yfunb_1,\Yfunb_1),\ (\Yfun_2,\Yfun_1)\\
\hline
95_3,\ 5_15_2\ \sus 
     &(\Yfun_1,\Yfun_2),\ (\Yfunb_1,\Yfunb_1),\ (\Yfun_2,\Yfun_1)\\
\hline\hline
\multicolumn{2}{|c|}{\hbox{closed string spectrum}}\\ \hline
\rm sector       &\hbox{$\cN=1$ multiplets}\\ \hline
\rm untw.        &\hbox{gravity, 1 lin., 3 chir.} \\ \hline
\hbox{order-two} &\hbox{15 chir.} \\ \hline
\rm remaining    &\hbox{12 lin.} \\ \hline
\end{array}$$
\caption{Spectrum of the $\bZ_2\times\bZ_6'$ orientifold with discrete
         torsion and $(\mu_3,\mu_1)=(-1,-1)$. The notation is explained
         in the tables of section \ref{Z2Z2}. 
         \label{Z2Z6'spec1}}
\end{table}

In the cases without discrete torsion, the tadpole conditions for the $(0,2)$
sector are identical to eq.\ (\ref{tadpZ2Z6'}) if all $D5_i$-branes sit at
the origin. Now, the remaining twisted sectors give additional conditions
because, in general, the $\gamma$ matrices are not traceless. The tadpole
equations have many solutions, 1590 for the $(-1,-1,-1)$ model and 1295
for the $(+1,+1,-1)$ model. In general, the three sets of $D5_i$-branes
have different gauge groups. However, if we impose an additional symmetry
between the $D5_i$-branes, \eg $\Tr(\gamma_{\bar k,5_1})=
\Tr(\gamma_{\bar k,5_2})=\Tr(\gamma_{\bar k,5_3})$ for odd $\bar k$ without
fixed planes, then there is a unique solution.
We display the 99 and $5_i5_i$ spectrum of these models in tables
\ref{Z2Z6'spec3} and \ref{Z2Z6'spec4}. We determined also the $95_i$
and $5_i5_j$ spectrum of these models and verified that the complete 
spectrum is free of \nonab gauge anomalies.

\begin{table}[htp]
\renewcommand{\arraystretch}{1.25}
$$\begin{array}{|c|c|}
\hline
\multicolumn{2}{|c|}{\IZ_2\times\IZ_6',\ 
       (\alpha_1,\alpha_2,\alpha_3)=(-1,+1,-1),\ \eps=-1}
   \\ \hline\hline
\multicolumn{2}{|c|}{\hbox{open string spectrum}}\\ \hline
\rm sector &\hbox{gauge group / matter fields}\\ \hline
99,\ 5_25_2    &SO(4)_1\times U(6)_2\\
\sus  &3\,(\Yfun_1,\Yfunb_2),\ 2\,\Yasym_2,\ \Ysym_2\\
\hline
5_15_1,\ 5_35_3  &USp(4)_1\times U(6)_2\\
\nsus &\smult 3\,(\Yfun_1,\Yfunb_2)\\
      &\spin \Yasym_1,\ adj_2,\ 2\,\Yasym_2,\ \Ysym_2,\quad  
       \scal 2\,\Ysym_2,\ \Yasym_2\\
\hline  
95_1,\ 5_25_3
      &\spin (\Yfun_1,\Yfun_2),\ (\Yfun_2,\Yfun_1),\ (\Yfunb_2,\Yfunb_2)\\
\nsus &\scal (\Yfun_1,\Yfun_1),\ (\Yfun_2,\Yfunb_2),\ (\Yfunb_2,\Yfun_2)\\
\hline
95_2,\ 5_35_1\ \sus 
      &(\Yfun_1,\Yfunb_2),\ (\Yfun_2,\Yfun_2),\ (\Yfunb_2,\Yfun_1)\\ 
\hline
95_3,\ 5_15_2 
      &\spin (\Yfun_1,\Yfun_2),\ (\Yfunb_2,\Yfunb_2),\ (\Yfun_2,\Yfun_1)\\  
\nsus &\scal (\Yfun_1,\Yfun_1),\ (\Yfun_2,\Yfunb_2),\ (\Yfunb_2,\Yfun_2)\\
\hline\hline
\multicolumn{2}{|c|}{\hbox{closed string spectrum}}\\ \hline
\rm sector       &\hbox{$\cN=1$ multiplets}\\ \hline
\rm untw.        &\hbox{gravity, 1 lin., 3 chir.} \\ \hline
\hbox{order-two} &\hbox{5 chir., 10 vec.} \\ \hline
\rm remaining    &\hbox{12 lin.} \\ \hline
\end{array}$$
\caption{Spectrum of the $\bZ_2\times\bZ_6'$ orientifold with discrete
         torsion and $(\mu_3,\mu_1)=(+1,+1)$. The notation is explained
         in the tables of section \ref{Z2Z2}. 
         \label{Z2Z6'spec2}}
\end{table}

\begin{table}[htp]
\renewcommand{\arraystretch}{1.25}
$$\begin{array}{|c|c|}
\hline
\multicolumn{2}{|c|}{\IZ_2\times\IZ_6',\ 
       (\alpha_1,\alpha_2,\alpha_3)=(-1,-1,-1),\ \eps=1}
   \\ \hline\hline
\multicolumn{2}{|c|}{\hbox{open string spectrum}}\\ \hline
\rm sector &\hbox{gauge group / matter fields}\\ \hline
99     &SO(2)_1\times U(3)_2\times U(3)_3\times SO(2)_4\\
       &\qquad \times SO(2)_5\times U(3)_6\times U(3)_7\times SO(2)_8\\
\sus   &(\Yfun_1,\Yfunb_2),\ (\Yfun_1,\Yfunb_6),\ (\Yfun_1,\Yfun_7),\ 
        (\Yfun_2,\Yfunb_3),\ (\Yfunb_2,\Yfun_5),\ (\Yfun_2,\Yfun_6),\\
       &(\Yfun_2,\Yfunb_7),\ (\Yfunb_2,\Yfun_8),\ (\Yfun_3,\Yfun_4),\ 
        (\Yfun_3,\Yfun_5),\ (\Yfunb_3,\Yfunb_7),\ (\Yfun_3,\Yfun_8),\\
       &(\Yfun_4,\Yfunb_6),\ (\Yfun_4,\Yfun_7),\ (\Yfun_5,\Yfunb_6),\ 
        (\Yfun_6,\Yfunb_3),\ (\Yfun_6,\Yfunb_7),\ (\Yfun_7,\Yfun_8)\\
\hline
5_i5_i &U(3)_1\times U(2)_2\times U(3)_3\times 
        U(3)_4\times U(2)_5\times U(3)_6\\
\nsus &\smult (\Yfun_1,\Yfunb_2),\ (\Yfunb_1,\Yfunb_2),\ (\Yfun_1,\Yfun_3),\
       (\Yfunb_1,\Yfunb_5),\ (\Yfun_1,\Yfun_6),\ (\Yfun_2,\Yfunb_3),\\
      &\qquad (\Yfunb_2,\Yfunb_4),\ (\Yfun_2,\Yfun_5),\ (\Yfun_3,\Yfun_4),\ 
       (\Yfun_3,\Yfunb_4),\ (\Yfunb_3,\Yfunb_6),\ (\Yfun_4,\Yfunb_5),\\
      &\qquad (\Yfunb_4,\Yfunb_5),\ (\Yfun_4,\Yfun_6),\ (\Yfun_5,\Yfunb_6),\
       (\Yfun_6,\Yfunb_1)\\
      &\spin adj_1,\ \ldots,\ adj_6,\ 
             \Yasym_2,\ \Yasymb_3,\ \Yasym_5,\ \Yasymb_6\\ 
      &\scal \Ysym_2,\ \Ysymb_3,\ \Ysym_5,\ \Ysymb_6\\ 
\hline  
95_i,\ 5_i5_j &\hbox{many bifundamentals}\\
\hline\hline
\multicolumn{2}{|c|}{\hbox{closed string spectrum}}\\ \hline
\rm sector       &\hbox{$\cN=1$ multiplets}\\ \hline
\rm untw.        &\hbox{gravity, 1 lin., 3 chir.} \\ \hline
\hbox{order-two} &\hbox{18 lin.} \\ \hline
\rm remaining    &\hbox{15 lin.} \\ \hline
\end{array}$$
\caption{Spectrum of the $\bZ_2\times\bZ_6'$ orientifold without discrete
         torsion and $(\mu_3,\mu_1)=(+1,+1)$. In general, the tadpole
         equations have many solutions for the ranks of the gauge group
         factors. The most symmetric one is shown. The notation is explained
         in the tables of section \ref{Z2Z2}. 
         \label{Z2Z6'spec3}}
\end{table}

\begin{table}[htp]
\renewcommand{\arraystretch}{1.25}
$$\begin{array}{|c|c|}
\hline
\multicolumn{2}{|c|}{\IZ_2\times\IZ_6',\ 
       (\alpha_1,\alpha_2,\alpha_3)=(+1,+1,-1),\ \eps=1}
   \\ \hline\hline
\multicolumn{2}{|c|}{\hbox{open string spectrum}}\\ \hline
\rm sector &\hbox{gauge group / matter fields}\\ \hline
99,\ 5_15_1,\ 5_25_2  &U(3)_1\times U(2)_2\times U(3)_3\times 
                       U(3)_4\times U(2)_5\times U(3)_6\\
\sus   &(\Yfun_1,\Yfun_2),\ (\Yfun_1,\Yfunb_2),\ (\Yfunb_1,\Yfunb_4),\ 
        (\Yfun_1,\Yfun_5),\ (\Yfun_2,\Yfunb_3),\ (\Yfunb_2,\Yfunb_3),\\
       &(\Yfun_2,\Yfun_4),\ (\Yfunb_2,\Yfunb_6),\ (\Yfun_3,\Yfunb_4),\ 
        (\Yfunb_3,\Yfunb_5),\ (\Yfun_3,\Yfun_6),\ (\Yfun_4,\Yfun_5),\\
       &(\Yfun_4,\Yfunb_5),\ (\Yfun_5,\Yfunb_6),\ (\Yfunb_5,\Yfunb_6),\ 
        (\Yfun_6,\Yfunb_1),\ \Yasymb_1,\ \Yasym_3,\ \Yasymb_4,\ \Yasym_6\\
\hline
5_35_3 &USp(2)_1\times U(3)_2\times U(3)_3\times USp(2)_4\\
       &\qquad \times USp(2)_5\times U(3)_6\times U(3)_7\times USp(2)_8\\
\nsus &\smult (\Yfun_1,\Yfunb_2),\ (\Yfun_1,\Yfunb_6),\ (\Yfun_1,\Yfun_7),\ 
       (\Yfun_2,\Yfunb_3),\ (\Yfunb_2,\Yfun_5),\ (\Yfun_2,\Yfun_6),\\
      &\qquad (\Yfun_2,\Yfunb_7),\ (\Yfunb_2,\Yfun_8),\ (\Yfun_3,\Yfun_4),\ 
       (\Yfun_3,\Yfun_5),\ (\Yfunb_3,\Yfunb_7),\ (\Yfun_3,\Yfun_8),\\
      &\qquad (\Yfun_4,\Yfunb_6),\ (\Yfun_4,\Yfun_7),\ (\Yfun_5,\Yfunb_6),\ 
       (\Yfun_6,\Yfunb_3),\ (\Yfun_6,\Yfunb_7),\ (\Yfun_7,\Yfun_8)\\
      &\spin \Yasym_1,\  adj_2,\  adj_3,\  \Yasym_4,\  \Yasym_5,\  
               adj_6,\ adj_7,\ \Yasym_8\\
\hline  
95_i,\ 5_i5_j &\hbox{many bifundamentals}\\
\hline\hline
\multicolumn{2}{|c|}{\hbox{closed string spectrum}}\\ \hline
\rm sector       &\hbox{$\cN=1$ multiplets}\\ \hline
\rm untw.        &\hbox{gravity, 1 lin., 3 chir.} \\ \hline
\hbox{order-two} &\hbox{6 lin., 12 chir.} \\ \hline
\rm remaining    &\hbox{15 lin.} \\ \hline
\end{array}$$
\caption{Spectrum of the $\bZ_2\times\bZ_6'$ orientifold without discrete
         torsion and $(\mu_3,\mu_1)=(+1,-1)$. In general, the tadpole
         equations have many solutions for the ranks of the gauge group
         factors. The most symmetric one is shown. The notation is explained
         in the tables of section \ref{Z2Z2}. 
         \label{Z2Z6'spec4}}
\end{table}

\subsection{$\bZ_6\times\bZ_6$, $v=\frac16(1,-1,0)$, $w=\frac16(0,1,-1)$}
\label{Z6Z6}

This orientifold is symmetric under a permutation of the three sets 
of $D5_i$-branes. Thus, only four of the eight possible $\IZ_6\times\IZ_6$ 
models are inequivalent. The $(+1,+1,+1)$ model has already been 
constructed in \cite{z}.

The only \nontriv tadpole conditions for the models with discrete
torsion are (\ref{tad_untw}), (\ref{tad_fixeven1}) and (\ref{tad_evenk}),
corresponding to the untwisted sector, the the three sectors twisted by
$\bar k_1=(0,2)$, $\bar k_2=(2,2)$, $\bar k_3=(2,0)$ and the $(2,4)$ sector
respectively.\footnote{In contrast to the notation used in the rest of this
paper, the three twists $\bar k_i$ defined in this paragraph are not of
 order two.} The untwisted tadpoles fix the rank of the gauge group to 
be 8 for each of the four sectors 99, $5_i5_i$.
The tadpole conditions of the $\bar k_i$ and $(2,4)$ sectors read
\bea  \label{tadpZ6Z6}
&& \Tr(\gamma_{\bar k_i,9})=\Tr(\gamma_{\bar k_i,5_i})=
  -\Tr(\gamma_{\bar k_i,5_j})=-\Tr(\gamma_{\bar k_i,5_k})=
  -\alpha_i\,\eps^{-k_{i,1}k_{i,2}/4}\,8,\nonumber\\
&& \Tr(\gamma_{(2,4),9})=-\Tr(\gamma_{(2,4),5_1})=
  -\Tr(\gamma_{(2,4),5_2})=\Tr(\gamma_{(2,4),5_3})=
   \alpha_3\,4,
\eea
where $k_{i,1}$, $k_{i,2}$ are the components of the two-vector $\bar k_i$
and $(ijk)$ is a permutation of $(123)$.
We assumed that all $D5_i$-branes are at the origin.
The sign $\eps^{-k_{i,1}k_{i,2}/4}$ is $-1$ for the $(2,2)$ sector in the
models with discrete torsion and $+1$ else.
In the case without discrete torsion, these equations are still valid.
But in addition, one has to take into account the tadpole conditions for
all the remaining twisted sectors.
The tadpole equations can be solved with a computer algebra program. 
The complete spectrum of the two models with discrete torsion
is displayed in tables \ref{Z6Z6spec1} and \ref{Z6Z6spec2}.
For the two models without discrete torsion, we restrict ourselves to
give the 99 and $5_i5_i$ spectrum in tables \ref{Z6Z6spec3} -- 
\ref{Z6Z6spec4b}. We determined also the $95_i$ and $5_i5_j$ spectrum 
of these models and verified that the complete spectrum is free of \nonab 
gauge anomalies. In the case without discrete torsion, the tadpole equations
have many solutions. However, if we impose an additional symmetry
between the $D5_i$-branes, \eg $\Tr(\gamma_{\bar k,5_1})=
\Tr(\gamma_{\bar k,5_2})=\Tr(\gamma_{\bar k,5_3})$ for odd $\bar k$ without
fixed planes, then there is a unique solution. The ranks of the gauge
group factors for this solution are as follows:
\bea  \label{Z6Z6solution}
[n_1,n_2,\ldots,n_{20}] &=&[2,1,1,2,1,0,1,1,0,1,1,1,0,1,1,0,2,1,1,2]
\nonumber\\{}
[m_1,m_2,\ldots,m_{18}] &=&[1,1,0,1,1,0,2,1,1,2,1,1,1,0,1,1,0,1]
\eea

\begin{table}[htp]
\renewcommand{\arraystretch}{1.25}
$$\begin{array}{|c|c|}
\hline
\multicolumn{2}{|c|}{\IZ_6\times\IZ_6,\ 
       (\alpha_1,\alpha_2,\alpha_3)=(+1,+1,+1),\ \eps=-1}
   \\ \hline\hline
\multicolumn{2}{|c|}{\hbox{open string spectrum}}\\ \hline
\rm sector &\hbox{gauge group / matter fields}\\ \hline
99,\ 5_i5_i    &U(2)_1\times U(2)_2\times U(2)_3\times USp(4)_4\\
\sus  &(\Yfun_1,\Yfun_4),\ (\Yfun_2,\Yfunb_1),\ (\Yfun_2,\Yfun_3),\ 
       (\Yfunb_2,\Yfun_4),\ (\Yfun_3,\Yfunb_1), (\Yfunb_3,\Yfun_4)\\
      & \Yasymb_1,\ \Yasym_2,\ \Yasym_3\\
\hline
95_1,\ 5_25_3\ \sus 
     &(\Yfun_1,\Yfun_4),\ (\Yfunb_1,\Yfunb_1),\ (\Yfun_2,\Yfun_3),\ 
      (\Yfun_3,\Yfun_2),\ (\Yfun_4,\Yfun_1)\\
\hline
95_2,\ 5_35_1\ \sus 
     &(\Yfunb_1,\Yfun_3),\ (\Yfunb_2,\Yfun_4),\ (\Yfun_2,\Yfun_2),\ 
      (\Yfun_3,\Yfunb_1),\ (\Yfun_4,\Yfunb_2)\\
\hline
95_3,\ 5_15_2\ \sus 
     &(\Yfunb_1,\Yfun_2),\ (\Yfun_2,\Yfunb_1),\ (\Yfunb_3,\Yfun_4),\ 
      (\Yfun_3,\Yfun_3),\ (\Yfun_4,\Yfunb_3)\\
\hline\hline
\multicolumn{2}{|c|}{\hbox{closed string spectrum}}\\ \hline
\rm sector       &\hbox{$\cN=1$ multiplets}\\ \hline
\rm untw.        &\hbox{gravity, 1 lin., 3 chir.} \\ \hline
\hbox{order-two} &\hbox{3 chir.} \\ \hline
\rm remaining    &\hbox{36 lin., 12 chir.} \\ \hline
\end{array}$$
\caption{Spectrum of the $\bZ_6\times\bZ_6$ orientifold with discrete
         torsion and $(\mu_3,\mu_1)=(-1,-1)$. The notation is explained
         in the tables of section \ref{Z2Z2}. 
         \label{Z6Z6spec1}}
\end{table}

\begin{table}[htp]
\renewcommand{\arraystretch}{1.25}
$$\begin{array}{|c|c|}
\hline
\multicolumn{2}{|c|}{\IZ_6\times\IZ_6,\ 
       (\alpha_1,\alpha_2,\alpha_3)=(-1,+1,-1),\ \eps=-1}
   \\ \hline\hline
\multicolumn{2}{|c|}{\hbox{open string spectrum}}\\ \hline
\rm sector &\hbox{gauge group / matter fields}\\ \hline
99,\ 5_25_2    &SO(4)_1\times U(2)_2\times U(2)_3\times U(2)_4\\
\sus  &(\Yfun_1,\Yfun_2),\ (\Yfun_1,\Yfunb_3),\ (\Yfun_1,\Yfun_4),\ 
       (\Yfunb_2,\Yfunb_4), (\Yfun_3,\Yfunb_2), (\Yfun_3,\Yfunb_4)\\ 
      &\Yasymb_2,\ \Yasym_3,\ \Ysymb_4\\
\hline
5_15_1,\ 5_35_3 &USp(4)_1\times U(2)_2\times U(2)_3\times U(2)_4\\
\nsus &\smult \hbox{identical to bifundamentals of 99}\\
      &\spin \Yasym_1,\ adj_2,\ adj_3,\ adj_4,\ 
             \Yasymb_2,\ \Yasym_3,\ \Ysymb_4\\
      &\scal \Ysymb_2,\ \Ysym_3,\ \Yasymb_4\\
\hline  
95_1,\ 5_25_3
      &\spin (\Yfun_1,\Yfun_3),\ (\Yfun_2,\Yfun_4),\ (\Yfunb_3,\Yfunb_3),\ 
             (\Yfun_3,\Yfun_1),\ (\Yfun_4,\Yfun_2)\\
\nsus &\scal (\Yfun_2,\Yfun_3),\ (\Yfunb_2,\Yfun_4),\ (\Yfunb_3,\Yfunb_4),\ 
             (\Yfun_3,\Yfun_2),\ (\Yfunb_4,\Yfunb_3),\ (\Yfun_4,\Yfunb_2)\\
\hline
95_2,\ 5_35_1\ \sus 
      &(\Yfun_1,\Yfun_4),\ (\Yfunb_2,\Yfun_3),\ (\Yfun_3,\Yfunb_2),\ 
       (\Yfunb_4,\Yfunb_4),\ (\Yfun_4,\Yfun_1)\\
\hline
95_3,\ 5_15_2 
      &\spin (\Yfun_1,\Yfunb_2),\ (\Yfun_2,\Yfun_2),\ (\Yfunb_2,\Yfun_1),\ 
             (\Yfunb_3,\Yfun_4),\ (\Yfun_4,\Yfunb_3)\\
\nsus &\scal (\Yfunb_2,\Yfun_4),\ (\Yfun_2,\Yfun_3),\ (\Yfun_3,\Yfun_2),\ 
             (\Yfunb_3,\Yfunb_4),\ (\Yfun_4,\Yfunb_2),\ (\Yfunb_4,\Yfunb_3)\\
\hline\hline
\multicolumn{2}{|c|}{\hbox{closed string spectrum}}\\ \hline
\rm sector       &\hbox{$\cN=1$ multiplets}\\ \hline
\rm untw.        &\hbox{gravity, 1 lin., 3 chir.} \\ \hline
\hbox{order-two} &\hbox{1 chir., 2 vec.} \\ \hline
\rm remaining    &\hbox{36 lin., 12 chir.} \\ \hline
\end{array}$$
\caption{Spectrum of the $\bZ_6\times\bZ_6$ orientifold with discrete
         torsion and $(\mu_3,\mu_1)=(+1,+1)$. The notation is explained
         in the tables of section \ref{Z2Z2}. 
         \label{Z6Z6spec2}}
\end{table}

\begin{table}[htp]
\renewcommand{\arraystretch}{1.25}
$$\begin{array}{|c|c|}
\hline
\multicolumn{2}{|c|}{\IZ_6\times\IZ_6,\ 
       (\alpha_1,\alpha_2,\alpha_3)=(-1,-1,-1),\ \eps=1}
   \\ \hline\hline
\multicolumn{2}{|c|}{\hbox{open string spectrum}}\\ \hline
\rm sector &\hbox{gauge group / matter fields}\\ \hline
99 &SO(n_{1})\times U(n_{2})\times U(n_{3})\times SO(n_{4})
       \times U(n_{5})\\
   &\qquad \times U(n_{6})\times U(n_{7})\times U(n_{8})\times U(n_{9})
       \times U(n_{10})\\
   &\qquad \times U(n_{11})\times U(n_{12})\times U(n_{13})\times U(n_{14})
       \times U(n_{15})\\
   &\qquad \times U(n_{15})\times SO(n_{17})\times U(n_{18})\times U(n_{19})
       \times SO(n_{20})\\
\sus  &(\Yfun_{1},\Yfun_{2}),\ (\Yfun_{1},\Yfunb_{5}),\ 
       (\Yfun_{1},\Yfun_{10}),\ (\Yfun_{2},\Yfunb_{6}),\ 
       (\Yfun_{2},\Yfun_{9}),\ (\Yfunb_{2},\Yfunb_{10}),\\
      &(\Yfun_{3},\Yfunb_{2}),\ (\Yfunb_{3},\Yfun_{4}),\ 
       (\Yfun_{3},\Yfunb_{7}),\ (\Yfun_{3},\Yfun_{8}),\ 
       (\Yfunb_{3},\Yfunb_{9}),\ (\Yfun_{4},\Yfun_{7}),\\
      &(\Yfun_{4},\Yfunb_{8}),\ (\Yfun_{5},\Yfunb_{2}),\ 
       (\Yfun_{5},\Yfunb_{10}),\ (\Yfun_{5},\Yfunb_{11}),\ 
       (\Yfun_{6},\Yfunb_{3}),\ (\Yfun_{6},\Yfunb_{5}),\\
      &(\Yfun_{6},\Yfunb_{12}),\ (\Yfun_{7},\Yfunb_{6}),\ 
       (\Yfun_{7},\Yfunb_{13}),\ (\Yfun_{8},\Yfunb_{7}),\ 
       (\Yfun_{8},\Yfunb_{14}),\ (\Yfun_{9},\Yfunb_{8}),\\
      &(\Yfun_{9},\Yfunb_{15}),\ (\Yfun_{10},\Yfunb_{9}),\ 
       (\Yfun_{10},\Yfunb_{15}),\ (\Yfun_{11},\Yfunb_{6}),\ 
       (\Yfun_{11},\Yfunb_{15}),\ (\Yfun_{11},\Yfun_{17}),\\
      &(\Yfunb_{11},\Yfunb_{18}),\ (\Yfun_{12},\Yfunb_{7}),\ 
       (\Yfun_{12},\Yfunb_{11}),\ (\Yfunb_{12},\Yfun_{17}),\ 
       (\Yfun_{12},\Yfunb_{18}),\ (\Yfun_{13},\Yfunb_{8}),\\ 
      &(\Yfun_{13},\Yfunb_{12}),\ (\Yfun_{13},\Yfunb_{19}),\ 
       (\Yfun_{14},\Yfunb_{9}),\ (\Yfun_{14},\Yfunb_{13}),\ 
       (\Yfun_{14},\Yfun_{20}),\ (\Yfun_{15},\Yfunb_{10}),\\ 
      &(\Yfun_{15},\Yfunb_{14}),\ (\Yfun_{15},\Yfun_{19}),\ 
       (\Yfunb_{15},\Yfun_{20}),\ (\Yfun_{15},\Yfunb_{5}),\ 
       (\Yfun_{15},\Yfunb_{15}),\ (\Yfun_{15},\Yfun_{18}),\\ 
      &(\Yfunb_{15},\Yfunb_{19}),\ (\Yfun_{17},\Yfun_{18}),\ 
       (\Yfun_{18},\Yfunb_{13}),\ (\Yfun_{19},\Yfunb_{14}),\ 
       (\Yfun_{19},\Yfunb_{18}),\ (\Yfunb_{19},\Yfun_{20})\\
\hline
\end{array}$$
\caption{Spectrum of the $\bZ_6\times\bZ_6$ orientifold without discrete
         torsion and $(\mu_3,\mu_1)=(+1,+1)$. The numbers $n_i$ are 
         determined by solving the tadpole equations. There are many
         solutions. The most symmetric one is given in the main text,
         eq.\ (\ref{Z6Z6solution}). In general, several of the $n_i$
         vanish because the total rank of the 99 gauge group must be 16.
         \label{Z6Z6spec3}}
\end{table}

\begin{table}[htp]
\renewcommand{\arraystretch}{1.25}
$$\begin{array}{|c|c|}
\hline
\multicolumn{2}{|c|}{\IZ_6\times\IZ_6,\ 
       (\alpha_1,\alpha_2,\alpha_3)=(-1,-1,-1),\ \eps=1}
   \\ \hline\hline
\multicolumn{2}{|c|}{\hbox{open string spectrum}}\\ \hline
\rm sector &\hbox{gauge group / matter fields}\\ \hline
5_i5_i &U(m_1)\times U(m_{2})\times U(m_{3})\times 
        U(m_{4})\times U(m_{5})\times U(m_{6})\\
       &\qquad \times U(m_{7})\times U(m_{8})\times U(m_{9})\times 
                      U(m_{10})\times U(m_{11})\times U(m_{12})\\
       &\qquad \times U(m_{13})\times U(m_{14})\times U(m_{15})\times 
                      U(m_{16})\times U(m_{17})\times U(m_{18})\\
\nsus &\smult (\Yfun_{1},\Yfun_{6}),\ (\Yfun_{1},\Yfunb_{6}),\ 
              (\Yfun_{1},\Yfunb_{7}),\ (\Yfun_{2},\Yfunb_{1}),\ 
              (\Yfun_{2},\Yfun_{5}),\ (\Yfunb_{2},\Yfunb_{6}),\\ 
      &\qquad (\Yfun_{2},\Yfunb_{8}),\ (\Yfun_{3},\Yfunb_{2}),\ 
              (\Yfun_{3},\Yfun_{4}),\ (\Yfunb_{3},\Yfunb_{5}),\ 
              (\Yfun_{3},\Yfunb_{9}),\ (\Yfun_{4},\Yfunb_{3}),\\
      &\qquad (\Yfun_{4},\Yfunb_{10}),\ (\Yfun_{5},\Yfunb_{4}),\ 
              (\Yfun_{5},\Yfunb_{11}),\ (\Yfun_{6},\Yfunb_{5}),\ 
              (\Yfun_{6},\Yfunb_{12}),\ (\Yfun_{7},\Yfunb_{2}),\\
      &\qquad (\Yfun_{7},\Yfunb_{12}),\ (\Yfun_{7},\Yfunb_{13}),\ 
              (\Yfun_{8},\Yfunb_{3}),\ (\Yfun_{8},\Yfunb_{7}),\ 
              (\Yfun_{8},\Yfunb_{14}),\ (\Yfun_{9},\Yfunb_{4}),\\
      &\qquad (\Yfun_{9},\Yfunb_{8}),\ (\Yfun_{9},\Yfunb_{15}),\ 
              (\Yfun_{10},\Yfunb_{5}),\ (\Yfun_{10},\Yfunb_{9}),\ 
              (\Yfun_{10},\Yfunb_{16}),\ (\Yfun_{11},\Yfunb_{6}),\\
      &\qquad (\Yfun_{11},\Yfunb_{10}),\ (\Yfun_{11},\Yfunb_{17}),\ 
              (\Yfun_{12},\Yfunb_{1}),\ (\Yfun_{12},\Yfunb_{11}),\ 
              (\Yfun_{12},\Yfunb_{18}),\ (\Yfun_{13},\Yfunb_{8}),\\
      &\qquad (\Yfunb_{13},\Yfunb_{14}),\ (\Yfun_{13},\Yfunb_{18}),\ 
              (\Yfun_{14},\Yfunb_{9}),\ (\Yfun_{14},\Yfunb_{13}),\ 
              (\Yfun_{14},\Yfun_{18}),\ (\Yfun_{15},\Yfunb_{10}),\\
      &\qquad (\Yfun_{15},\Yfunb_{14}),\ (\Yfun_{15},\Yfun_{17}),\ 
              (\Yfunb_{15},\Yfunb_{18}),\ (\Yfun_{16},\Yfunb_{11}),\ 
              (\Yfun_{16},\Yfunb_{15}),\ (\Yfunb_{16},\Yfunb_{17}),\\
      &\qquad (\Yfun_{17},\Yfunb_{12}),\ (\Yfun_{17},\Yfunb_{16}),\ 
              (\Yfun_{18},\Yfunb_{7}),\ (\Yfun_{18},\Yfunb_{17})\\
      &\spin adj_{1},\ \ldots,\ adj_{18},\ 
             \Yasymb_{1},\ \Yasymb_{4},\ \Yasym_{13},\ \Yasym_{16}\\
      &\scal \Ysymb_{1},\ \Ysymb_{4},\ \Ysym_{13},\ \Ysym_{16}\\
\hline  
95_i,\ 5_i5_j &\hbox{many bifundamentals}\\
\hline\hline
\multicolumn{2}{|c|}{\hbox{closed string spectrum}}\\ \hline
\rm sector       &\hbox{$\cN=1$ multiplets}\\ \hline
\rm untw.        &\hbox{gravity, 1 lin., 3 chir.} \\ \hline
\hbox{order-two} &\hbox{12 lin.} \\ \hline
\rm remaining    &\hbox{54 lin., 15 chir.} \\ \hline
\end{array}$$
\caption{Spectrum of the $\bZ_6\times\bZ_6$ orientifold without discrete
         torsion and $(\mu_3,\mu_1)=(+1,+1)$. (continued) 
         \label{Z6Z6spec3b}}
\end{table}

\begin{table}[htp]
\renewcommand{\arraystretch}{1.25}
$$\begin{array}{|c|c|}
\hline
\multicolumn{2}{|c|}{\IZ_6\times\IZ_6,\ 
       (\alpha_1,\alpha_2,\alpha_3)=(+1,+1,-1),\ \eps=1}
   \\ \hline\hline
\multicolumn{2}{|c|}{\hbox{open string spectrum}}\\ \hline
\rm sector &\hbox{gauge group / matter fields}\\ \hline
99,\ 5_15_1,\ 5_25_2 &U(m_{1})\times U(m_{2})\times U(m_{3})\times 
                      U(m_{4})\times U(m_{5})\times U(m_{6})\\
       &\qquad \times U(m_{7})\times U(m_{8})\times U(m_{9})\times 
                      U(m_{10})\times U(m_{11})\times U(m_{12})\\
       &\qquad \times U(m_{13})\times U(m_{14})\times U(m_{15})\times 
                      U(m_{16})\times U(m_{17})\times U(m_{18})\\
\sus   &(\Yfun_{1},\Yfun_{6}),\ (\Yfun_{1},\Yfunb_{6}),\ 
        (\Yfun_{1},\Yfunb_{7}),\ (\Yfun_{2},\Yfunb_{1}),\ 
        (\Yfun_{2},\Yfun_{5}),\ (\Yfunb_{2},\Yfunb_{6}),\\ 
       &(\Yfun_{2},\Yfunb_{8}),\ (\Yfun_{3},\Yfunb_{2}),\ 
        (\Yfun_{3},\Yfun_{4}),\ (\Yfunb_{3},\Yfunb_{5}),\ 
        (\Yfun_{3},\Yfunb_{9}),\ (\Yfun_{4},\Yfunb_{3}),\\ 
       &(\Yfun_{4},\Yfunb_{10}),\ (\Yfun_{5},\Yfunb_{4}),\ 
        (\Yfun_{5},\Yfunb_{11}),\ (\Yfun_{6},\Yfunb_{5}),\ 
        (\Yfun_{6},\Yfunb_{12}),\ (\Yfun_{7},\Yfunb_{2}),\\ 
       &(\Yfun_{7},\Yfunb_{12}),\ (\Yfun_{7},\Yfunb_{13}),\ 
        (\Yfun_{8},\Yfunb_{3}),\ (\Yfun_{8},\Yfunb_{7}),\ 
        (\Yfun_{8},\Yfunb_{14}),\ (\Yfun_{9},\Yfunb_{4}),\\
       &(\Yfun_{9},\Yfunb_{8}),\ (\Yfun_{9},\Yfunb_{15}),\ 
        (\Yfun_{10},\Yfunb_{5}),\ (\Yfun_{10},\Yfunb_{9}),\ 
        (\Yfun_{10},\Yfunb_{16}),\ (\Yfun_{11},\Yfunb_{6}),\\ 
       &(\Yfun_{11},\Yfunb_{10}),\ (\Yfun_{11},\Yfunb_{17}),\ 
        (\Yfun_{12},\Yfunb_{1}),\ (\Yfun_{12},\Yfunb_{11}),\ 
        (\Yfun_{12},\Yfunb_{18}),\ (\Yfun_{13},\Yfunb_{8}),\\ 
       &(\Yfunb_{13},\Yfunb_{14}),\ (\Yfun_{13},\Yfunb_{18}),\ 
        (\Yfun_{14},\Yfunb_{9}),\ (\Yfun_{14},\Yfunb_{13}),\ 
        (\Yfun_{14},\Yfun_{18}),\ (\Yfun_{15},\Yfunb_{10}),\\ 
       &(\Yfun_{15},\Yfunb_{14}),\ (\Yfun_{15},\Yfun_{17}),\ 
        (\Yfunb_{15},\Yfunb_{18}),\ (\Yfun_{16},\Yfunb_{11}),\ 
        (\Yfun_{16},\Yfunb_{15}),\ (\Yfunb_{16},\Yfunb_{17}),\\ 
       &(\Yfun_{17},\Yfunb_{12}),\ (\Yfun_{17},\Yfunb_{16}),\ 
        (\Yfun_{18},\Yfunb_{7}),\ (\Yfun_{18},\Yfunb_{17}),\ 
        \Yasymb_{1},\ \Yasymb_{4},\ \Yasym_{13},\ \Yasym_{16}\\ 
\hline
\end{array}$$
\caption{Spectrum of the $\bZ_6\times\bZ_6$ orientifold without discrete
         torsion and $(\mu_3,\mu_1)=(+1,-1)$. The numbers $m_i$ are 
         determined by solving the tadpole equations. There are many
         solutions. In general, the three sectors 99, $5_15_1$ and $5_25_2$
         have different gauge groups. The most symmetric solution is given 
         in the main text, eq.\ (\ref{Z6Z6solution}). In general, several 
         of the $m_i$ vanish because the total rank of the each of the 99 
         and $5_i5_i$ gauge groups must be 16.
         \label{Z6Z6spec4}}
\end{table}

\begin{table}[htp]
\renewcommand{\arraystretch}{1.25}
$$\begin{array}{|c|c|}
\hline
\multicolumn{2}{|c|}{\IZ_6\times\IZ_6,\ 
       (\alpha_1,\alpha_2,\alpha_3)=(+1,+1,-1),\ \eps=1}
   \\ \hline\hline
\multicolumn{2}{|c|}{\hbox{open string spectrum}}\\ \hline
\rm sector &\hbox{gauge group / matter fields}\\ \hline
5_35_3 &USp(n_{1})\times U(n_{2})\times U(n_{3})\times USp(n_{4})
           \times U(n_{5})\\
       &\qquad \times U(n_{6})\times U(n_{7})\times U(n_{8})
           \times U(n_{9})\times U(n_{10})\\
       &\qquad \times U(n_{11})\times U(n_{12})\times U(n_{13})
           \times U(n_{14})\times U(n_{15})\\
       &\qquad \times U(n_{16})\times USp(n_{17})\times U(n_{18})
            \times U(n_{19})\times USp(n_{20})\\
\nsus &\smult (\Yfun_{1},\Yfun_{2}),\ (\Yfun_{1},\Yfunb_{5}),\ 
              (\Yfun_{1},\Yfun_{10}),\ (\Yfun_{2},\Yfunb_{6}),\ 
              (\Yfun_{2},\Yfun_{9}),\ (\Yfunb_{2},\Yfunb_{10}),\\ 
      &\qquad (\Yfun_{3},\Yfunb_{2}),\ (\Yfunb_{3},\Yfun_{4}),\ 
              (\Yfun_{3},\Yfunb_{7}),\ (\Yfun_{3},\Yfun_{8}),\ 
              (\Yfunb_{3},\Yfunb_{9}),\ (\Yfun_{4},\Yfun_{7}),\\ 
      &\qquad (\Yfun_{4},\Yfunb_{8}),\ (\Yfun_{5},\Yfunb_{2}),\ 
              (\Yfun_{5},\Yfunb_{10}),\ (\Yfun_{5},\Yfunb_{11}),\ 
              (\Yfun_{6},\Yfunb_{3}),\ (\Yfun_{6},\Yfunb_{5}),\\
      &\qquad (\Yfun_{6},\Yfunb_{12}),\ (\Yfun_{7},\Yfunb_{6}),\ 
              (\Yfun_{7},\Yfunb_{13}),\ (\Yfun_{8},\Yfunb_{7}),\ 
              (\Yfun_{8},\Yfunb_{14}),\ (\Yfun_{9},\Yfunb_{8}),\\ 
      &\qquad (\Yfun_{9},\Yfunb_{15}),\ (\Yfun_{10},\Yfunb_{9}),\ 
              (\Yfun_{10},\Yfunb_{16}),\ (\Yfun_{11},\Yfunb_{6}),\ 
              (\Yfun_{11},\Yfunb_{16}),\ (\Yfun_{11},\Yfun_{17}),\\ 
      &\qquad (\Yfunb_{11},\Yfunb_{18}),\ (\Yfun_{12},\Yfunb_{7}),\ 
              (\Yfun_{12},\Yfunb_{11}),\ (\Yfunb_{12},\Yfun_{17}),\ 
              (\Yfun_{12},\Yfunb_{18}),\ (\Yfun_{13},\Yfunb_{8}),\\
      &\qquad (\Yfun_{13},\Yfunb_{12}),\ (\Yfun_{13},\Yfunb_{19}),\ 
              (\Yfun_{14},\Yfunb_{9}),\ (\Yfun_{14},\Yfunb_{13}),\ 
              (\Yfun_{14},\Yfun_{20}),\ (\Yfun_{15},\Yfunb_{10}),\\ 
      &\qquad (\Yfun_{15},\Yfunb_{14}),\ (\Yfun_{15},\Yfun_{19}),\ 
              (\Yfunb_{15},\Yfun_{20}),\ (\Yfun_{16},\Yfunb_{5}),\ 
              (\Yfun_{16},\Yfunb_{15}),\ (\Yfun_{16},\Yfun_{18}),\\ 
      &\qquad (\Yfunb_{16},\Yfunb_{19}),\ (\Yfun_{17},\Yfun_{18}),\ 
              (\Yfun_{18},\Yfunb_{13}),\ (\Yfun_{19},\Yfunb_{14}),\ 
              (\Yfun_{19},\Yfunb_{18}),\ (\Yfunb_{19},\Yfun_{20})\\
      &\spin \Yasym_1,\  adj_2,\  adj_3,\  \Yasym_4,\  adj_5,\ \ldots,\ 
             adj_{16},\ \Yasym_{17},\ adj_{18},\ adj_{19},\ \Yasym_{20}\\
\hline  
95_i,\ 5_i5_j &\hbox{many bifundamentals}\\
\hline\hline
\multicolumn{2}{|c|}{\hbox{closed string spectrum}}\\ \hline
\rm sector       &\hbox{$\cN=1$ multiplets}\\ \hline
\rm untw.        &\hbox{gravity, 1 lin., 3 chir.} \\ \hline
\hbox{order-two} &\hbox{4 lin., 8 chir.} \\ \hline
\rm remaining    &\hbox{54 lin., 15 chir.} \\ \hline
\end{array}$$
\caption{Spectrum of the $\bZ_6\times\bZ_6$ orientifold without discrete
         torsion and $(\mu_3,\mu_1)=(+1,-1)$. (continued)
         \label{Z6Z6spec4b}}
\end{table}

\section{Conclusions}
We have constructed a large number of compact $\IZ_N\times\IZ_M$ orientifold
models, with $N$, $M$ even. These models are labelled by three signs: the
vector structure associated to the two generators of the orbifold group and 
the discrete torsion, \ie eight models for each $\IZ_N\times\IZ_M$.

Some of these models can be supersymmetric, \ie only $D$-branes need to be
added in order to cancel the tadpoles. All supersymmetric models have 
discrete torsion and no vector structure in both generators of the 
orbifold group. These models are the $\IZ_2\times\IZ_2$, 
$\IZ_2\times\IZ_6$, $\IZ_2\times\IZ_6'$, $\IZ_6\times\IZ_6$.

We have extended the list of consistent orientifolds adding some models
that need antibranes to cancel the RR tadpoles, along the lines of 
\cite{aaads}. For most of these models there is at least one solution.
However in one of the $\IZ_2\times\IZ_4$ models and in one of the
$\IZ_4\times\IZ_4$ models without Wilson lines, there is an 
incompatibility between the twisted tadpoles related to different 
fixed points.

\vskip3cm

\centerline{\bf Acknowledgements}
It is a pleasure to thank Luis~Ib\'a\~nez, Augusto Sagnotti and 
Angel Uranga for many helpful discussions. 
The work of M.K.\ is supported by a TMR network of the European Union, 
ref. FMRX-CT96-0090. The work of R.R.\ is supported by the MEC through 
an FPU Grant.

\newpage
\begin{center}\huge\bf Appendix \end{center}
\begin{appendix}

\section{How to obtain the Hodge numbers of orbifolds with and 
         without discrete torsion} 
\label{app_hodge}

In this section, we explain how to compute the Hodge numbers of compact
orbifold spaces $T^6/\Gamma$, where the six-torus $T^6$ is of the form
$\IC^3/\Lambda$ and $\Gamma$ is an Abelian finite group. The Hodge number
$h^{p,q}$ is defined as the number of independent harmonic $(p,q)$-forms
that can be defined on this space. 

An element $g\in\Gamma$ acts on the three complex coordinates of 
$\IC^3/\Lambda$ as
\be  \label{g_action}
g:\ (z_1,z_2,z_3)\ \to\ (e^{2\pi iv_g^{(1)}}z_1,e^{2\pi iv_g^{(2)}}z_2,
                                               e^{2\pi iv_g^{(3)}}z_3),
\qquad{\rm with\ } 0\le v_g^{(i)}<1.
\ee

We will be interested in the separate contributions of each twisted sector
to $h^{p,q}$:
\be \label{hodge_contrib}
h^{p,q}\ =\ h^{p,q}_{\rm untw}\ +\ \sum_{g\in\Gamma\setminus\{e\}}h^{p,q}_g.
\ee
The contribution of the untwisted sector is just the number of 
$\Gamma$-invariant harmonic $(p,q)$-forms that can be defined on $T^6$.
The forms on $T^6$ are generated by
$$\matrix {& & & &1& & &  \cr
           & & &dz_i&& d\bar{z}_i& &  \cr
           & &dz_idz_j& &dz_id\bar{z}_j& &d\bar{z}_id\bar{z}_j&  \cr
           &dz_idz_jdz_k& &dz_idz_jd\bar{z}_k& &dz_id\bar{z}_jd\bar{z}_k& 
                 &d\bar{z}_id\bar{z}_jd\bar{z}_k \cr
           & & & &\vdots& & &   }$$           
The contributions of the twisted sectors are due to the singularities of 
the orbifold which arise because $T^6$ has fixed points or fixed planes 
under the action of the elements of $\Gamma$. We split the twisted sectors 
into two sets:

a) Sectors twisted by $g$, where $g$ is a group element that only has fixed
points, but no fixed planes. This sector only contributes to $h^{1,1}$ or
$h^{2,2}$. If $\sum_{i=1}^3 v_g^{(i)}=1$, then \cite{imnq,vw}
\beq  \label{hg11}
h_g^{1,1}\ =\ {1\over|\Gamma|} \sum_{h\in\Gamma} \beta_{g,h}\, 
                 \chi(g,h),\qquad h_g^{2,2}\ =\ 0,
\eeq
where $|\Gamma|$ is the number of elements of the discrete group, 
$\beta_{g,h}$ is the discrete torsion phase and $\chi(g,h)$ is the 
Euler characteristic of the subspace (i.e.\ the set of fixed points)
left simultaneously fixed by $g$ and $h$. In our case, $\chi(g,h)$ is
the number of points that are simultaneously fixed by $g$ and $h$.
If $\sum_{i=1}^3 v_g^{(i)}=2$, then this sector contributes the same 
value, but to $h_g^{2,2}$, and $h_g^{1,1}=0$.

The sector twisted by $g^{-1}$ gives the same contribution to 
$h^{2,2}$ as the $g$-twisted sector to $h^{1,1}$ and vice versa:
\beq  \label{hginverse}
h_{g^{-1}}^{2,2}\ =\ h_g^{1,1},\qquad h_{g^{-1}}^{1,1}\ =\ h_g^{2,2}.
\eeq

b) Sectors twisted by $g$, where $g$ is a group element that has fixed
planes. This sector gives a contribution to  $h^{1,1}$, $h^{1,2}$, $h^{2,1}$ 
and $h^{2,2}$ of the following form \cite{imnq,vw}:
\beqa  \label{hg12}
h_g^{2,2} \ =\ h_g^{1,1} &= &{1\over|\Gamma|} \sum_{h\in\Gamma} \beta_{g,h}\, 
                             \widetilde\chi(g,h), \nonumber\\
h_g^{1,2} &= &{1\over|\Gamma|} \sum_{h\in\Gamma} \beta_{g,h}\,  
               \widetilde\chi(g,h)\, e^{2 \pi iv_h(g)},  \\
h_g^{2,1} &= &{1\over|\Gamma|} \sum_{h\in\Gamma} \beta_{g,h}\,
               \widetilde\chi(g,h)\, e^{-2 \pi iv_h(g)}. \nonumber   
\eeqa
where $v_h(g)=v_h^{(i)}$ if the $i\th$ plane is fixed by $g$, and
$\widetilde\chi(g,h)$ is the Euler characteristic of the subspace that
is simultaneously fixed under $g$ and $h$ and that is contained in the 
two internal complex planes which are not fixed under $g$ (see \cite{imnq} 
for a proper definition).
The phase that appears in the  formulae for $h_g^{1,2}$ and  $h_g^{2,1}$ 
corresponds to the phase acquired by the forms $dz_i$ and $d\bar{z}_i$ 
(defined on the plane that is fixed by $g$) under a twist by $h$. 
Note that in the case with discrete torsion the 
contributions to the Hodge numbers  $h^{1,2}$ and  $h^{2,1}$ can be different.
But one always has: $h_g^{1,2}+h_{g^{-1}}^{1,2}=h_g^{2,1}+h_{g^{-1}}^{2,1}$.

To illustrate this method, let us analyse the $\IZ_2\times\IZ_4$ orbifold. 
From the untwisted sector, one has the universal contribution (i.e.\ present 
in any orbifold): $h^{0,0}=h^{3,3}=h^{3,0}=h^{0,3}=1$. The contribution 
of each sector to $h^{1,1}$, $h^{2,2}$, $h^{1,2}$ and  $h^{2,1}$
in the cases with and without discrete torsion is shown in table
\ref{hodge_example}. 

\begin{table}[ht]
\renewcommand{\arraystretch}{1.25}
$$\ba{|c|c||c|c|c|c||c|c|c|c|}
\hline
\multicolumn{2}{|c||}{\IZ_2\times\IZ_4}
&\multicolumn{4}{|c||}{\eps=1}
&\multicolumn{4}{|c|}{\eps=-1}\\
\hline
g &\hbox{fixed plane} &h_g^{1,1} &h_g^{2,2} &h_g^{1,2} &h_g^{2,1}
                      &h_g^{1,1} &h_g^{2,2} &h_g^{1,2} &h_g^{2,1}\\
\hline\hline
(0,0) &{\rm all}  &3  &3  &1  &1    &3  &3  &1  &1\\
(0,1) &1^{\rm st} &4  &4  &-  &-    &-  &-  &4  &4\\
(0,2) &1^{\rm st} &10 &10 &-  &-    &10 &10 &-  &-\\
(0,3) &1^{\rm st} &4  &4  &-  &-    &-  &-  &4  &4\\
(1,0) &3^{\rm rd} &12 &12 &-  &-    &4  &4  &-  &-\\
(1,1) &-          &-  &16 &-  &-    &-  &-  &-  &-\\
(1,2) &2^{\rm nd} &12 &12 &-  &-    &4  &4  &-  &-\\
(1,3) &-          &16 &-  &-  &-    &-  &-  &-  &-\\
\hline\hline
{\rm total} &     &61 &61 &1  &1    &21 &21 &9  &9\\
\hline
\ea$$
\caption{Hodge numbers of the $\bZ_2\times\bZ_4$ orbifold with and without
discrete torsion. \label{hodge_example}}
\end{table}

\section{Closed string spectrum from shifts}  
\label{app_closed}

In this appendix we shall confirm the result on the closed string spectrum
of type IIB and type I orbifolds that was obtained by analysing the 
cohomology as described in section \ref{closed}. Let us consider the RR part 
of the spectrum.

\noindent We divide the twisted sectors into two sets:

\noindent a) Sectors without fixed tori:

Let $v=(0,v_1,v_2,v_3)$ be the shift vector corresponding to this sector, with 
$0\le v_i<1$ . The vacuum energy corresponding to this sector is
\beq
E_0 \ =\ \frac{1}{2} \sum_i v_i (1-v_i).
\eeq
The RR states are characterised by $SO(8)$ weight vectors of the form
$r=(\pm\half,\pm\half,\pm\half,\pm\half)$, with an odd number of minus
signs because of the GSO projection. The mass of a state characterised
by the weight vector $r_v$ is given by
\beq
M^2 \ =\ \frac{(r_v+v)^2}{2} - \frac{1}{2} +E_0.  
\eeq
There is only one massless state (having imposed the GSO projection):
$r_v=(\half,-\half,-\half,-\half)$. 

In type IIB string theory, there is only one RR state in this sector 
$r_{v,L} \otimes r_{v,R}$, with helicity $\chi = r_{v,L}^0 - r_{v,R}^0 = 0$. 
This is a scalar or its dual 2-form. 
From the sector with shift $-v$, one obtains another solution: 
$r_{-v}=(-\half,+\half,+\half,+\half)$. 
This gives another scalar (or a 2-form). 
The degeneracy of these states can be obtained from the partition function,
as explained in \cite{imnq,fiq}.
For the total number of states, one finds:
\beq  \label{nscal}
n_{\rm scal}(v) \ =\ \frac{2}{|\Gamma|} \sum_{h\in\Gamma}  \beta_{v,h}\, 
  \chi(v,h),  
\eeq
where it is understood that we only sum over half the shifts to get
the total number of scalars (because we combined $v$ and $-v$).
This corresponds to the $h^{1,1}_v$ scalars and 2-forms (see the formula
(\ref{hg11}) for the Hodge numbers $h^{1,1}_g$) coming from the reduction of 
the RR 2-form and 4-form. Adding the scalars from the NSNS sector and the
corresponding fermions, we obtain $h^{1,1}_v$ \N2 tensor multiplets, as 
predicted by the cohomology computation.

In type I string theory, there is an $\Omega$ projection that exchanges 
left-movers and right-movers. It acts together with a $J$ operation that 
exchanges $v$ and $-v$. As we are in the RR sector, we must take the 
antisymmetric combinations: 
$r_{v,L} \otimes r_{v,R} - r_{-v,L} \otimes r_{-v,R}$. 
Only a scalar (or a 2-form) remains. The degeneracy is again $h^{1,1}_v$,
i.e.\ half the value of (\ref{nscal}), and coincides with the result from 
the cohomology computation. Together with the NSNS and fermionic states,
one finds $h^{1,1}_v$ \N1 linear multiplets.

\noindent b) Sectors with a fixed torus:

Let $v=(0,0,v_2,v_3)$ be the shift corresponding to this sector, with 
$0\le v_i<1$ . The first torus is fixed.

There are two massless states: 
$r_{+}=(\half,-\half,\half,\half)$
and $r_{-}=(-\half,\half,\half,\half)$. 
(The GSO projection has been imposed.)

In type IIB string theory, there are four RR states in this sector:
\begin{itemize}
\item $r_{+,L} \otimes r_{+,R}$ \quad
      with helicity: $\chi = r_L^0 - r_R^0 = 0$, 
\item $r_{-,L} \otimes r_{-,R}$ \quad
      with helicity: $\chi = r_L^0 - r_R^0 = 0$,
\item $r_{+,L} \otimes r_{-,R}$ \quad
      with helicity: $\chi = r_L^0 - r_R^0 = 1$,
\item $r_{-,L} \otimes r_{+,R}$ \quad
      with helicity: $\chi = r_L^0 - r_R^0 = -1$.
\end{itemize}
This is a pair of scalars (or 2-forms) and a vector. The number of scalars is:
\beq  \label{nscal_fix}
n_{\rm scal}(v) \ =\ \frac{2}{|\Gamma|} \sum_{h\in\Gamma} \beta_{v,h}\, 
  \widetilde{\chi}(v,h).
\eeq
This corresponds to the $h^{1,1}_v$ scalars and 2-forms (see the formula 
(\ref{hg12}) for the Hodge numbers $h^{1,1}_g$) coming from the reduction 
of the RR 2-forms and 4-forms, as predicted by the cohomology computation.
The number of vectors is:
\beq  \label{nvec_fix}
n_{\rm vec}(v) \ =\ \frac{1}{|\Gamma|} \sum_{h\in\Gamma} \beta_{v,h}\,
\widetilde{\chi}(v,h)\, e^{2 \pi i v_h^{(1)}},  
\eeq
where $v_h^{(1)}$ is the first component of the shift vector corresponding
to the twist $h$ (in general one has to take $v_h^{(i)}$ if the $i\th$ torus
is fixed). These are $h^{1,2}_v$ vectors from the reduction of the RR 4-form.

In type I string theory, one must distinguish between the sectors of order 
different from two and of order two. The latter are mapped onto themselves
under $J$. For sectors that are not of order two, the sectors $v$ and $-v$ 
are combined under the $J$ operation, leading to four linear combinations
of type IIB RR states:
\begin{itemize}
\item $r_{+,L}^v \otimes r_{+,R}^v\ -\ r_{-,L}^{-v} \otimes r_{-,R}^{-v}$ 
  \quad with helicity: $\chi = r_L^0 - r_R^0 = 0$,
\item $r_{-,L}^v \otimes r_{-,R}^v\ -\ r_{+,L}^{-v} \otimes r_{+,R}^{-v}$ 
  \quad with helicity: $\chi = r_L^0 - r_R^0 = 0$,
\item $r_{+,L}^v \otimes r_{-,R}^v\ -\ r_{+,L}^{-v} \otimes r_{-,R}^{-v}$ 
  \quad with helicity: $\chi = r_L^0 - r_R^0 = 1$,
\item $r_{-,L}^v \otimes r_{+,R}^v\ -\ r_{-,L}^{-v} \otimes r_{+,R}^{-v}$ 
  \quad with helicity: $\chi = r_L^0 - r_R^0 = -1$.
\end{itemize}

Note that, because of the $J$ operation, the pairing of states we have 
performed above is only possible if $h^{1,2}_v=h^{2,1}_v$. From 
(\ref{nvec_fix}), one finds that this is equivalent to the statement 
that the discrete torsion $\beta_{v,h}$ only takes real
values, i.e.\ $\beta_{v,h}=\pm 1$.

If the sector is of order two, then $v$ is identical to $-v$. Of the four 
states that survive in the general case only one remains:
\begin{itemize}
\item $r_{+,L}^v\otimes r_{+,R}^v\ -\ r_{-,L}^{-v}\otimes r_{-,R}^{-v}$ 
  \quad with helicity: $\chi = r_L^0 - r_R^0 = 0$.
\end{itemize}
This corresponds to the $h^{1,1}_v$ scalars from the reduction of the 
RR 2-form. 

We see that the results from the shift formalism coincide with those obtained 
from the cohomology computation.

\end{appendix}

\end{document}